\documentclass[11pt]{article}
\usepackage{amssymb,dsfont,amsmath,amsfonts,hyperref,mdframed,mathrsfs,stmaryrd}
\usepackage[active]{srcltx}
\usepackage{slashed}
\usepackage{color}
\usepackage{amsthm}
\usepackage{scrtime}
\usepackage[normalem]{ulem}
\usepackage{framed}
\usepackage{cancel}
\usepackage{comment}
\usepackage{empheq}
\usepackage{rotating}
\usepackage{pdflscape}
\usepackage{afterpage}

\advance \textheight by \topskip
\let\oldsqrt\sqrt

\def\sqrt{\mathpalette\DHLhksqrt}
\def\DHLhksqrt#1#2{%
\setbox0=\hbox{$#1\oldsqrt{#2\,}$}\dimen0=\ht0
\advance\dimen0-0.2\ht0
\setbox2=\hbox{\vrule height\ht0 depth -\dimen0}%
{\box0\lower0.4pt\box2}}

\def\be{\begin{equation}}
\def\ee{\end{equation}}
\def\bea{\begin{eqnarray}}
\def\eea{\end{eqnarray}}

\def\a{\alpha}
\def\b{\beta}

\def\d{\delta}
\def\e{\epsilon}
\def\k{\kappa}
\def\l{\lambda}
\def\m{\mu}
\def\n{\nu}
\def\s{\sigma}
\def\Comp{\mathbb{C}}
\def\Real{\mathbb{R}}

\def\ua{\underline{\alpha}}
\def\ub{\underline{\beta}}
\def\uc{\underline{\gamma}}
\def\ud{\underline{\delta}}

\def\para{\paragraph}

\input{glyphtounicode}
\newcommand{\eq}[1]{(\ref{#1})}
\newcommand{\ww}[1]{\\[0.#1cm]}
\newcommand{\nn}{\nonumber}

\newcommand{\bse}{\begin{subequations}}
\newcommand{\ese}{\end{subequations}}

\usepackage{everypage}

\newcommand{\Lpagenumber}{\ifdim\textwidth=\linewidth\else\bgroup
	\dimendef\margin=0
	\ifodd\value{page}\margin=\oddsidemargin
	\else\margin=\evensidemargin
	\fi
	\raisebox{\dimexpr -\topmargin-\headheight-\headsep-0.5\linewidth}[0pt][0pt]{%
		\rlap{\hspace{\dimexpr \margin+\textheight+\footskip}%
			\llap{\rotatebox{90}{\thepage}}}}%
	\egroup\fi}
\AddEverypageHook{\Lpagenumber}%

\begin{document}

\begin{flushright}
 MI-TH-1772\\
\end{flushright}

\vspace{10pt}

\begin{center}


\Large{\textbf {FRW and Domain Walls in Higher Spin Gravity}}

\end{center}

\vspace{0.1cm}

\begin{center}

\large{ R. Aros$^{\rm a}$, C. Iazeolla$^{\rm b}$,  J. Nore\~na$^{\rm c}$, E. Sezgin$^{d}$, P. Sundell$^{\rm a}$ and Y. Yin$^{\rm a}$  }\\
\vskip 1cm

\small{
\textit{ $^{\rm a}$ Departamento de Ciencias F\'isicas, Universidad Andres Bello\\ Republica 220, Santiago de Chile} 
}

\small{
\textit{ $^{\rm b}$  NSR Physics Department , G. Marconi University \\  via Plinio 44, Rome, Italy} 
} 

\small{
\textit{   $^{\rm c}$ Instituto de F\'isica, Pontificia Universidad Cat\'olica de Valpara\'iso, \\
Casilla 4059, Valpara\'iso, Chile} 
} 

\small{
\textit{   $^{\rm d}$   Mitchell Institute for Fundamental Physics and Astronomy\\ Texas A\&M University
College Station, TX 77843, USA } 
} 

{\let\thefootnote\relax\footnote{\tt raros@unab.cl, c.iazeolla@gmail.com, jorge.norena@pucv.cl, sezgin@tamu.edu, per.anders.sundell@gmail.com, yinyihao@gmail.com}}
\setcounter{footnote}{0}

\vspace{.1cm}

\end{center}


\centerline{\bf  Abstract}

\bigskip

We present exact solutions to Vasiliev's bosonic higher spin gravity
equations in four dimensions with positive and negative cosmological constant that admit an interpretation in terms of domain walls, quasi-instantons 
and Friedman-Robertson-Walker (FRW) backgrounds. 
Their isometry algebras are infinite dimensional higher-spin extensions of spacetime isometries generated by six Killing vectors. 
The solutions presented are obtained by using  a method of holomorphic 
factorization in noncommutative twistor space and gauge functions. 
In interpreting the solutions in  terms of Fronsdal-type fields in spacetime, 
a field-dependent higher spin transformation is required, which is implemented at leading order. 
To this order, the scalar field solves Klein-Gordon equation with conformal mass in $(A)dS_4.$
We interpret the FRW solution with de Sitter asymptotics in the context of inflationary cosmology and we expect that the domain wall and FRW solutions are associated with spontaneously broken scaling symmetries in their holographic description.
We observe that the factorization method provides a convenient framework for setting up a perturbation theory around the exact solutions, and we propose that the nonlinear completion of particle excitations over FRW and domain wall solutions requires black hole-like states.

\bigskip
\bigskip
\bigskip

\newpage

{\tableofcontents }

\numberwithin{equation}{section}

\newpage

\section{Introduction}

Vasiliev's theory in four dimensions \cite{vasiliev} has so far been studied mainly 
around its maximally symmetric anti-de Sitter vacuum.
%
%
The perturbations around the anti-de Sitter spacetime describe an unbroken phase of the theory, with spectrum given by infinite towers of massless fields, corresponding to conserved higher spin currents of dual free conformal field theories in three dimensions \cite{Sezgin:2002rt,Klebanov:2002ja,Sezgin:2003pt}.
Higher spin gravity is well known to admit a cosmological term of positive sign and de Sitter vacuum solution as well. It has been proposed that the parity invariant minimal version of  higher spin  $dS_4$ gravity is holographically dual to the three dimensional conformal field theory of an Euclidean $Sp(N)$ vector model with anticommuting scalars residing at the boundary of $dS_4$ at future timelike infinity \cite{Anninos:2011ui}. For further developments in this direction, see \cite{Anninos:2012ft,Banerjee:2013mca,Anninos:2013rza,Anninos:2014hia,Hertog:2017ymy,Neiman:2017zdr,Anninos:2017eib}. 
These studies mostly exploit the higher spin symmetries.  On the other hand, a detailed bulk description of the early universe physics, including the inflationary era, requires understanding of 
accelerating solutions of Vasiliev theory and cosmological perturbations around them. Such solutions have isometries forming a subgroup of the de Sitter spacetime symmetries.

Higher spin gauge symmetries can be broken by quantum \cite{Girardello:2002pp} as well as classical effects.
In the latter case, a simple mechanism is to replace the maximally symmetric vacuum
by vacua with six Killing symmetries forming a Lie algebra $\mathfrak g_6$, as summarized in 
Table \ref{Table:1}\footnote{The various vacua possess unbroken higher spin symmetries;
the unbroken symmetry algebra of the $\mathfrak g_6$-invariant 
vacua is the intersection of the enveloping algebra of $\mathfrak g_6$ with the
unbroken symmetry algebra of the maximally symmetric vacua (see \cite{Sezgin:2005pv} for the case of an $\mathfrak{so}(1,3)$-invariant solution), which is given by the quotient of
the enveloping algebra of the $(A)dS_4$ Killing symmetry algebra over the 
two-sided ideal given by the singleton annihilator.}. 
These correspond to the isometries of domain walls, FRW-like solutions 
and quasi-instantons\footnote{The quasi-instantons are Lorentzian counter
parts of instantons, which can be viewed as the results of gluing together
a domain wall and a FRW-like geometry \cite{Sezgin:2005pv,Sezgin:2005hf}.}. 
While we shall leave to a future work an analysis of the the holographic aspects 
of the exact  solutions that we present here, we propose to interpret 
the domain walls as bulk duals of vacua of three-dimensional massive quantum 
field theories arising through spontaneous breaking of conformal (higher spin) 
symmetries; for a relatively recent  study of spontaneous breaking of scale 
invariance in certain CFTs in $D=3$, see \cite{Bardeen:2014paa}.

In this paper, we shall use a solution generating 
technique \cite{Iazeolla:2011cb,Iazeolla:2017vng,Iazeolla:2017dxc}  
to build  $\mathfrak g_6$-invariant solutions to Vasiliev's bosonic theory 
with non-vanishing (positive or negative) cosmological constant from gauge 
functions, representing large gauge transformations that alter the asymptotics 
of the gauge fields, and $\mathfrak g_6$-invariant scalar field 
profiles in the maximally symmetric background. Solutions of Vasiliev's equations 
with $\mathfrak g_3, \mathfrak g_4$ and $\mathfrak g_6$ symmetries, which are subgroups of the
$AdS_4$ symmetry group, were constructed only at the linearized level in \cite{Sezgin:2005pv} 
(see \cite{Iazeolla:2017dxc} for a review) by using a 
different technique. The fully non-linear solutions presented in this paper are instead obtained by using 
a different method based on a holomorphic factorization ansatz, and in what 
we refer to as the holomorphic and $L$-gauges, described in Section \ref{SecConstrSoln}. 
In furnishing an interpretation of the solution in 
terms of Fronsdal-type fields in spacetime, however, a higher spin transformation 
needs to be implemented order by order in weak fields to reach what we refer 
to as the Vasiliev gauge, also discussed in Section \ref{SecConstrSoln}. We have implemented this 
gauge transformation only at leading order in this paper, leaving the computation 
of higher order terms to a future work. As we shall see in Section \ref{SecCosm}, an important advantage of the 
method we have used to obtain the exact solutions in the holomorphic gauge is the 
validity of linear superposition principle in constructing solutions, thus facilitating 
the description of fluctuations around an exact solution. 
Even though we leave to future work the analysis of a cosmological perturbation theory around our solutions, an inspection of the star product algebra among the master field will lead us to propose that the nonlinear completion of particle excitations over FRW and domain wall solutions requires black hole-like states (see \cite{Iazeolla:2017vng} for the study of scalar particle fluctuations over higher-spin black hole modes).

Among all solutions we have found, we shall, in particular, take a closer look at
the FRW-like solution with $\mathfrak{iso}(3)$ symmetry and positive cosmological constant.
We will provide a perturbative procedure for obtaining the solutions in 
the Vasiliev gauge mentioned above, to any order in a  suitable perturbation parameter that breaks the de Sitter symmetry to $\mathfrak{iso}(3)$. 
On the solutions, the scalar field, whose value is vanishing in de Sitter vacuum, is turned on at first order in the symmetry-breaking parameter, and the metric gets corrected at the second order. Moreover, at linear order the fields with spins $s>2$ vanish in  the background solution. 
Whether they arise in higher orders remains to be determined.  At linear order the scalar field behaves similarly to a 
conformally coupled scalar  field in $dS_4$. In Section \ref{SecCosm}, we shall compare its behaviour 
with that of the inflaton in the standard cosmological scenarios.

The FRW-like solutions are intriguing because if higher 
spin fluctuation fields are suppressed by the background, then they may yield
cosmologically viable models based on Vasiliev's theory, opening up a new 
window for embedding the standard models of particles and cosmology into 
higher spin theory, which may be viewed as the unbroken phase of string theory 
in which the string is tensionless \cite{sundborg,Sezgin:2002rt,perjohan,Gaberdiel:2014cha,minwalla}. This setting will inevitably involve the 
coupling of an infinite number of (massive) higher spin multiplets. One may envisage 
a scenario in which their presence will play a role in the resolution of the 
initial singularity, and near the end or after the inflation when the breaking 
of higher spin symmetry is expected to take place. 
A much bolder proposal would be the consideration of only massless higher spin 
theory with its matter couplings furnished through the Konstein-Vasiliev or 
supersymmetric extension of Vasiliev theory  (see \cite{Sezgin:2012ag} for a 
survey). Such a proposal is motivated by the high degree of symmetry that may 
yield a UV finite theory, and by the availability of a mechanism \cite{Girardello:2002pp} 
for breaking  of higher spin symmetries by quantum effects without the need to introduce 
fields other than those present in the theory, whose spectrum consists of the 
two-fold product of the singleton representation of the $AdS_4$ group. 
Thus it is natural to consider the (matter coupled) higher spin theory as the candidate for a tensionless limit of string theory, in which all the massive trajectories are decoupled completely, and to investigate its consequences for the early universe physics. 
There are very powerful no-go theorems 
that forbid accelerating spacetimes in string theory in its tensile phase 
(see \cite{Kutasov:2015eba} and references therein), inviting the considerations 
of non-perturbative and string loop effects in a full-fledged formulation of 
string field theory, and finding its vacuum solutions. On the other hand, higher 
spin theory can be viewed as a much simpler  version of string field theory, 
in which finding asymptotically de Sitter vacua is a more amenable problem. 

The introduction of matter and higher spin symmetry breaking  remain a largely 
uncharted terrain. These aspects are expected to play key roles either for 
reheating in an inflationary  scenario or an analogous mechanisms in 
non-inflationary scenarios. 
In the simplest inflation model in standard cosmology, Einstein gravity 
and a single real scalar field with a suitable potential dominate the 
early inflationary phase. 
Here we instead envisage a scenario in which the Einstein plus scalar system is 
replaced by the bosonic Vasiliev higher spin theory, which consists of 
a coupled set of massless fields with all integer spins $s=0,1,2,3,..\infty$. 
One can then try to employ the well-known mechanism whereby rapidly inflated 
fluctuation modes with wavelengths larger than the Hubble length freeze and 
subsequently re-enter the cosmological horizon after inflation has ended. 
Assuming that higher spin symmetry breaking and reheating take place
at around the same time, one can compute the effects of higher spin 
fluctuations on the CMB observations at large scales. 
In these scenarios, it is important to keep in mind that while the 
higher spin modes may dissipate in time, their couplings to and 
mixing with the gravitational field may have observable effects. 
Some studies have already been done along these lines, see e.g. \cite{Arkani-Hamed:2015bza,Lee:2016vti,Kehagias:2017cym}, but 
based on assumptions on higher spin dynamics not born out of Vasiliev's theory.
Let us also note that the analog of the $\mathfrak{so}(1,3)$ invariant solution, referred 
to as the ``instanton" solution in Table \ref{Table:1},
was obtained as an exact solution for $\Lambda<0$ in \cite{Sezgin:2005hf} and for $\Lambda>0$ as well in
\cite{Iazeolla:2007wt}. In the case of $\Lambda<0$, a cosmological implications of the solution has been 
discussed in \cite{Sezgin:2005hf} where it has been argued that it leads to a bouncing cosmology, in some respects reminiscent of the work of \cite{Hertog:2004rz} based on supergravity considerations\footnote{An analogue Lorentz-invariant instanton solutions, with additional twisted sectors of the theory excited, and the characteristic extra deformation parameter $\l$ that allows to vary the mass of the scalar, was also found for the Prokushkin-Vasiliev theory in $D=3$ in \cite{Iazeolla:2015tca}, where also its cosmological interpretation was discussed and its holographic study initiated.}.

This paper is organized as follows: 
In Section \ref{SecVasModel}, we review Vasiliev's  higher spin gravity 
equations. They are formulated in terms of master one-form $A$ and master Weyl zero-form $\Phi$ which live on a base manifold ${\cal X}_4\times {\cal Z}_4$ with coordinates $(x^\mu, Z^{\ua})$ where ${\cal Z}_4$ is a non-commutative real four manifold. The master fields also depend on the coordinates of the fiber space ${\cal Y}_4$ with coordinates $Y^{\ua}$. In Section \ref{SecConstrSoln}, we describe the construction of the exact solutions 
with ${\mathfrak g}_6$ symmetries. 
For the reader's convenience we summarize the solutions here. The master fields are the zero-form $\Phi$ and one-form $A$ whose components are displayed in \eq{1f}. In holomorphic gauge, $\Phi'$ is given in Table \ref{Table:1}, and $A'_\a$ in \eq{s1} and \eq{aa}. In the $L$-gauge, $\Phi^{(L)}$ is given in \eq{defO} and \eq{c123}, $A^{(L)}_\a$ is given in \eq{tcs1} and \eq{scodd}, and $W_\mu^{(L)}$ in 
\eq{WLG}. In Vasiliev gauge, the linear order results for 
$\Phi^{(G,1)}$ is given in \eq{Gsc}, $A_\a^{(G,1)}$ in \eq{Gsc2} and $A_\mu^{(G,1)}$ is given by $(A)dS$ connection with a detailed discussion of $G$-gauge transformations given in \ref{STC} and Appendix \ref{App:H1}.
In Section \ref{SecRegular}, we examine the regularity of the Weyl zero-form.
The scalar field profiles $\phi(x)$ are described in a unified manner by using stereographic coordinate system. In studying their regularity, one needs to distinguish between singularities that are gauge artifacts and genuine singularities in the full $(x,Y,Z)$ space, sometimes referred to as the correspondence space. To this end, one needs to study the solution $\Phi(x,Y,Z)$ for the Weyl zero-form, and associated higher spin invariant and the on-shell conserved zero-form charges, as we shall discuss further in Section \ref{SecRegular}.
In Section \ref{SecCosm}, we take a closer look at the $\mathfrak{iso}(3)$ invariant solution and 
compare with the  standard cosmological backgrounds. In Section \ref{SecConclusions}, we summarize our  results and comment on future directions. 
Frequently used symbols and notation are summarized in Appendix \ref{SecConvention}. Various coordinates systems used to describe $(A)dS$ and the associated Killing vectors are given in Appendix \ref{SecCoord}. The gauge functions used in the construction of the exact solutions are described in Appendix \ref{SecGaugeFunc}. Details of the passage to Vasiliev gauge in leading order are given in Appendix \ref{App:H1}, and useful formula in the description of twistor space distributions and the star products of relevant projector operators are provided in Appendix \ref{App:Lemmas}.

\section{Bosonic Vasiliev model}
\label{SecVasModel}

In what follows, we review the basic properties of Vasiliev's equations \cite{Vasiliev:1999ba}
and their classical solution spaces, including boundary conditions in spacetime and twistor space suitable for asymptotically (anti-)de Sitter solutions. For a recent review of the exact solutions see 
\cite{Iazeolla:2017dxc}.


\subsection{Review of the full equations of motion}

\subsubsection{Non-commutative space}

Vasiliev's theory is formulated in terms of  
horizontal forms on a non-commutative fibered 
space ${\cal C}$ with four-dimensional non-commutative symplectic 
fibers and eight-dimensional base manifold equipped
with a non-commutative differential Poisson structure.
On the total space, the differential form algebra $\Omega({\cal C})$ is
assumed to be equipped with an associative degree preserving 
product $\star$, a differential $d$, and an Hermitian conjugation 
operation $\dagger$, that are assumed to be mutually compatible 
in the sense that if $f,g,h\in\Omega({\cal C})$, then
\be (f\star g) \star h=f\star (g\star h)\ ,\ee
\be d(df)=0\ ,\qquad d(f\star g)=(df)\star g
+(-1)^{|f|}f\star (dg)\ ,\ee
\be (df)^\dagger=d(f^\dagger)\ ,\quad
(f\star g)^\dagger=(-1)^{|f||g|}(g^\dagger)\star
(f^\dagger)\ ,\ee
where $|f|$ denotes the form degree of $f$.
We shall also assume that\footnote{More generally, $\dagger\circ\dagger$
can be an automorphism of $\Omega({\cal C})$, which is of relevance, for example,
in the case of models in de Sitter signature with fermions.}
\be (f^\dagger)^\dagger=f\ .\ee
It is furthermore assumed that $\Omega({\cal C})$ contains 
a horizontal subalgebra, $\Omega_{\rm hor}({\cal C})$, consisting
of equivalence classes defined using a globally defined closed
and central hermitian top-form on the fiber space, and whose product, 
differential and hermitian conjugation operation we shall denote by
$\star$, $d$ and $\dagger$ as well. 

The base manifold is assumed to be the direct product of a 
commuting real four-manifold ${\cal X}_4$ with coordinates $x^\mu$,
and a non-commutative real four-manifold ${\cal Z}_4$ with coordinates
$Z^{\ua}$; the fiber space and its coordinates are denoted by ${\cal Y}_4$
and $Y^{\ua'}$, respectively.
The non-commutative coordinates are assumed to obey
\be 
[Y^{\ua'},Y^{\ub'}]_\star=2iC^{\underline{\a'\b'}}\ ,\qquad 
[Z^{\ua},Z^{\ub}]_\star=-2iC^{\underline{\a\b}}\ ,\qquad
[Y^{\ua},Z^{\ub'}]_\star=0\ ,
\label{yza}
\ee
and the differential Poisson structure is assumed to be
trivial in the sense that
\be [Y^{\ua'},dY^{\ub}]_\star=[Z^{\ua},dY^{\ub'}]_\star= [Z^{\ua},dZ^{\ub}]_\star=[Z^{\ua},dZ^{\ub}]_\star=0\ .\ee
The star product is defined in \eq{star}. 
The non-commutative space is furthermore assumed to have
a compatible complex structure, such that 
\be Y^{\ua'}=(y^{\alpha'},\bar y^{\dot\alpha'})\ ,\qquad
Z^{\ua}=(z^{\alpha},\bar z^{\dot\alpha})\ ,\ee
\be (y^{\alpha'})^\dagger=\bar y^{\dot\alpha'}\ ,\qquad
(z^{\alpha})^\dagger=-\bar z^{\dot\alpha}\ ,\ee
where the complex doublets obey
\be
[y^{\alpha'},y^{\beta'}]_\star=2i \epsilon^{\alpha'\beta'}
\ ,\qquad
[z^{\alpha},z^{\beta}]_\star=-2i \epsilon^{\alpha\beta}
\ .\ee
The horizontal forms can be represented as sets of
locally defined forms on ${\cal X}_4\times {\cal Z}_4$ valued 
in oscillator algebras ${\cal A}({\cal Y}_4)$ generated by the fiber 
coordinates glued together by transition functions.
Assuming the latter to be defined locally on ${\cal X}_4$
yields a bundle over ${\cal X}_4$ with fibers given by the differential graded 
associative algebra $\Omega({\cal Z}_4)\otimes {\cal A}({\cal Y}_4)$, whose
elements can be given represented using symbols defined using
various ordering schemes, which correspond to choosing different
bases for the operator algebra.
In what follows, we shall assume that it is possible to describe
the field configurations using symbols defined in the Weyl ordered basis, which is 
manifestly $Sp(4;\Real)\times Sp(4;\Real)'$ invariant, as well
as the normal ordered basis consisting of monomials in\footnote{
The normal order reduces to Weyl order for elements that 
are independent of either $Y$ or $Z$.}
\be 
a^{\ua}:=Y^{\ua}+ Z^{\ua}\ ,\qquad b^{\ua}:=Y^{\ua}- Z^{\ua}\ ,
\ee
with $a$- and $b$-oscillators standing to the left and right, respectively,
which breaks $Sp(4;\Real)\times Sp(4;\Real)'\rightarrow
(Sp(4;\Real)\times Sp(4;\Real)')_{\rm diag}$.
Equivalently, we shall assume that the elements in 
$\Omega({\cal Z}_4)\otimes {\cal A}({\cal Y}_4)$
have well-defined symbols in normal order, which
can be composed using the Fourier transformed 
twisted convolution formula \eq{star}, and that
they can furthermore be expanded over the Weyl
ordered basis of ${\cal A}({\cal Y}_4) $ with
coefficients in $\Omega({\cal Z}_4)$, using the 
aforementioned star product.

As for the fiber algebra ${\cal A}({\cal Y}_4) $,
it is assumed to be an associative algebra closed under 
the star product and the hermitian conjugation operation 
$\dagger$ defined above.
As we shall describe in more detail in Sections \ref{gf} and \ref{SecPertExact}, 
the algebra ${\cal A}({\cal Y}_4) $ will furthermore be assumed to contain
certain nonpolynomial elements and distributions playing a role in 
constructing higher spin background and fluctuation fields\footnote{
In order to construct higher spin invariants playing a role
as classical observables, the algebra needs to be furthermore
equipped with a trace operation that provides it with a Hilbert 
space structure or other suitable inner product structure
}. 

\subsubsection{Master fields}
%
The model is formulated  in terms of a zero-form $\Phi$, a one-form 
\be
A=dx^\mu A_\mu + dz^\alpha A_\alpha + d\bar z^{\dot\alpha} A_{\dot\alpha}\ ,
\label{1f}
\ee
and a non-dynamical holomorphic two-form 
\be
J:=-\frac{ib}4 dz^\alpha \wedge dz_\alpha \kappa\ ,\ee
with Hermitian conjugate $\overline J=(J)^\dagger$, 
where $b$ is a complex parameter and 
\be \kappa:=\kappa_y \star \kappa_z\ ,\qquad 
\kappa_y:=2\pi \delta^2(y)\ ,\qquad 
\kappa_z:=2\pi \delta^2(z)\ ,\ee
are inner Klein operators obeying 
\be 
\kappa_y\star f \star \kappa_y=\pi_y(f) \ ,\qquad \kappa_z\star f \star \kappa_z=\pi_z(f) \ ,
\ee
for any zero-form $f$, where 
$\pi_y$ and $\pi_z$ are the
automorphisms of $\Omega({\cal Z}_4)\otimes {\cal A}({\cal Y}_4)$ defined
in Weyl order by
\be
\pi_y:(x;z,\bar z; y,\bar y) \mapsto (x;z,\bar z;- y,\bar y)\ ,\qquad 
\pi_z:(x;z,\bar z; y,\bar y) \mapsto (x;-z,\bar z;y,\bar y)\ ,
\ee
and $\pi_y\circ d= d\circ \pi_y$ and 
$\pi_z\circ d= d\circ \pi_z$.
It follows that 
\be dJ=0\ ,\qquad
J\star f=\pi(f)\star J\ ,\qquad \pi(J)=J\ ,\qquad 
\pi:=\pi_y \circ \pi_z\ ,
\ee
for any form $f$, \emph{idem} $\overline J$ and $\bar\pi:=\pi_{\bar y}\circ \pi_{\bar z}$.

\subsubsection{Kinematic conditions} 

Higher spin gravities consisting
of Lorentz tensor gauge fields can be obtained
by imposing the integer-spin projection
$$ \pi\circ \bar\pi(\Phi)=\Phi\ ,\qquad 
\pi\circ \bar\pi(A)=A\ .$$
Models in Lorentzian spacetimes with cosmological
constants $\Lambda$ are obtained
by imposing reality conditions as follows 
\cite{Iazeolla:2007wt}:
\be
\rho(\Phi^\dagger)=\pi(\Phi)\ ,\qquad \rho(A^\dagger)=-A\ ,\qquad
\rho:=\left\{ \begin{array}{ll} \pi\ ,&\Lambda>0\ ,\\
{\rm Id}\ ,&\Lambda<0\ ,\end{array}\right.
\ee
that is, the real form of the 
$\mathfrak{sp}(4;\Comp)$ realized in
terms of bilinears in $Y^{\ua}$ 
is chosen by the Hermitian
conjugation operation $\rho\circ \dagger$;
the consistency follows from $\bar\pi(f)=( \pi(f^\dagger))^\dagger$
and the fact that if $\Lambda>0$, then 
$(\rho\circ \dagger)^2\equiv \pi\circ\bar{\pi}$, 
which reduces to the identity modulo
the integer-spin projection.

\subsubsection{Equations of motion}
%
Introducing the curvature and twisted-adjoint 
covariant derivative defined by
\be
F :=dA+A\star A\ ,\qquad D\Phi :=d\Phi+[A,\Phi]_\pi
\ ,
\label{curvatures}
\ee
respectively, one has the Bianchi identities
\be DF:=dF+[A,F]_\star\equiv 0\ ,\qquad DD\Phi:=d(D\Phi)+[A,D\Phi]_\pi\equiv [F,\Phi]_\pi\ ,\ee
where ordinary and $\pi$-twisted star commutators
\be [f,g]_\star :=f\star g-(-1)^{|f||g|}g\star f\ ,\qquad
[f,g]_\pi:=f\star g-(-1)^{|f||g|}g\star \pi(f)\ ,\ee
respectively.
The Vasiliev equations of motion are given by 
\be
F+ \Phi\star (J- {\overline J})=0\ ,\qquad D\Phi=0\ ,
\label{ve}
\ee
which are compatible with the kinematic 
conditions and the Bianchi identities,
implying that the classical solution space 
is invariant under the following infinitesimal 
gauge transformations:
\be \delta A=D\e:=d\e+[A,\e]_\star\ ,\qquad
\delta\Phi=-[\e,\Phi]_\pi\ ,\ee
for parameters obeying the same kinematic
conditions as the connection, \emph{viz.}
\be \pi\bar\pi(\e)=\e\ ,\qquad \rho(\e^\dagger)=-\e\ .\ee

\subsubsection{Component form} 
Decomposition of the equations of motion under
the coordinate basis $(\vec \partial_\mu,\vec\partial_{\ua})$
for the tangent space of the base manifold, yields the Vasiliev equations
\be
F_{\m\n}= 0\ ,\qquad D_\m \Phi=0\ ,\qquad F_{\mu\ua}=0\ ,\qquad 
\qquad D_{\ua}\Phi=0\ ,\ee
\be F_{\a\b} + \frac{ib}{2}\, \Phi\star \k \e_{\a\b} =0\ ,\qquad  F_{\a\dot\b} =0\ ,\qquad 
F_{\dot\a\dot\b} +\frac{i\bar b}{2}\, \Phi\star \bar\k \e_{\dot\a\dot\b} =0\ ,\ee
\be D_\m \Phi= \imath_{\vec \partial_\mu} D\Phi=\partial_\mu \Phi + A_\mu \star \Phi - 
\Phi \star \pi(A_\mu)\ ,\ee
\bea
D_\a \Phi &=&\imath_{\vec \partial_\a} D\Phi=
\partial_\a \Phi + A_\a \star \Phi - \Phi \star \bar\pi(A_\a)\ ,\\
D_{\dot\a} \Phi&=&\imath_{\vec \partial_{\dot\a}} D\Phi=\partial_{\dot\a} \Phi + A_{\dot\a} \star \Phi - 
\Phi \star \pi(A_{\dot\a})\ ,
\eea
using $\pi(A)=\bar\pi(A)$, $\bar{\pi}(dz^\a)=dz^\a$ and $\pi(d\bar z^{\dot\a})=d\bar z^{\dot\a}$,
and the one-form components obey the following kinematic conditions:
\be
\pi\bar\pi(A_\mu,A_\alpha,A_{\dot\a})=(A_\mu,-A_\alpha,-A_{\dot\a})\ ,
\ee
\be
(A_\mu,A_\alpha,A_{\dot\a})^\dagger=\left\{\begin{array}{ll}
(-\pi(A_\mu),\pi(A_{\dot\a}),\bar\pi(A_\alpha))\ ,&\Lambda>0\ ,\\[5pt] 
(-A_\mu,A_{\dot\a},
A_\alpha)\ ,&\Lambda<0\ .\end{array}\right.
\ee

\subsubsection{Deformed oscillators} 
%
Alternatively, introducing 
\be 
S_\a:= z_\a-2i A_\a\ ,\qquad S_{\dot\a}=z_{\dot\a}-2iA_{\dot\a}\ ,
\ee
the equations of motion involving twistor-space derivatives can be written as
\be
D_\mu S_\a=0\ ,\qquad D_\mu S_{\dot\a}=0\ ,
\ee
\be
[S_\a,S_\b]_\star=-2i\e_{\a\b}(1-b \Phi\star \k )\ ,\qquad [S_\a,S_{\dot\b}]_\star =0\ ,\qquad 
[S_{\dot\a},S_{\dot\b}]_\star=-2i\e_{\dot\a\dot\b}(1-\bar b \Phi\star \bar \k )\ ,
\ee
\be
S_\a\star \Phi+ \Phi\star \pi(S_\a)=0\ ,\qquad S_{\dot\a}\star \Phi
+ \Phi\star \bar\pi(S_{\dot\a})=0 \label{SPhi}\ ,
\ee
that is, the master fields $(S_\a,S_{\dot\a})$ define a covariantly
constant set of Wigner-deformed oscillators with deformation 
parameter given by $\Phi$.
The deformed oscillators obey reality conditions and integer-spin conditions as follows:
\be
\pi\bar\pi(S_\alpha,S_{\dot\a})=(-S_\alpha,-S_{\dot\a})\ ,
\ee
\be
(S_\alpha,S_{\dot\a})^\dagger=\left\{\begin{array}{ll}
(-S_{\dot\a},
-S_\alpha) &\mbox{for $\Lambda<0$}\ ,
\\[5pt] 
(-\pi(S_{\dot\a}),-\bar\pi(S_\alpha))&\mbox{for $\Lambda>0$}\ .\end{array}\right.
\ee
Besides being useful in constructing exact solutions,
observables and exhibiting certain discrete symmetries, 
the deformed oscillators facilitate the casting of the
equations of motion into a manifestly Lorentz covariant 
form.
\subsubsection{Discrete symmetries}
The equations of motion and the gauge transformations 
exhibit the following discrete symmetries:
\begin{itemize}
\item[i)] Holomorphic parity transformation
\be (\Phi,A;\epsilon)\mapsto (\pi(\Phi),\pi(A);\pi(\epsilon))\ ;\ee
\item[ii)] Deformed oscillator parity transformation
\be (\Phi,A_\mu,S_{\ua};\epsilon)\mapsto (\Phi,A_\mu,-S_{\ua};\epsilon)\ ,\ee
which is equivalent to $A_{\ua}\mapsto -iZ_{\ua}-A_{\ua}$;
\item[iii)]  \emph{Vectorial parity transformation}
\be (\Phi,A;\epsilon)\mapsto (P(\Phi),P(A);P(\epsilon))\ ,\ee
where $P$ is the star product algebra automorphism
\be P(y^\alpha,\bar y^{\dot\a};z^\alpha,\bar z^{\dot\a}):=
(\bar y^{\dot\a},y^\alpha;\bar z^{\dot\a},z^\alpha)
\ ,\qquad 
d\circ P:=P\circ d\ ,\ee
from which it follows that $P\circ\dagger=\dagger\circ P$
and $P\circ\pi=\bar\pi\circ P$
\end{itemize}
From 
\be P(J)=-(b/\bar b)\, \overline J\ ,\ee
it follows that (iii) exchanges a solution
to the equations with parameter $b$ to a 
solution to the equations with parameter 
$\bar b$.
In particular, if $\bar b=\pm b$, then one can 
extend $P$ to 
\be \widehat P=P\circ P'\ ,\ee
where $P'$ is an internal parity map acting on the
component fields, and project the spectrum of the 
theory by demanding 
\be \widehat P(A,\Phi)=\left\{\begin{array}{ll} (A,\Phi)& \mbox{$b=1$ (A model)}\ ,\\
(A,-\Phi)& \mbox{$b=i$ (B model)}\ ,\end{array}\right.\ee
which thus correlates the internal parity with the
vectorial parity in twistor space.

\subsubsection{Manifest Lorentz covariance} 
%
To cast the equations on a manifestly Lorentz covariant form, 
one introduces the field-dependent generators \cite{Vasiliev:1999ba,Sezgin:2002ru}
\be
M^{({\rm tot})}_{\a\b}:= y_{(\a}\star y_{\b)}-z_{(\a}\star z_{\b)}
+S_{(\a}\star S_{\b)}\ ,\qquad
M^{({\rm tot})}_{\dot\a\dot\b}:= \bar y_{(\dot\a}\star \bar y_{\dot\b)}
-\bar z_{(\dot\a}\star \bar z_{\dot\b)}
+S_{(\dot\a}\star S_{\dot\b)}\ ,
\label{LT}
\ee
and redefines
\be
A_\mu=W_\mu+\frac{1}{4i}\left(\omega_\mu^{\a\b} M^{({\rm tot})}_{\a\b}+
\omega_\mu^{\dot\a\dot\b}M^{({\rm tot})}_{\dot\a\dot\b}\right)\ ,
\label{Wdef}
\ee
where $(\omega_\mu^{\a\b}(x),\omega_\mu^{\dot\a\dot\b}(x))$ is a bona fide
canonical Lorentz connection on ${\cal X}_4$, 
after which the equations of motion involving spacetime 
derivatives can be re-written on the following 
manifestly Lorentz covariant form\footnote{The 
closure of the algebra generated by $M^{(\rm tot)}$
contains additional Lorentz transformations on acting
on the component fields; for details, see \cite{Sezgin:2011hq,Didenko:2014dwa}.} 
\cite{Sezgin:2011hq,Iazeolla:2011cb,Colombo:2012jx}:
\be
\nabla W+W\star W+ \frac{1}{4i}\left(r^{\a\b} M^{({\rm tot})}_{\a\b}+
r^{\dot\a\dot\b}M^{({\rm tot})}_{\dot\a\dot\b}
\right)=0\ ,
\ee
\be
\nabla \Phi+W\star \Phi-\Phi\star \pi(W)=0\ ,\qquad
\nabla S_\a+[W,S_\a]_\star=0\ ,
\ee
where 
\be
\nabla W:= dW+[\omega^{(0)},W]_\star\ ,\qquad \nabla \Phi:= d\Phi+[\omega^{(0)},\Phi]_\star\ ,
\ee
\be
\nabla S_\a:=dS_\a-\omega_\a{}^\b S_\b+[\omega^{(0)},S_\a]_\star\ ,
\ee
\be
r^{\a\b}:= d\omega^{\a\b}-\omega^{\a\gamma}\wedge \omega_{\gamma}{}^\b\ ,\qquad
r^{\dot\a\dot\b}:= d\omega^{\dot\a\dot\b}-\omega^{\dot\a\dot\gamma}\wedge \omega_{\dot\gamma}{}^{\dot\b}\ ,
\ee
with
\be 
\omega^{(0)}:=\frac{1}{4i}\left(\omega^{\a\b} M^{(0)}_{\a\b}+
\omega^{\dot\a\dot\b}M^{(0)}_{\dot\a\dot\b}\right)\ ,\ee
\be 
M^{(0)}_{\a\b}:= y_{(\a}\star y_{\b)}-z_{(\a}\star z_{\b)}
\ ,\qquad
M^{(0)}_{\dot\a\dot\b}:= \bar y_{(\dot\a}\star \bar y_{\dot\b)}
-\bar z_{(\dot\a}\star \bar z_{\dot\b)}\ .
\ee
The Lorentz connection is defined, as usual, up to tensorial shifts, 
that can be fixed by requiring that the projection of $W$ onto 
$M^{(0)}_{\a\b}$ vanish at $Z=0$.
%

\subsection{Vacuum solutions}

\para{Flat connections.}
%
The equations of motion admit solutions 
\be  \Phi=0\ ,\qquad A=\Omega\ ,\ee
where $\Omega$ is a locally defined one-form on ${\cal X}_4\times {\cal Z}_4$
valued in ${\cal A}({\cal Y}_4)$ that is flat, \emph{viz.}
\be d\Omega+\Omega\star \Omega=0\ .\ee
If $\Omega\in \Omega({\cal X}_4)\otimes {\cal A}({\cal Y}_4)$, then 
there exists locally defined gauge functions $L$ on ${\cal X}_4$
such that
\be \Omega=L^{-1}\star dL\ ,\ee
that we shall refer to as vacuum connections, as they 
preserve higher symmetries with rigid parameters
\be \e=L^{-1}\star \e'\star L\ ,\qquad d\e'=0\ ,\qquad \e'\in{\cal A}({\cal Y}_4)\ ;\ee
the space $\Omega({\cal Z}_4)\otimes {\cal A}({\cal Y}_4)$, on the other
hand, contains flat connections constructed from projector algebras 
that cannot be described using gauge functions and that break some of the
vacuum symmetries \cite{Sezgin:2005pv}.

\para{Maximally symmetric spaces.} 
%
The $(A)dS_4$ vacua are described by gauge functions 
valued in the real form $G_{10}$ of $Sp(4;\Comp)$ selected by the reality 
condition introduced above. Thus, $G_{10}$ refers to $AdS$ group for $\lambda^2 >0$ and $dS$ group for $\lambda^2 <0$, with the commutation rules for the ${\mathfrak g}_{10}$ algebra given by 
\be
[M_{AB}, M_{CD}]= 4i\eta_{[C|[B} M_{A]|D]}\ , 
\qquad \eta_{AB}:= \left(\eta_{ab}, -{\rm sign} (\lambda^2) \right)\ ,\qquad \eta_{ab} = {\rm diag} (-+++)\ .
\ee
and they can be realized in terms of the $Y$-oscillators as  
\be
-\ell^{-1} M_{a5} \equiv P_{a}=\frac{\lambda}{4}\left( \sigma_{a}\right) _{\alpha \dot{\beta}}y^{\alpha }
\bar{y}^{\dot{\beta}}\ \ ,\ \ \ 
M_{ab}=-\frac{1}{8}\left[\left( \sigma_{ab}\right)_{\alpha\beta }y^{\alpha }y^{\beta} +\left(\bar{\sigma}_{ab}\right)_{\dot{\alpha}\dot{\beta}}\bar{y}^{\dot{\alpha}}
\bar{y}^{\dot{\beta}}\right]\ ,
\label{pm}
\ee
where 
\be
\lambda =\left\{\begin{array}{ll} \ell^{-1}&\mbox{for $\Lambda<0$}\\[5pt]
i\ell^{-1}& \mbox{for $\Lambda>0$}\ ,\end{array}\right.
\ee
It follows that
\be
[M_{ab},M_{cd}]_\star = i\eta_{bc} M_{ad} + \mbox{3 more}\ ,\qquad 
[M_{ab},P_c]_\star = 2i\eta_{c[b} P_{a]}\ ,\ee
\be  [P_a,P_b]_\star = i\lambda^2 M_{ab}\ 
\ee
with reality conditions
\be 
\rho((P_a)^\dagger)=P_a\ ,\qquad 
\rho((M_{ab})^\dagger)=(M_{ab})^\dagger=M_{ab}\ .
\ee
Introducing coset elements
\be L: G_{10}/SO(1,3)\to G_{10}\ ,\qquad
\rho(L^\dagger)=L^{-1}\ ,\ee
the Maurer--Cartan form
decomposes into a frame field and a 
Lorentz connection as follows:
\be
\Omega =\frac{1}{4i} \Omega_{\underline{\a\b}} Y^{\ua} Y^{\ub}
=\frac{1}{4i}\left(2\Omega_{\alpha \dot{\alpha}}y^{\alpha }\bar{y}^{\dot{\alpha}}
+\Omega_{\alpha \beta }y^{\alpha }y^{\beta}
+\Omega_{\dot{\alpha}\dot{\beta}}\bar{y}^{\dot{\alpha}}
\bar{y}^{\dot{\beta}}\right)
= i\Omega_a P^a + 
\frac{1}{2i}\Omega_{ab} M^{ab}
\label{omega}
\ee
where thus  
\be
\Omega_{\alpha\dot\alpha} =
-\frac{\lambda}{2}\left( \sigma _{a}\right)
_{\alpha \dot{\alpha}}\Omega^{a}\ ,\qquad
\Omega_{\alpha \beta } := -\frac{1}{4}\left(\sigma _{ab}\right)
_{\alpha \beta } \Omega^{ab}
\ ,\qquad \Omega_{\dot\alpha \dot\beta } = -\frac{1}{4}\left( \bar{\sigma}
_{ab}\right) _{\dot\alpha \dot\beta }\Omega^{ab}
\ .
\ee
In these bases, the flatness condition reads
\be 
d\Omega^{\underline{\a\b}}-\Omega^{\underline{\a\gamma}}\wedge
\Omega_{\uc}{}^{\ub}=0\ ,
\ee
that is
\be 
d\Omega_{\alpha\dot\alpha}-\Omega_\alpha{}^\beta 
\wedge\Omega_{\beta\dot\alpha}-\Omega_{\dot\alpha}{}^{\dot\beta} 
\wedge\Omega_{\alpha\dot\beta}=0\ ,
\ee
\be 
R_{\alpha\beta}- \Omega_{\alpha}{}^{\dot\alpha} 
\wedge\Omega_{\dot\alpha\beta}=0\ ,\qquad 
 R_{\dot\alpha\dot\beta}-\Omega_{\dot\alpha}{}^{\alpha} 
\wedge\Omega_{\alpha\dot\beta}=0\ ,
\ee
or
\be 
d\Omega_a+\Omega_a{}^b \wedge\Omega_b=0\ ,\qquad
R_{ab}+\lambda^2 \Omega_a \wedge\Omega_b=0\ ,
\ee
where the Riemann two-form
\begin{align}
R^{\a\b}&:= d\Omega^{\a\b}-\Omega^{\a\gamma}\wedge \Omega_{\gamma}{}^\b=-\frac{1}{4}\left(\sigma _{ab}\right)
^{\alpha \beta } R^{ab}\ ,\cr
R^{\dot\a\dot\b}&:= d\Omega^{\dot\a\dot\b}-\Omega^{\dot\a\dot\gamma}\wedge \Omega_{\dot\gamma}{}^{\dot\b}= -\frac{1}{4}\left( \bar{\sigma}
_{ab}\right)^{\dot\alpha \dot\beta }R^{ab}\ ,
\end{align}
with $R_{ab}:= d\Omega_{ab}+\Omega_{a}{}^{c}\Omega_{cb}$.

The full equations of motion can be solved in two dual fashions, one involving normal ordered scheme and perturbatively defined Fronsdal fields, and the other based on a topological field theory approach, which we describe below. 

\subsection{Normal ordered perturbation scheme}
%

In the normal order, defined by the star product formula \eq{star},
the inner Klein operators become real analytic in $Y$ and $Z$ space,
\emph{viz.}
\be
\kappa=\kappa_y\star \kappa_z= \exp(iy^\alpha z_\alpha)\ ,\qquad
\bar\kappa=\kappa_{\bar y}\star \kappa_{\bar z}= 
\exp(i\bar y^{\dot\alpha} \bar z_{\dot \alpha})\ .
\ee
Assuming that the full field configurations are real-analytic on
${\cal Z}_4$ for generic points in ${\cal X}_4$, one may thus 
impose initial conditions 
\be \Phi|_{Z=0}=C\ ,\qquad A_\mu|_{Z=0}=a_\mu\ .\ee
Assuming furthermore that $A_{\ua}|_{C=0}$ is a trivial flat connection 
on ${\cal Z}_4$, that one may choose to be $A_{\ua}|_{C=0}=0$,
and choosing a homotopy contractor for the de Rham differential
on ${\cal Z}_4$, which entails imposing a gauge condition 
on $A_{\ua}$, one may solve the constraints on $D_{\ua}\Phi$, 
$F_{\underline{\a\b}}$ and $F_{{\ua}\mu}$ on ${\cal Z}_4$ in a
perturbative expansion of the form:
\begin{align} \Phi&=\sum_{n\geqslant1}\Phi^{(n)}(C,\dots,C)\ ,\qquad
\Phi^{(1)}(C)\equiv C\ ,\cr
A_{\ua}&=\sum_{n\geqslant1}A_{\ua}^{(n)}(C,\dots,C)\ ,\cr
A_\mu&=\sum_{n\geqslant 0 }A^{(n)}(a_\mu;C,\dots,C)\ ,\qquad
A^{(0)}(a_\mu)\equiv a_\mu\ ,\end{align}
where $\Phi^{(n)}(C,\dots,C)$ is an $n$-linear functional
in $C$ \emph{idem} $A_{\ua}^{(n)}(C,\dots,C)$ and 
$A^{(n)}(a_\mu;C,\dots,C)$, and the latter 
is linear in $a_\mu$.
These quantities are real-analytic in ${\cal Y}_4\times {\cal Z}_4$
provided that $C$ and $a_\mu$ are real analytic in $Y$-space 
and all star products arising along the perturbative expansion 
are well-defined.

From the Bianchi identities, it follows that the remaining equations, 
that is, $F_{\mu\nu}=0$ and $D_\mu\Phi=0$, are perturbatively equivalent
to $F_{\mu\nu}|_{Z=0}=0$ and $D_\mu\Phi|_{Z=0}=0$,
which form a perturbatively defined Cartan integrable system
on ${\cal X}_4$ for $C$ and $a_\mu$.

To Lorentz covariantize, one imposes
\be W|_{Z=0}=w\ ,\ee
and substitutes
\be a_\mu=w_\mu+\frac1{4i}\left.(\omega^{\a\b}_\mu 
M^{({\rm tot})}_{\a\b}+\omega^{\dot\a\dot\b}_\mu 
M^{({\rm tot})}_{\dot\a\dot\b})\right|_{Z=0}\ ,\ee
into $A^{(n)}(a_\mu;C,\dots,C)$.
Due to the manifest Lorentz covariance, the quantities $F_{\mu\nu}|_{Z=0}$ and $D_\mu\Phi|_{Z=0}$ 
depend on the Lorentz connection only via the Lorentz covariant 
derivative $\nabla$ and the Riemann two-form $(r^{\alpha\beta},r^{\dot\a\dot\b})$;
it follows that 
\be \nabla w + \frac1{4i}\left.(r^{\a\b} 
M^{({\rm tot})}_{\a\b}+r^{\dot\a\dot\b} 
M^{({\rm tot})}_{\dot\a\dot\b})\right|_{Z=0}+ 
\sum_{\tiny\begin{array}{c}n_1+n_2\geqslant 0\\ n_{1,2}\geqslant 0
\end{array}}
A^{(n_1)}(w;C,\dots,C)\star 
A^{(n_2)}(w;C,\dots,C)=0\ ,\ee
\be \nabla C +  
\sum_{\tiny \begin{array}{c}n_1+n_2\geqslant 1\\ n_1\geqslant 0,\ n_2\geqslant 1\end{array}}[A^{(n_1)}(w;C,\dots,C),\Phi^{(n_2)}(C,\dots,C)]_\pi=0\ ,\ee
where $w$ can be chosen to not contain any component
field proportional to $y_{\a}y_{\b}$ and $\bar y_{\dot\a}
\bar y_{\dot\b}$.

\para{Perturbatively defined Fronsdal fields.}
%
Expanding the differential algebra around the (anti-)de Sitter vacuum 
\be
\Phi^{(0)}=0\ ,\qquad A^{(0)}= \Omega\ ,
\ee
and assuming that the homotopy contraction in $Z$-space is performed
such that
\be
z^{\ua} A_{\ua}^{(1)}=0\ ,
\ee
referred to as the Vasiliev gauge \cite{Vasiliev:1992av},
the resulting linearized system on $X$-space provides an unfolded 
description of a dynamical scalar field 
\be
\phi=\Phi\mid_{Y=0=Z}\ ,
\ee
and a tower of spin-$s$ Fronsdal fields 
\be
\phi_{a(s)}=\left(((e^{-1})^\mu_{a})^{(0)} \left((\sigma_a)^{\alpha\dot\alpha} \frac{\partial^2}{\partial y^\alpha \partial {\bar y}^{\dot\alpha}}\right)^{s-1} w_\mu\right)\Bigg\vert_{Y=0=Z}\ ,
\ee
where we use the convention that repeated indices are symmetrized.

At the nonlinear level, the Cartan integrable system on $X$-space provides
a deformation of the equations of motion for these fields, which is consistent
as a set of partial differential equations but that depends on the choice of
initial data for $\Phi$ and $W_\mu$ as well as the gauge for $A_\alpha$ (which
enters via the homotopy contractor in $Z$ space).
Whether there exists a choice that yields a formulation of higher 
spin gravity in $X$-space that lends itself to a standard path
integral formulation remains an open problem\footnote{To our best understanding, the standard classical Noether procedure breaks down \cite{Sleight:2017pcz}, while there exists a quantum effective action in ${\cal X}_4$ for asymptotically (anti-)de Sitter boundary conditions. In order to obtain a path integral measure, one may instead follow the approach proposed in \cite{Boulanger:2011dd,Sezgin:2011hq,Boulanger:2015kfa}.}.

\subsection{Gauge function method}\label{gf}

\subsubsection{Topological field theory approach}

Alternatively, one may treat the system as an infinite set of topological fields on 
${\cal X}_4\times {\cal Z}_4$ packaged into master fields valued 
in ${\cal A}({\cal Y}_4)$ represented by symbols in 
Weyl order, that is, as expansions in terms of the 
generators of ${\cal A}({\cal Y}_4)$ star multiplied 
by differential forms on $\Omega({\cal X}_4\times {\cal Z}_4)$,
referred to as mode forms.

The field configurations are assigned a bundle structure, 
whereby a projection of $A$ is assumed to define a
connection valued in a Lie subalgebra of ${\cal A}({\cal Y}_4)$.
The complementary projection of $A$, referred to as the
generalized frame field, together with the Weyl zero-form $\Phi$ 
are taken to belong to adjoint and twisted adjoint sections, respectively, 
over ${\cal X}_4\times {\cal Z}_4$, which is treated as a base manifold; the two-forms $J$ 
and $\overline J$ by their definition belong to twisted adjoint
sections.
The bundle connection is assumed to act faithfully on the sections,
and the bundle curvature is assumed to be an adjoint section, as 
required by the equations of motion.

As for boundary conditions, the base manifold is assumed to be compact,
and the sections, \emph{i.e.} their mode forms, are assumed to be bounded
away from a set of marked points representing boundaries.
In a generic coordinate chart $U\subset{\cal X}_4$, the sections are described
by an integration constant for the Weyl zero-form, a flat connection 
on ${\cal Z}_4$ and a gauge function on $U\times {\cal Z}_4$.
At the marked points, the initial data instead consists of prescribed 
singularities in the generalized frame field and related fall-off in
the Weyl zero-form (including the physical scalar field).
For asymptotically (anti-)de Sitter solutions, we take ${\cal X}_4$ to 
have the topology of $S^1\times S^3$ with a marked $S^1$, such that,
to all orders in classical perturbation theory,
the leading terms in the master fields at the marked $S^1\times{\cal Z}_4$ 
describe a set of free Fronsdal fields,
which one may view as a condition at the boundary of $(A)dS_4$
times ${\cal Z}_4$.
To impose boundary conditions on ${\cal Z}_4$, we assume that 
$\Omega({\cal Z}_4)$ 
\begin{itemize}
\item[i)] is closed under star products, which can be achieved by taking 
the Fourier transforms of the zero-forms in $\Omega({\cal Z}_4)$ to be $L^1$
in momentum space (\emph{i.e.} to be expandable in terms of plane waves that 
generate a twisted abelian group algebra), which requires the zero-form
sections on ${\cal Z}_4$ to be bounded at $Z=0$;
\item[ii)] has a graded trace operation given by integration 
of the top-forms on ${\cal Z}_4$, which requires these 
to fall off at $Z=\infty$ so as to belong to $L^1({\cal Z}_4)$. 
\end{itemize}
Finally, ${\cal A}({\cal Y}_4)$ is taken to be a set of operators
in a quantum-mechanical system equipped with a (possibly regularized) 
trace operation that is dual to the boundary conditions at the
marked $S^1$.

The above geometries can be characterized by functionals,
playing the role of classical observables (including on-shell actions), given 
by combined traces over ${\cal A}({\cal Y}_4)$
and integrations over ${\cal X}_4\times {\cal Z}_4$ (possibly with insertions
of delta functions localized to submanifolds).
These gauge transformations that leave these functionals 
invariant are referred to as proper, or small, gauge 
transformations, as opposed to large gauge transformations 
that alter the asymptotics of the fields and hence the value 
of the observables.
The resulting moduli space is thus sliced into (proper) gauge orbits 
labelled by the observables, each of which defines a microstate of the
theory\footnote{A subset of the classical
observables are extensive; keeping these fixed defines a higher
spin ensemble consisting of a large number of microstates.
In \cite{Colombo:2010fu,Colombo:2012jx}, it has been proposed that the extensive
variables are the zero-form charges \cite{Sezgin:2005pv,Sezgin:2011hq}, and in \cite{Bonezzi:2017vha} it 
has been proposed that a complete set of classical observables
in the case that ${\cal X}_4$ has trivial topology is given by
the space of twisted open Wilson lines in ${\cal Z}_4$.
According to these proposals, the rigid symmetries of the vacuum 
should leave the extensive variables invariant while acting
nontrivially on the microscopic variables; for related remarks,
see \cite{Bonezzi:2017vha,Iazeolla:2017vng}.}.

\subsubsection{Gauge functions}\label{gf2}

In the topological field theory approach, solution spaces 
are obtained starting from a reference solution $(\Phi',A')\in \Omega(\{p_0\}\times{\cal Z}_4)\otimes {\cal A}({\cal Y}_4)$ at a base point $p_0\in {\cal X}_4$, constructed from an integration constant $C'$ for
$\Phi'$ at, say, $Z=0$, and an flat connection on ${\cal Z}_4$, that we shall trivialize in most
of what follows.
Moduli associated to the connection and generalized frame field on ${\cal X}_4$ 
are then introduced by means of a large gauge transformation
\be
A^{(G)}=G^{-1}\star (A'+d)\star G\ ,\qquad 
\Phi^{(G)}=G^{-1}\star \Phi'\star \pi(G)\ ,\qquad G=L\star H\ ,
\ee
where $L$ is the vacuum gauge function, and $H$ is a gauge function 
determined perturbatively by the requirements that
\begin{itemize}
\item[a)] in Weyl order, $\Phi^{(G)}$ and 
the twisted open Wilson lines $V(M):=\exp_\star( i M^{\ua} S^{(G)}_{\ua})$,
where $M^{\ua}\in \Comp^4$ (for details, see \cite{Sezgin:2011hq,Colombo:2012jx,Bonezzi:2017vha}), 
are sections in $\Omega({\cal X}_4\times {\cal Z}_4)\otimes {\cal A}({\cal Y}_4)$
in form degree zero; and 
\item[b)] in normal order, $(\Phi^{(G)},A^{(G)}_{\ua},W^{(G)}_\mu)$ asymptote 
to configurations describing free Fronsdal fields\footnote{The full field 
configurations are thus assumed to contain contain asymptotically (anti-)de Sitter 
regions where the \emph{full}
tensor gauge fields $\phi_{a(s)}$ approach Fronsdal fields give on
shell in terms of polarization tensors that are non-linear functionals
of the zero-form initial data $C$.}
in accordance with the central on mass-shell theorem 
close to the marked $S^1\times {\cal Z}_4$.
\end{itemize}
We shall refer refer to \emph{(a)} and \emph{(b)} as dual boundary conditions, as \emph{(a)} 
requires factorization of the master fields in Weyl order, 
whereas \emph{(b)} requires normal order.
We thus propose to fix $H^{(n)}$ by requiring
\begin{itemize}
\item[i]) Manifest Lorentz covariance and real analyticity in $Y$ of the normal
ordered symbols of $(\Phi^{(G)}, A^{(G)})$ at the origin of ${\cal Y}_4\times {\cal Z}_4$,
so that the field configurations are expandable in terms of Lorentz tensorial 
component fields on ${\cal X}_4$ defined by Taylor expansion in $Y$ at $Y=0=Z$.
\item[ii)] the Weyl ordered symbols of $(\Phi^{(G)}, V(M))$ to be traceable over 
$\Omega({\cal Z}_4)\otimes {\cal A}({\cal Y}_4)$, for there to exist higher
spin invariants playing the role of classical observables; 
\item[iii)] Perturbatively stable asymptotic Fronsdal fields in weak-coupling 
regions of ${\cal X}_4$ (where the Weyl zero-form goes to zero), for the 
classical observables to admit perturbative expansions in terms of 
parameters related to sources for weakly coupled higher spin gauge fields.
\end{itemize}
The following additional remarks are in order:

\noindent 1. \emph{Zig-zagging self-consistency:} At $n$th order, the 
quantity $\Phi^{(G,n)}$ is a functional of $H^{(n')}$ with $1\leqslant n'\leqslant n-1$ 
and initial data $C^{\prime(n')}$ with $1\leqslant n'\leqslant n$, which means that condition (a),
which must hold for finite $Z$, is in effect a non-trivial admissibility condition on the 
$Y$-dependence of the initial data $C'$, \emph{i.e.} on ${\cal A}({\cal Y}_4)$.

\noindent 2. \emph{Residual small gauge transformations:} The above conditions do not determine 
the $\mathfrak{hs}_1(4)$ part of $H^{(n)}$, which is real analytic in $Y$, and
which can thus be used for small gauge transformations inside the bulk.

\noindent 3. \emph{Deformed oscillators:} Although the master fields $S^{(G)}$ are not sections, 
one can require that $\Phi^{(G)}$ and the twisted open Wilson loops $V(M))$ 
form an associative algebra with traces, which
can be use to construct a complete set of higher spin invariant observables
that one may think of as substitutes for the standard ADM-like charges that 
can be used to define higher spin ensembles in unbroken phases; for further details,
see \cite{Iazeolla:2017vng}.

\noindent 4. \emph{Residual symmetries:} The full solution $(\Phi^{(G)},A^{(G)})$ 
is left invariant under gauge transformations with parameters 
\be \e^{(G)}= G^{-1}\star \e'\star G\ ,\ee
where $\e'$ are constant parameters stabilizing $\Psi$, \emph{viz.} 
\be [\e',\Psi]_\star =0\ .\label{symmetries}\ee
Conversely, given a set of symmetries forming a Lie algebra
$\mathfrak g$, spaces of $\mathfrak g$-invariant solutions
can be found by solving the linear constraint \eq{symmetries}
on $\Psi$ together with the conditions that $\Psi$ belongs to 
an associative algebra that is left invariant under star
multiplication by the inner Klein operators, \emph{i.e.}
$\Psi\star\kappa_y$ and $\Psi\star \Psi$ should belong to
the algebra, which is the approach that we shall employ.

In summary, the dual boundary conditions are physically well-motivated and
non-trivial; in this paper, we shall focus on their implementation at the
linearized level, leaving higher orders, starting with the issue of whether 
$\Phi^{(G,2)}$ obeys (a), for a forthcoming publication including various
types of boundary conditions.

\subsubsection{A universal particular solution in holomorphic gauge} 
%
For all vector fields $\vec{v}$ tangent to $X$-space, we have $\imath_{\vec{v}} A'=0$,
and hence $\imath_{\vec v}dA'=0$ and $\imath_{\vec v}d\Phi'=0$, \emph{i.e.}
\be
A'=dz^\a A'_\a +d\bar z^{\dot\a} A'_{\dot\a}\ ,\qquad \partial_\mu \Phi'=0=\partial_\mu A'_\alpha\ ,
\ee
and 
\be
F'_{\a\b} + \frac{ib}{2}\, \Phi'\star \k \e_{\a\b} =0 ~,\qquad F'_{\a\dot\b} =0~,
\ee
\be
\partial_\a \Phi' + A'_\a \star \Phi' - \Phi' \star \bar\pi(A'_\a) =0~.
\ee
Thus, prior to switching on the gauge function $G$, we need to find
a \emph{particular} solution to the above system subject to a generic 
zero-form initial datum.
To this end, we observe that the Ansatz \footnote{Note that this ansatz for $A_\a^\prime$
is holomorphic in $z$, and hence the terminology of \emph{holomorphic gauge}; see  \cite{Iazeolla:2017dxc} for a review.} 
\be
\Phi'=\Psi(y,\bar y)\star \kappa_y\ ,\qquad A'_\alpha=A'_\alpha(z;\Psi)=\sum_{n\geqslant 1}
a_\alpha^{(n)}(z) \star \Psi^{\star n}\ ,
\label{s1}
\ee
where thus both $\Psi$ and $\Psi\star\kappa_y$ are assumed to be elements in 
${\cal A}({\cal Y}_4)$, and
\be
\Psi^\dagger = \rho(\Psi)\star \kappa_y\bar\kappa_{\bar y}\ ,
\ee
solves the fully non-linear equations provided that
\be
\pi_z(a_\alpha^{(n)}(z) )=-a_\alpha^{(n)}(z)\ ,
\ee
and that 
\be
s_\alpha:=z_\alpha-2i a_\alpha\ ,\qquad a_\alpha:=\sum_{n\geqslant 1}
a_\alpha^{(n)}(z) \nu^{n}\ ,
\ee
obeys the deformed oscillator algebra 
\be
[s_\alpha,s_\beta]_\star=-2i\epsilon_{\alpha\beta}(1-b\nu \kappa_z)\ ,
\qquad \kappa_z\star s_\alpha=- s_\alpha\star \kappa_z\ . \label{defred}
\ee
One class of solutions is given by \cite{Iazeolla:2011cb} 
\be
a_\alpha = -\frac{ib\nu}2  z_\alpha \int_{-1}^{+1} \frac{d\tau}{(\tau+1)^2}  
\exp \left(i\frac{\tau-1}{\tau+1} z^+ z^-\right){}
_1 F_1 (\tfrac12;1;b\nu \log \tau^2 )\ ,
\label{aa}
\ee
where we have introduced a spinor frame $(u^+_\alpha,u^-_\alpha)$ obeying
$$ u^{+\alpha} u^-_{\alpha}=1\ ,$$
and $z^\pm$ is defined in \eq{symbols}. The introduction of these variables is 
required in order to integrate the delta function in Weyl order.

Thus, in order to construct solution spaces with desired physical properties,
we need to expand $\Psi$ over suitable subalgebras of ${\cal A}({\cal Y}_4)$;
for the cases of particle fluctuation modes and black hole-like generalized
Type D modes, see \cite{Iazeolla:2011cb,Iazeolla:2017vng}.
In what follows, we shall examine a new type of subalgebras related to 
solutions with six Killing symmetries inside the isometry algebra of $(A)dS_4$.

\section{Construction of the exact solutions with six symmetries}
\label{SecConstrSoln}

In this section, we shall begin by describing the factorization method that will be used to construct the solutions. We shall than construct domain walls (DW), instantons\footnote{The instantons break all transvection isometries of the
vacuum solution, \emph{i.e.} $\mathfrak{g}_6$ coincides
with the Lorentz algebra of the vacuum solution.} (I) and
FRW-like solutions (FRW) given by foliations of a four-dimensional
spacetime $M_4$ with three-dimensional foliates $M_3$ that are maximally 
symmetric metric spaces, we shall first choose embeddings of the corresponding
six-dimensional isometry algebras $\mathfrak{g}_6$ into the ten-dimensional
isometry algebra $\mathfrak{g}_{10}$ of the vacuum solution. 
We then switch on $\mathfrak{g}_6$-invariant Weyl zero-forms and 
gauge functions.


\subsection{Initial data for Weyl zero-form with six Killing symmetries}

\subsubsection{Unbroken symmetries} 

In order to describe foliations with maximally symmetric foliates,
we embed $\mathfrak{g}_6$ into $\mathfrak{g}_{10}$ as follows
\cite{Sezgin:2005pv}:
\be 
M_{rs} = L_{r}{}^{a} L_{s}{}^{b} M_{ab}\ ,\qquad 
T_r =L_{r}{}^{a}\left(\a M_{ab}L^b + \b P_a\right)\ ,
\label{embed}
\ee
where\footnote{One can always choose $\a\geqslant 0$ 
by redefining $L_r^a$, after which one may take
$\beta\geqslant 0$ using the global $\mathbb Z_2$-symmetry 
generated by the $\pi$-map, which exchanges $\beta$ with $-\beta$.}
\be 
\a,\b\in \mathbb R\ ,\qquad \a,\b\geqslant 0\ ,\qquad  (\a,\b)\neq (0,0)\ ,
\label{condab}
\ee
and the representatives of the cosets $\mathfrak{so}(3,1)/\mathfrak{so}(2,1)$ for $\e=1$, and the coset $\mathfrak{so}(3,1)/\mathfrak{so}(3)$ for $\e=-1$ obey
\be \label{coset}
\begin{split}
& L_{r}{}^{a} L_{s}{}^{b} \eta_{ab} =\eta_{rs}\ , \qquad  L^aL_a=\e\ , \qquad L_{r}{}^{a}L_a=0\ ,
\ww2
& \eta_{ab} = {\rm diag} (-+++)\ ,\qquad \eta_{rs} = {\rm diag} (++,-\e)\ ,\qquad \e=\pm 1 
\end{split}
\ee
where we have introduced the parameter $\epsilon$.
The resulting symmetry algebra reads as follows\footnote{The symmetry of the full solutions 
is generated by the parameters given by the conjugation of the linearized symmetry 
parameters by the gauge function.}:
\be
[M_{rs},M_{pq}] = i\eta_{sp} M_{rq}+ \mbox{3 more}\ ,\qquad 
[M_{rs}, T_p] = 2i\eta_{p[s} T_{r]}\ ,\ee
\be [T_r, T_s]= -i (\e\a^2 -\lambda^2\b^2) M_{rs}\ ,
\ee
giving rise to the cases listed in Table \ref{Table:1}.
%
\afterpage{
\begin{landscape}
    \pagestyle{empty}
\begin{table}
\centering
\bgroup
\def\arraystretch{1.3}
\begin{tabular}{|ll|llll|ll|l|}
\hline
Type & $M_3$ &
$\mathfrak{g}_6$ & $\e$ & $\l^2$
& Condition on $(\a,\b)$  & $\gamma:=\frac{i\a}{\lambda\beta}$ & 
$(\eta_+,\eta_-)$ 
& $\Phi'$\ ,\quad $\mu\in \Comp$\ ,\quad $\nu,\tilde \nu,\nu_\pm\in \Real$\\[-5pt]
&&&&& for $\mathfrak g_6$ closure &{\small (mod $G_{10})$} &{\small $\eta_\pm\equiv -\gamma\pm\sqrt{\e+\gamma^2}$} &
{\small $P=L^aP_a$\ ,\quad $(\lambda^{-1}P)^\dagger=\lambda^{-1}P$} \\
\hline
DW$_+^{\rm \tiny{(dS)}}$ & $dS_3$ & $\mathfrak{so}(1,3)$& $+1$ & $<0$ & $\a^2-\lambda^2\b^2> 0$\,,\ $\beta\neq 0$ 
&  $\gamma=0$  & $(1,-1)$  & $\nu_+ e^{-4\lambda^{-1} P}+ \nu_- e^{4\lambda^{-1}P}$ \\
FRW$_+$ & $S^3$ & $\mathfrak{so}(4)$  &   $-1$ & $<0$ & $-\lambda^2\b^2 > \a^2$ 
& $\gamma=0$& $(i,-i)$ 
& $\mu e^{-4i \l^{-1} P}+ \bar \mu e^{4i \l^{-1} P}$ \\
FRW$_0$ & $Eucl_3$ & $\mathfrak{iso}(3)$ &  $-1$ & $<0$ & $-\lambda^2\b^2 = \a^2>0$ & 
$\gamma=1$ & $(-1,-1)$ & $(\nu -4 \tilde \nu \lambda^{-1} P )e^{4\l^{-1} P} $ \\
FRW$_-^{\rm \tiny{(dS)}}$ & $H_3$ & $\mathfrak{so}(1,3)$ & $-1$ & $<0$ & $\a^2>-\lambda^2\b^2$ & 
$\gamma >1$ & $\eta_-<-1<\eta_+<0$
& $\nu_+ e^{-4\eta_+\lambda^{-1} P}+ \nu_- e^{-4\eta_-\lambda^{-1}P}$\\
I & $dS_3$,\ $H_3$ & $\mathfrak{so}(1,3)$ & $\pm 1$ & $\neq 0$ & $\alpha^2>0$\ , $\beta=0$  & 
$\gamma=\infty$ 
& $(0,\infty)$ & $\nu$ \\
DW$_+^{\rm \tiny{(AdS)}}$ & $dS_3$ & $\mathfrak{so}(1,3)$& $+1$ & $>0$ & $\a^2>\lambda^2\b^2$
&  $-i\gamma>1$  &  $0<i\eta_-<1<i\eta_+$&  $\nu_+ e^{-4\eta_+\lambda^{-1} P}+ \nu_- e^{-4\eta_-\lambda^{-1}P}$\\
DW$_0$ & $Mink_3$ & $\mathfrak{iso}(1,2)$&$+1$ & $>0$ & $\lambda^2\b^2 = \a^2>0$ &  
$-i\gamma=1$ &  $(-i,-i)$  & $(\nu -4 i\tilde \nu \lambda^{-1} P )e^{4i\l^{-1} P}$\\
DW$_-$ & $AdS_3$ & $\mathfrak{so}(2,2)$& $+1$ & $>0$ & $\lambda^2\b^2 > \a^2$
&  $\gamma =0$ & $(1,-1)$&   $\mu e^{-4 \l^{-1} P}+ \bar \mu e^{4 \l^{-1} P}$\\
FRW$_-^{\rm \tiny{(AdS)}}$ & $H_3$ & $\mathfrak{so}(1,3)$& $-1$ & $>0$ & $\a^2+\lambda^2\b^2> 0$,\ $\beta\neq 0$
 & $\gamma=0$  & $(i,-i)$ & $\nu_+ e^{-4i\lambda^{-1} P}+ \nu_- e^{4i\lambda^{-1}P}$\\
\hline
\end{tabular}
\egroup
\caption{$\mathfrak{g}_6$-invariant $M_3$-foliations
arising in the bosonic models, with I 
standing for instantons, and FRW$_k$ and DW$_k$, respectively,
standing for FRW-like solutions ($\e=-1$) and domainwalls ($\e=+1$)
with foliates with curvatures of sign 
$k={\rm sign} (\e\a^2 -\lambda^2\b^2)$.
The embeddings of $\mathfrak{g}_6$ into the isometry algebra
of the $(A)dS_4$ vacua are governed by a vector $L^a$ with $L^2=\e$ 
and two real parameters $\alpha,\beta>0$. 
The last column contains the corresponding $\mathfrak{g}_6$-invariant
initial data for the Weyl zero-form.
Two families of foliations with $k=-1$ 
interpolate between the cases with $k=0$
and the instantons.}
\label{Table:1}
\end{table}
\end{landscape}
}
%

\subsubsection{Invariant Weyl-zero form integration constant}

Imposing $\mathfrak{g}_6$-invariance of zero-form initial data, \emph{viz.}
\be
[M_{rs},\Phi']_\pi=0\ ,\qquad  [T_{r},\Phi']_\pi=0\ ,
\label{ic}
\ee
it follows from the first condition that
$$\Phi'=\Phi'(P)\ ,\qquad P:=L^aP_a\ ,$$
and from the second condition that 
\begin{equation}
\left(-\frac{\epsilon\beta\lambda^2}{8}\frac{d^2}{dP^2}+i\epsilon\alpha
\frac{d}{dP}+2\beta\right)\Phi'(P)=0\ .
\label{ce}
\end{equation}
where we have used 
\begin{eqnarray}
L_r{}^a L^b \left[M_{ab},P^n\right]_{\star}
&=&in\epsilon L_r{}^a P_a P^{n-1}
\ ,\\
L_r{}^a\left\{P_a,P^n\right\}_{\star}
&=&L_r{}^a P_a 
\left(2P^n
-\frac{n(n-1)\epsilon\lambda^2}{8}P^{n-2}\right)\ .
\end{eqnarray}
\subsubsection{Regular presentation} 
%
To solve \eq{ce}, we Laplace transform $\Phi'$ as  
\be 
\Phi'=\oint_{C} \frac{d\eta}{2\pi i} \widetilde{\Phi}'(\eta)\exp(-4\eta\lambda^{-1} P)
\equiv {\cal O}\exp(-4\eta\lambda^{-1} P)\ .
\label{Phicontour}
\ee
This gives a characteristic equation for $\eta$ solved by
\be
\eta_\pm= -\gamma\pm\sqrt{\e+\gamma^2}\ ,\qquad
\gamma:=\frac{i\a}{\lambda\beta}\ ,\qquad 
\eta_+ \eta_- = -\e\ ,
\ee
that are either real or purely imaginary. Thus, for $\beta>0$ we have
\bse
\label{c123}
\begin{align}
\mathfrak{so}(1,3)\ :&\quad \widetilde{\Phi}'(\eta)= \frac{\nu_+}{\eta-\eta_+}+ \frac{\nu_-}{\eta-\eta_-}\ ,
\\[5pt]
\mathfrak{iso}(1,2)\,,\ \mathfrak{iso}(3)\ :&\quad \widetilde{\Phi}' (\eta) = \frac{\nu}{\eta+\sqrt{-\epsilon}}+\frac{\sqrt{-\epsilon}\widetilde\nu}{(\eta+\sqrt{-\epsilon})^2}\ ,\label{c2}
\\[5pt]
\mathfrak{so}(4)\,,\ \mathfrak{so}(2,2)\ :&\quad \widetilde{\Phi}'(\eta) 
= \frac{\mu}{\eta-\eta_+}+ \frac{\bar\mu}{\eta-\eta_-}\ .
\end{align}
\ese
The small contours $C$ encircle the poles of $\widetilde{\Phi}'$ counterclockwise.
The corresponding solutions for $\Phi'$ are listed in Table \ref{Table:1} modulo 
rigid $G_{10}$ transformations, that can be used to set $\alpha=0$ for $\e k=-1$
and ${\rm sign}(\e\Lambda)=+1$, \emph{i.e.} the FRW$_+$, the DW$_-$, the
the DW$_+$ in $dS_4$ and the FRW$_-$ in $AdS_4$.

\subsubsection{Limits}
%
The flat solutions with $k=0$ arise from the 
$\mathfrak{so}(1,3)$-invariant families in the limit
\be 
\gamma\rightarrow \sqrt{-\epsilon}\ ,
\label{L1}
\ee
keeping 
\be 
\nu:=\nu_++\nu_-  \ ,\qquad
\widetilde\nu:=\frac{\gamma-\sqrt{\gamma^2+\epsilon}-\sqrt{-\e}}
{\sqrt{-\e}}(\nu_- -\nu_+)\ ,
\label{L2}
\ee
fixed.
In the limit $\beta\rightarrow 0$, one has 
$\eta_+\rightarrow 0$ and $\eta_-\rightarrow\infty$, 
and hence 
\be
\widetilde{\Phi}'= \frac{\nu}{\eta} \ ,\qquad 
\nu\equiv \nu_+\ ,
\ee
and $C$ is a small contour encircling $\eta=0$
counterclockwise\footnote{
Alternatively, the decoupling of the $\eta_+$-mode
in the limit $\beta\rightarrow0$ can be achieved 
using a twisted-adjoint $G_{10}$ conjugation of 
$\Phi'$ that annihilates the $\eta_+$-mode 
using the regularization procedure.}.

\subsubsection{Regularization of star products} \label{regs}
%
In order to compute 
\be
\Psi\star \Psi= \Phi'\star \pi(\Phi')\ ,
\ee
we use the lemma
\bea
e^{-4\eta \lambda^{-1}P}\star e^{-4\eta' \lambda^{-1}P} &=&\frac{1}{(1-\e  \eta\eta')^2}\exp 
\left(-4{\frac{\eta+\eta'}{1-\e  \eta\eta'}\lambda^{-1}P}\right)\ ,\label{3.17}
\eea
and the regularization procedure spelled out in \cite{Iazeolla:2011cb},
\emph{viz.} 
\bea
e^{-4s \lambda^{-1}P}\star e^{4\frac{\e}{ s}\lambda^{-1}P}|_{\rm reg} &=&\oint_s \frac{d\eta}{2\pi i(\eta-s)} e^{-4\eta \lambda^{-1}P}\star 
e^{4\frac{\e}{ s}\lambda^{-1}P}
\nonumber\\
&=& \oint_s \frac{d\eta}{2\pi i(\eta-s)^3} 
\exp\left[4\frac{\eta s-\epsilon}{\eta-s}\lambda^{-1} P 
\right]=0\ ,
\label{regular}
\eea
which suffices to handle the cases with $k\neq 0$.
In the case of $\mathfrak{g}_6=\mathfrak{iso}(3)$, we have 
\begin{eqnarray}
&&\left.\left. \Psi \star \Psi \right\vert _{\mathfrak{iso}(3)}\right|_{\rm reg}  \nonumber \\
&=&\left[ \left( \nu +\tilde{\nu}\lambda ^{-1}P\right) e^{4\lambda ^{-1}P}
\right] \star \left[ \left( \nu -\tilde{\nu}\lambda ^{-1}P\right)
e^{-4\lambda ^{-1}P}\right]   \nonumber \\
&=&\oint_{-1}\oint_{-1}\frac{d\eta d\xi }{\left( 2\pi i\right) ^{2}}\left( 
\frac{\nu }{\eta +1}+\frac{\tilde{\nu}}{\left( \eta +1\right) ^{2}}\right)
\left( \frac{\nu }{\xi +1}+\frac{\tilde{\nu}}{\left( \xi +1\right) ^{2}}
\right) e^{-4\eta \lambda ^{-1}P}\star e^{4\xi \lambda ^{-1}P}  \nonumber \\
&=&\oint_{-1}\oint_{-1}\frac{d\eta d\xi }{\left( 2\pi i\right) ^{2}}\left( \frac{\nu}{ \eta+1 } +\frac{\tilde{\nu}}{(\eta+1)^2}
\right) \left( \frac{\nu}{ \xi+1 } +\frac{\tilde{\nu}}{(\xi+1)^2}
\right) \frac{1}{(1-\eta\xi)^2}e^{4\lambda^{-1}\frac{\xi-\eta}{1-\eta\xi}P}
\nn\\
&=&\oint_{-1}\frac{d\eta }{2\pi i}\left[ \frac{2\tilde{\nu}^{2}\left(
4P-\lambda \right) }{\lambda \left( \eta +1\right) ^{5}}+\frac{\lambda
\left( 2\tilde{\nu}^{2}-\nu \tilde{\nu}\right) +4\left( 2\nu\tilde{\nu} -\tilde{\nu}^{2}
\right)  P}{\lambda \left( \eta +1\right) ^{4}}+\frac{\lambda \left( \nu
^{2}+2\nu \tilde{\nu}\right) -4\nu \tilde{\nu}P}{\lambda \left( \eta
+1\right) ^{3}}\right] e^{-4\lambda ^{-1}P}  \nonumber \\
&=&0\ ;
\end{eqnarray}
and the case of $\mathfrak{g}_6=\mathfrak{iso}(1,2)$ is similar\footnote{Crucial for the regularization is that, after evaluating (any) one of the two integrals, the exponential that results from the star product $e^{-4\eta \lambda ^{-1}P}\star e^{4\xi \lambda ^{-1}P}$ becomes independent of the other auxiliary contour-integral variable. Supposing, for concreteness, that one evaluates the integral over $\xi$ first, as above, the calculation shows that the only assumption one uses in this kind of regularization is that $|\xi +1| << |\eta + 1| << 1$, in order to keep $\xi = -1$ as the only pole encircled by the contour \cite{Iazeolla:2011cb}.}.
It follows that
\be
\Psi\star \Psi|_{\rm reg}={\cal C}^2\ ,
\label{cc}
\ee
where ${\cal C}^2$ is a constant given by
\begin{align} 
\epsilon k = -1\ &:\qquad {\cal C}^2= \frac{\mu^2+\bar\mu^2}4\ ,\\
\epsilon k = 0\ &:\qquad {\cal C}^2= 0\ ,\\
\epsilon k = +1\ &:\qquad {\cal C}^2= 
\frac{(\nu_+)^2}{(1+\epsilon(\eta_+)^2)^2}+
\frac{(\nu_-)^2}{(1+\epsilon(\eta_-)^2)^2}\ ,
\end{align}
where the last case contains the instantons, for which  
${\cal C}^2=\nu^2$.

\subsection{Twistor space connection in holomorphic gauge}

\para{The general solution in integral form.}
In the expression for the twistor space connection, it is convenient to express the 
hypergeometric function in an integral representation as follows
\be
{}_1 F_1(a;b;w)=\frac{\Gamma(b)}{\Gamma(a)\Gamma(b-a)} \int_0^1 \frac{ds}{s(1-s)} s^a (1-s)^{b-a} e^{ws}\ .
\label{1F1int}
\ee
From \eq{s1}, \eq{aa}, \eq{1F1int} and \eq{cc}, one finds that
\bea
A'_\alpha&=&-\frac{ib}\pi  \int_{-1}^{+1} \frac{d\tau}{(\tau+1)^2} 
\int_0^1 ds \sqrt{\frac{1-s}s} 
\nonumber\\ 
&&\left[ z_\alpha  \exp\left(i\frac{\tau-1}{\tau+1} z^+ z^-\right) \right] 
\star\left[\Psi \cosh\left(\frac{b\,{\cal C}s}2 \log\tau^2\right)+
{\cal C}\sinh\left(\frac{b\,{\cal C}s}2 \log\tau^2\right)\right]
\nn\\
&=& \sum_{n\ge 1} A^{\prime (n)}\ .
\label{Aprime}
\eea
Thus, expanding in powers of the deformation parameters, we find that all 
odd terms are linear in $\Psi$, while all even terms 
are $(y^\a,\bar{y}^{\dot{\a}})$-independent, \emph{viz.}
\bea
\left( A_{\alpha }^{\prime }\right) ^{(2k-1)}&=&\frac{-i\Gamma (2k-\frac{3}{2})%
}{\sqrt{\pi }\left( 2k-2\right) !\left( 2k-1\right) !}\left( \frac{b\mathcal{%
C}}{2}\right) ^{2k-2}\left( \frac{b\Psi }{2}\right) \star \left( z_{\alpha
}I_{2k-1}\right) \ ,
\label{odd}\ww2
\left( A_{\alpha }^{\prime }\right) ^{(2k)}&=&\frac{-i\Gamma (2k-\frac{1}{2})}{%
\sqrt{\pi }\left( 2k-1\right) !\left( 2k\right) !}\left( \frac{b\mathcal{C}}{%
2}\right) ^{2k}z_{\alpha }I_{2k}\ ,
\label{even}
\eea
where $k=1,2,\ldots $, and
\be
I_n = I_{n}(w) =\int_{-1}^{1}\frac{d\tau }{\left( \tau +1\right) ^{2}}
\ e^{-w \xi}  \left(\log\tau ^{2}\right) ^{n-1}\ ,
\ee
where
\be
w := iz^+z^-\ ,\qquad  \xi :=\frac{1-\tau}{1+\tau}\ . 
\ee
Let us proceed by looking into the internal connection order by order in its perturbative expansion.
%

\para{First order.}
%
The linearized twistor space connection is given by
\be
\left(A^\prime_\alpha\right)^{(1)}= a_\alpha^{(1)}(z)\star \Psi\
\ee
where  
\be
a_\alpha^{(1)} = -\frac{ib}2  z_\alpha \int_{-1}^{+1} \frac{d\tau}{(\tau+1)^2}  
\exp \left(i\frac{\tau-1}{\tau+1} z^+ z^-\right)
= -\frac{b}{4z^+z^-} z_\alpha\ .
\label{aa2}
\ee
For its basic distributional properties, see remark made above. 
Clearly, in Weyl order, the linearized twistor space connection is not real-analytic 
at the origin of $Z$-space; whether it becomes real analytic in normal order
depends on the details of $\Psi$, as we shall analyze in more detail below.
%

%
%


\para{Second order.} 
%
We have the second order
\be
\left(A'_\alpha\right)^{(2)}=-\frac{ib^2 {\cal C}^2}{16}  z_\alpha I_2(w)\ ,
\qquad I_2(w)=\int_{-1}^{+1} \frac{d\tau}{(\tau+1)^2} 
e^{-w \xi} \log \tau^2\ .
\ee
Thus we can split the integral into two pieces as follows:
\bea
I_2(w) &=&\frac12 \int_0^\infty d\xi e^{-w\xi} \log\left(\frac{1-\xi}{1+\xi}\right)^2
\nonumber\\
&=& \frac12\left(e^{-w} \int_{-1}^\infty d\xi e^{-w \xi}\log \xi^2-e^{w} \int_{1}^\infty d\xi e^{-w \xi}\log \xi^2\right)
\nonumber\\
&=& I^{>}_2(w)+I^{>}_2(-w)\ ,
\eea
where 
\be
I^{>}_2(w)=-\frac{e^{w}}{2w} \int_{w}^\infty d\xi e^{-\xi}\log (\xi/w)^2\ ,
\ee
which is convergent for all real $w$.
For $w>0$, we can integrate by parts and rewrite it as
\be
I^{>}_2(w)=-\frac{e^{w}}{w} {\rm E}_1(w)\ ,
\ee
where the exponential integral 
\be
{\rm E}_1(w)=\int_{w}^\infty \frac{dt}{t} e^{-t}\ ,\qquad w>0\ .
\ee
This function can be extended from the positive real axis to a complex function that is
analytic away from the negative real axis, where is has a Taylor expansion given by
\be
{\rm E}_1(w)=-\gamma_E-\log w-\sum_{p=1}^\infty \frac{(-w)^p}{ p p!}\ ,
\ee
where $\gamma_E$ is the Euler--Mascheroni constant; we note that $w\frac{d}{dw}{\rm E}_1(w)=-\exp(-w)$.
Thus, continuing $I^{>}_2(w)$ to complex $w$, and adding $I^{<}_2(-w)$, we find
\be 
I_2(w) = -\frac{ e^w {\rm E}_1(w)-e^{-w} {\rm E}_1(-w)}{w}\ ,
\ee
which can be rewritten as
\be
I_2(w)= R_1(w)+ R_2(w) \log w\ ,
\ee
where $R_{1,2}$ are real analytic at $w=0$:
\be
R_1(w)=\frac{2\gamma \sinh w}{w}
- e^w \sum_{p=1}^{\infty}\frac{(-w)^{p-1}}{p\ p!}
- e^{-w} \sum_{p=1}^{\infty}\frac{w^{p-1}}{p\ p!}
\ ,\qquad 
R_2(w)=\frac{2 \sinh w}{w}\ .
\ee
Therefore, in summary the second order correction $\left(A'_\alpha\right)^{(2)}$ is 
independent of $Y$ and bounded in $Z$, though it is not real analytic at $Z=0$. 
%

\subsection{Master fields in \texorpdfstring{$L$}{L}-gauge}

We recall that starting from the particular solution
obtained in the the holomorphic gauge, which incorporates 
the zero-form initial data, gauge inequivalent solutions 
can be reached by means of large gauge transformations
generated by gauge functions $G$ defined locally on patches.
In the case of asymptotically (anti-)de Sitter spacetimes,
we use $G=L\star H$, where $L$ is the vacuum gauge function,
which brings the master fields to what we refer to as 
the $L$-gauge, after which $H$ is constructed order by order 
by imposing the dual boundary conditions (a) and (b) 
specified in Section \ref{gf2}.
Finally, the patches are glued together using transition
functions belonging to a structure group.

\subsubsection{Weyl zero-form} 

The Weyl zero-form in L-gauge is given by
\be
\Phi^{(L)}(y,\bar y) = L^{-1} \star \Phi' \star \pi(L) \ .
\ee
Substituting $\Phi^\prime= \Psi\star \kappa_y$ according to the ansatz \eq{s1} we get
\be
\Phi^{(L)}(y,\bar y)=\Psi^L\star \kappa_y\ ,\qquad 
\Psi^L := L^{-1}\star \Psi\star L \ .
\label{phiL}\ee
We first compute  
\be
\Psi \equiv \Phi^\prime \star \kappa_y = \left( {\cal O} e^{-4\eta \lambda^{-1} P}\right) \star \kappa_y 
 = 2\pi {\cal O} \delta^2
 \left(y_\alpha-b_a (\sigma^a\bar y)_\alpha\right)\ ,\qquad b^a:=i\eta L^a\ ,
\ee
where we have used \eq{Phicontour}. 
The $L$-conjugate of $\Psi$ is given by  
\be
\Psi^L 
= 2\pi{\cal O} \delta^2\left(y^L_\alpha-b_a (\sigma^a\bar y^L)_\alpha\right)\ ,
\label{PsiL}
\ee
where 
\be
\left[\begin{array}{c} y^L_\alpha\\\bar y^L_{\dot\alpha}\end{array}\right] = 
L^{-1}\star \left[\begin{array}{c} y_\alpha\\\bar y_{\dot\alpha}\end{array}\right]\star L
= \left[\begin{array}{cc} L_\a{}^\beta& K_\a{}^{\dot\beta}\\ 
\overline K_{\dot\a}{}^\beta & \overline L_{\dot\alpha}{}^{\dot\beta}\end{array}\right] 
\left[\begin{array}{c} y_\b\\\bar y_{\dot\beta}\end{array}\right]\ 
\label{KL},
\ee
form a new set of canonical coordinates in which 
\be \rho( (y^L_\a)^\dagger)=\bar y^L_{\dot\alpha}\ ,\qquad 
\rho( (\bar y^L_{\dot\a})^\dagger)={\rm sign}(\l^2) y^L_{\dot\alpha}\ ,\qquad \kappa_y\bar\kappa_{\bar y}=\kappa_{y^L}\bar\kappa_{\bar y^L}\ .\ee
The matrices $K$ and $L$ are computed in stereographic and planar coordinate systems in Appendix \ref{SecGaugeFuncStereo} and \ref{SecGaugeFuncPlanar}, respectively. 
It follows that indeed
\be 
(\Psi^L)^\dagger=\Psi^L\star \kappa_y\bar\kappa_{\bar y}\ ,\qquad
\Psi^L\star \Psi^L|_{\rm reg}={\cal C}^2\ ,\ee
where ${\cal C}$ is the constant in \eq{cc}, as can be seen using the lemma
\be
(2\pi)^2 \delta^2\left(y^L_\alpha-b_a (\sigma^a\bar y^L)_\alpha\right)\star
\delta^2\left(y^L_\alpha-\tilde b_a (\sigma^a\bar y^L)_\alpha\right)
=\frac{1}{(1+\eta\tilde\eta \epsilon)^2}
\exp\left[\frac{\eta-\tilde\eta}{1+\eta\tilde\eta\epsilon} L_a (\sigma^a)^{\a\dot\a} y^L_\a \bar y^L_{\dot\a}\right]\ ,
\ee
where $\tilde b_a=i\tilde \eta L_a$, followed by contour integration.
Going back to the original canonical coordinates for ${\cal Y}_4$, we have
\be
\Psi^L = 2\pi{\cal O} \delta^2\left((Ay+B\bar y)_\alpha\right)\ ,
\label{psiL}
\ee
where
\be 
A_\a{}^\b:=L_\a{}^\b-b_a(\sigma^a \overline K)_\a{}^\b\ ,\qquad
B_\a{}^{\dot\b}:=K_\a{}^{\dot\b}-b_a(\sigma^a \overline L)_\a{}^{\dot\b}\ ,
\label{AB1}
\ee
Provided that $A$ is invertible, we can write
\be
\Psi^L = {\cal O} \left( \frac{2\pi}{\det A} \delta^2(\tilde y_\a)\right)\ ,\qquad 
\tilde y_\a := y_\alpha+M_\a{}^{\dot\beta}\bar{y}_{\dot \beta}\ ,\qquad
M = A^{-1} B\ .
\label{M}
\ee
We thus find
\be
\Phi^{(L)}(y,\bar y) = \Psi^L \star \kappa_y
= {\cal O} \left(\frac{2\pi}{\det A} \delta^2(\tilde y) \star \kappa_y\right)\ .
\label{scalar2}
\ee
which is readily computed with the result
\be
\Phi^{(L)}(y,\bar y) = 
{\cal O} \left(\frac{1}{\det A} e^{i y^\a M_\a{}^{\dot\a}\bar{y}_{\dot\a}}\right)= 
\oint_C \frac{d\eta}{2\pi i}  \frac{{\widetilde\Phi}^\prime(\eta)}{\det A} e^{i y^\a M_\a{}^{\dot\a}\bar{y}_{\dot\a}}\ ,
\label{defO}
\ee
with $\widetilde\Phi^\prime (\eta)$ from \eq{c123}.
The resulting Weyl zero-forms consist of scalar field profiles, 
that we shall analyze in more detail in Section \ref{SecRegular} using
stereographic coordinates, and in Appendix \ref{SecGaugeFunc} using
adapted coordinate systems.

\subsubsection{Twistor space connection at even orders} 
%
The even order terms are the same in the holomorphic gauge and the $L$-gauge, 
as they are independent of $Y$. From \eq{even}, the sum of all even orders is given by 
\be 
(A_\alpha^{(L)})^{({\rm even})}= -\frac{ib\,{\cal C}}\pi  
z_\alpha I_{\rm even}(w;b\,{\cal C}) \ ,
\label{tcs1}
\ee
where 
\be 
I_{\rm even}(w;\mu) =\int_{-1}^{+1} \frac{d\tau}{(\tau+1)^2} 
\int_0^1 ds \sqrt{\frac{1-s}s} e^{-w\xi}\, \sinh\left(\frac{\mu s}2 \log\tau^2\right)\ .
\label{wm}
\ee
We note that $(A_\alpha^{(L)})^{({\rm even})}$ is independent of $X$ and $Y$, and bounded in $Z$-space.

\subsubsection{Twistor space connection at odd orders} 

In the $L$-gauge, the sum of all odd-order terms from \eq{odd} is given by
\be 
(A_\alpha^{(L)})^{({\rm odd})}= -\frac{ib}\pi  \partial_\alpha^{(\rho)} V(\rho)\mid_{\rho=0}\ ,
\ee
where the generating function
\be
V(\rho) =\int_{-1}^{+1} \frac{d\tau}{(\tau+1)^2} 
\int_0^1 ds \sqrt{\frac{1-s}s} \exp\left(-\frac{i\xi}{2}  u^{\alpha\beta}z_\alpha z_\beta+\rho^\alpha z_\alpha \right)\star \Psi^{L}  
\cosh\left(\frac{b\,{\cal C} s}2 \log\tau^2\right)\ ,
\ee
and $u^{\a\b}:= 2u^{+(\a} u^{-\b)}$. Substituting for 
$\Psi_L= {\cal O}\left(e^{-4\eta \lambda^{-1} P_0} \star\kappa_y \right)^L$ gives
\be
(A_\alpha^{(L)})^{({\rm odd})}= -\frac{ib}\pi  \partial_\alpha^{(\rho)} {\cal O} V(\eta;\rho)\mid_{\rho=0}\ ,
\label{sc1}
\ee
where the extended generating function
\be
\begin{split}
V(\eta;\rho) &= \int_{-1}^{+1} \frac{d\tau}{(\tau+1)^2} 
\int_0^1 ds \sqrt{\frac{1-s}s} \times
\\
&\times \exp\left( -\frac{i\xi}{2}  u^{\alpha\beta}z_\alpha z_\beta+ \rho^\alpha z_\alpha\right) \star  
\left(e^{-4\eta \lambda^{-1} P_0} \star \kappa_y \right)^L  \cosh\left(\frac{b{\cal C} s}2 \log\tau^2\right)\ ,
\end{split}
\label{v2}
\ee

Next we perform the Gaussian star product
\bea
&& \exp\left(-\frac{i\xi}2  u^{\alpha\beta}z_\alpha z_\beta
+\rho^\alpha z_\alpha\right)\star (e^{-4\eta \lambda^{-1} P_0} \star \kappa_y )^L
\nonumber\\
&& =\frac{1}{\xi \det A } \exp\left(\frac{i}{2\xi} u^{\alpha\beta} (\rho_\alpha-i\tilde y_\alpha)(\rho_\beta-i\tilde y_\beta)-i z^\alpha \tilde y_\alpha\right)\ .
\label{v3}
\eea
It follows that $(A_\alpha^{(L)})^{({\rm odd})}$ is real analytic in $Z$-space,
which simplifies the construction of $H$, and that it has singularities in $Y$-space,
stemming from the divergence at $\tau=+1$ that arises as a result of the above
Gaussian integration.
Thus, we need to demonstrate that the latter singularities go away upon switching on $H$.
From \eq{sc1}, \eq{v2} and \eq{v3} we find 
\be
 (A_\alpha^{(L)})^{({\rm odd})}={\cal O}  V_\alpha(\eta)\ ,
\label{scodd}
\ee
where the generating function 
\be
V_\alpha(\eta)= \frac{ib}{\pi} u_\alpha{}^\beta\tilde y_\beta \frac{e^{i  \tilde y^\alpha z_\alpha }}{\det A}\int_{-1}^{+1} \frac{d\tau}{(\tau-1)^2} 
\int_0^1 ds \sqrt{\frac{1-s}s} \exp \left(\frac{i}{\xi} \tilde y^+ \tilde y^-\right)\,\cosh\left(\frac{b{\cal C} s}2 \log\tau^2\right)\ ,
\ee
consists of even orders in deformation parameters, and 
its $\eta$ dependence enters via $A$ and $\tilde y_\alpha$.

\subsubsection{Spacetime connection}

The spacetime connection is simply given by
\be
A_\mu^{(L)} = L^{-1}\star \partial_\mu L\ ,
\ee
while form \eq{Wdef} one finds
\be
W_\mu^{(L)} =  L^{-1}\star \partial_\mu L -\frac{1}{4i} \left( L^{-1}\star 
M^{({\rm tot})}_{\a\b} \star L +h.c.\right)\ ,
\label{WLG}
\ee
with $M^{({\rm tot})}_{\a\b}$ from \eq{LT}.

\subsubsection{Patching} 

The expressions given so far are defined in the
region of validity of the gauge function $L$, that is,
for $\lambda^2 x^2<1$, whereas a global formulation on the
the vacuum manifold $M_4^{(0)}$ requires the usage
of several coordinate charts.
A simple configuration consists of two gauge functions
$L_\pm$ defined on two stereographic coordinate charts 
$U_\pm$ with $1>\lambda^2 x_\pm^2\geqslant -1$ glued 
together along $\lambda^2 x_\pm^2=-1$, which implies
that the transition function is trivial since
\be
L_\pm|_{\lambda^2 x_\pm^2=-1}=L_\mp|_{\lambda^2 x_\mp^2=-1}\ ,
\ee
\emph{i.e.} this particular configuration can be
implemented for any choice of structure group,
that is, in any topological phase of the theory.
In these types of configurations, it follows from
the reflection symmetry that any singularity 
in the master fields that arises inside $U_\pm$ 
cannot be removed using patching.

\subsection{Reaching Vasiliev gauge at first order}

\subsubsection{The Weyl zero-form} At first order we observe that
\be
\Phi^{(G,1)}(y,\bar y) =\Phi^{(L)}(y,\bar y) 
= {\cal O} \left( \frac{1}{\det A} e^{i y^\a M_\a{}^{\dot\a}\bar{y}_{\dot\a}}\right) \ ,
\label{Gsc}
\ee

\subsubsection{Twistor space connection} At the linearized level, the role of $H^{(1)}$ is 
to ensure that 
\be
A^{(G,1)}_{\ua}:=A^{(L,1)}_{\ua}+\partial_{\ua} H^{(1)}
\ee
is real analytic in ${\cal Y}_4\times {\cal Z}_4$ and obeys
the Vasiliev gauge condition,
\be z^{\ua} A^{(G,1)}_{\ua}=0\ .\ee
Since $A^{(L,1)}_{\ua}$ is real analytic in ${\cal Z}_4$, 
it follows that
\be 
H^{(1)}=H^{(1)}|_{Z=0}-\frac1{{\cal L}_{\vec Z}} 
\left(z^\a A^{(L,1)}_{\alpha}+\bar z^{\dot\a} A^{(L,1)}_{\dot\alpha}\right)\ ,
\ee
where ${\cal L}_{\vec Z}=\{q,\imath_{\vec Z}\}$ is the 
Lie derivative along the Euler vector field
\be \vec Z=Z^{\ua}\vec\partial_{\ua}\ ,\ee
whose invserse can be represented (on real analytic functions) as
\be \frac1{{\cal L}_{\vec Z}}=\int_0^1 \frac{dt}{t} 
t^{{\cal L}_{\vec Z}}\ ,\ee
where $t^{{\cal L}_{\vec Z}}$ acting on differential forms implements 
the diffeomorphism $z^{\underline\alpha}\rightarrow t z^{\underline\alpha}$.
Thus, 
\be 
A^{(G,1)}_{\alpha}=\left(\delta_\a{}^\beta 
- \partial_\a \frac1{{\cal L}_{\vec Z}} z^\beta\right)A^{(L,1)}_{\beta}\ ,\ee
to which the initial datum $H^{(1)}|_{Z=0}$ does not contribute,
and we have taken into account the holomorphicity in $Z$ space.
Writing 
\be 
A^{(L,1)}_{\ua}={\cal O}V^{(0)}_{\ua}(\eta)\ ,\qquad 
V^{(0)}_\alpha (\eta)
=\frac{ib}{2} u_\alpha{}^\beta\tilde y_\beta \frac{e^{i\tilde y^\alpha z_\alpha }}{\det A}\int_{-1}^{+1} \frac{d\tau}{(\tau-1)^2} 
\exp\left(\frac i2\frac{\tau+1}{\tau-1}u^{\a\b}\tilde y_\a \tilde y_\b\right)\ ,
\ee
it follows that
\be 
H^{(1)}=H^{(1)}|_{Z=0}-{\cal O}\frac1{{\cal L}_{\vec Z}} 
\left(z^\a V^{(0)}_{\alpha}(\eta)+\bar z^{\dot\a} V^{(0)}_{\dot\alpha}(\eta)\right)\ ,
\ee
and
\be 
A^{(G,1)}_{\alpha}={\cal O}
\left(\delta_\a{}^\beta - \partial_\a \frac1{{\cal L}_{\vec Z}} z^\beta\right) 
V^{(0)}_{\beta}(\eta)\ .
\ee
Using
\be 
\int_0^1 dt t^{{\cal L}_{\vec Z}}e^{i\tilde y^\alpha z_\alpha}
=\frac{e^{i\tilde u}-1}{i\tilde u}\ ,\qquad \tilde u=\tilde y^\a z_\a\ ,
\ee
and
\be 
u^{\a\b}\tilde y_\a \tilde y_\b\int_{-1}^{+1} \frac{d\tau}{(\tau-1)^2} 
 \exp\left(\frac i2 \frac{\tau+1}{\tau-1}u^{\gamma\delta}
 \tilde y_\gamma \tilde y_\delta\right)=-i\ ,
 \ee
in accordance with Eq. \eq{adist}, 
one finds 
\be
A^{(G,1)}_{\alpha}= -\frac{ib}2 z_\alpha {\cal O} 
\frac{e^{i\tilde u}-1-i\tilde ue^{i\tilde u}}{\tilde u^2\det A}\ .
\label{Gsc2}
\ee
where the dependence on the auxiliary spinor frame $u^\pm_\alpha$ has dropped out.
Indeed, the above result agrees with that found working directly in normal order, \emph{viz.}
\be
A^{(G,1)}_{\alpha}= -\frac{ib}2 z_\alpha\int_0^1 dt \ t \ e^{i t y^\a z_\a} 
\left[\left. \Phi^{(G,1)}(y,\bar y)\right|_{y\rightarrow -tz}\right]\ ,
\label{ms1}
\ee
as can be seen using $\Phi^{(G,1)}(y,\bar y) =\Phi^{(L)}(y,\bar y)$
and 
\be
\int_0^1 dt \ t \ e^{i t y^\a z_\a} \left[\left. 
\Phi^{(G,1)}(y,\bar y)\right|_{y\rightarrow -tz}\right]= {\cal O}
\frac{e^{i\tilde u}-1-i\tilde ue^{i\tilde u}}{\tilde u^2\det A}\ .
\ee
%
\subsubsection{Spacetime connection} \label{STC}
%
In the Vasiliev gauge, the linearized spacetime connection 
\be
dx^\mu A_\mu^{(G,1)} = L^{-1}dL + D^{(0)} H^{(1)} \equiv 
L^{-1}dL + U^{(G,1)}\ ,
\label{AVG}
\ee
where the background covariant derivative
\be 
D^{(0)}=d+\frac{1}{4i} \Omega^{\underline{\a\b}}{\rm ad}^\star_{Y_{\ua}Y_{\ub}}\ .
\ee
As the linearized Weyl zero-form consists of a scalar mode,
it follows that there exists an initial datum $H^{(1)}|_{Z=0}$ 
such that $U^{(L,1)}|_{Z=0}$ is a pure (abelian) gauge field 
on ${\cal X}_4$ that is real-analytic on ${\cal Y}_4\times {\cal Z}_4$. 
To corroborate this fact, we first use $[\partial_\alpha,D^{(0)}]=0$ 
and $D^{(0)}A_\alpha^{(L,1)}=0$ to deduce that
\be \partial_\alpha U^{(G,1)}= \partial_\alpha( D^{(0)} H^{(1)})=
 D^{(0)} (\partial_\alpha H^{(1)})= D^{(0)} (A_\alpha^{(G,1)}-
 A_\alpha^{(L,1)})=D^{(0)} A_\alpha^{(G,1)}\ .\ee
Thus, as $A_\alpha^{(G,1)}$ is real analytic on ${\cal Y}_4\times {\cal Z}_4$ 
and independent on the auxiliary spin frame, it follows that 
these properties hold true as well for the 
$Z$-dependent part of $U^{(G,1)}$.
As for its $Z$-independent part, \emph{viz.}
\be 
U^{(G,1)}|_{Z=0} = D^{(0)} \left(H^{(1)}|_{Z=0}\right)-{\cal O}
\left.\left(D^{(0)} \frac1{{\cal L}_{\vec Z}} 
\left(z^\a V^{(0)}_{\alpha}(\eta)+\bar z^{\dot\a} 
V^{(0)}_{\dot\alpha}(\eta)\right)\right)\right|_{Z=0}\ ,
\ee
the requirement of real-analyticity on ${\cal Y}_4$ fixes 
$H^{(1)}|_{Z=0}$ up to a residual real analytic part 
(in $\mathfrak{hs}_1(4)$), with the desired result
\be U^{(G,1)}|_{Z=0}= 0\ ,\ee
modulo a pure gauge term;
for details, see Appendix \ref{App:H1}.

\subsection{Comments on residual symmetries, factorization
and Vasiliev gauge}

We recall from Section \ref{gf2} that as far as symmetry considerations 
are concerned in finding exact solutions, these are facilitated by the
the combined use of gauge functions and the holomorphic factorization 
method employed in \eq{s1}, which ensures that the symmetries of 
the initial datum $\Psi(Y)$, that can be imposed by means of undeformed generators, remain symmetries of the full master fields. 
We would like to contrast this approach to that followed in \cite{Sezgin:2005pv}, 
where an exact $\mathfrak{so}(1,3)$-invariant
solution was constructed for $\Lambda<0$ using the vacuum gauge function $L$ followed
by requiring the primed twistor space configuration 
to be invariant under the full field-dependent Lorentz 
generators (instead of using the holomorphic factorization method).  
In the same paper, the six-parameter symmetry groups considered 
here were also examined, but as the factorization method was not
used, the imposition of symmetry conditions involving translations
became problematic at the nonlinear level, and FRW-like 
and domain wall solutions were given explicitly at the linearized 
level, in agreement with Table \ref{Table:1}, and shown to exist 
at the second order of classical perturbation theory. 
It would be interesting to pursue the latter construction
to the second order and compare it to the second order
expressions obtained in the current paper in $L$-gauge.

The factorization method implies, however, that the linearized 
master gauge fields are not real analytic in 
$(y^\alpha,\bar y^{\dot\alpha})$ in $L$-gauge, but
as we have seen, these singularities can be removed 
by going to Vasiliev gauge by means of a large gauge 
transformation.
It remains to be shown whether this procedure can be
imposed to order by order in weak field perturbative 
expansion by imposing  dual boundary conditions as 
discussed in Section \ref{gf2}.

We conclude this section by explaining technically the reason for 
being able to impose equally the first of the conditions 
\eq{ic} via the full Lorentz generator \eq{LT}. First, defining $ 
\e_L^{(tot)} :=  \frac{1}{4i}\,\Lambda^{\a\b}M_{\a\b}^{(tot)}-\textrm{h.c.}  =  \e_L^{(0)}+\e_L^{(\textrm{extra})} 
 :=  \frac{1}{4i}\,\Lambda^{\a\b}M_{\a\b}^{(0)}-\textrm{h.c.} + 
 \frac{1}{4i}\,\Lambda^{\a\b}S_\a\star S_\b-\textrm{h.c.}$, one 
 can show \cite{Vasiliev:1990vu,Vasiliev:1999ba,Sezgin:2002ru} 
 that the fully nonlinear completion of the Lorentz generator is 
 exactly such that, on the solutions of the Vasiliev equations, 
 $\d_{L}\Phi(Y,Z) \ = \ -[\e_L^{(tot)},\Phi]_\pi \  = \  
 -[\e_L^{(0)},\Phi]_\pi$ .
The reason is that \eq{SPhi} implies $[\e_L^{(\textrm{extra})}, \Phi]_\pi=0$. 
It is then clear that, on a purely $Y$-dependent Weyl zero-form, such as 
that of our Ansatz \eq{s1}, the action of the Lorentz generators reduces 
to the one of their zeroth-order, purely $Y$-dependent piece, and we 
can therefore impose $\mathfrak{so}(3)$-symmetry as in \eq{ic}. In general, 
however, imposing invariance conditions on the master fields under 
subalgebras that include translations can only be done perturbatively, 
as it was suggested in \cite{Sezgin:2005pv}, since no fully non-linear 
completion of the $P_a$ is known. This limitation is not present on the 
subspace of the full solution space captured by the factorized Ansatz \eq{s1}, where $\Phi$ is first-order-exact. 

The virtue of the factorization (combined with the gauge function method) is that it gives us the possibility of solving exactly for the $Z$-dependence of the master fields irrespectively of the initial datum $\Phi'(Y)$, thereby dressing a solution of the linearized twisted-adjoint equation into a full solution of the Vasiliev equations. In particular, due to the factorized form, the equations for $S'_\a$ reduce to \eq{defred}, that do not involve $\Phi'$ and can be therefore solved once and for all (via the methods developed in \cite{Iazeolla:2011cb,Iazeolla:2017vng}). This allows us to impose the $\mathfrak{g}_r$ symmetries at full level, since the action of symmetry parameters $\e(Y)$ is sufficient to impose symmetry conditions on the full solution space \eq{s1}. Indeed, the symmetries $\e(Y)$ of $\Phi'$ (that is, the parameters such that $[\e(Y),\Phi'(Y)]_\pi=0$ ) are also symmetries of the \emph{full} $S'_\a$, since $[\e(Y),S'_\a]=0$ holds if $[\e(Y),\Psi]=0$, which in its turn is implied by $[\e(Y),\Phi'(Y)]_\pi=0$.

\section{Regularity of full master fields on correspondence space}
\label{SecRegular}

In this section, we examine the scalar field profiles and the
Weyl zero-form using the stereographic coordinates 
(see Appendix \ref{SecCoord} and \ref{SecGaugeFunc} for details), which facilitate a uniform treatment 
of all cases. 
We first study the linearized scalar field profile and 
then turn to the analysis of the regularity of the full Weyl zero-form 
in the correspondence space, by which we mean the twistor space extended 
spacetime with coordinates $(x,Y,Z)$.

\subsection{Linearized scalar field profile}

In stereographic coordinates, the linearized scalar field is given by
\be 
\phi^{(1)} = {\cal O} \frac1{\det A} \equiv {\cal O}\phi_\eta\ ,\qquad
\phi_\eta:= \frac{h^{2}}{1-2\lambda b^a x_a + \lambda^2 b^2 x^2}\ ,
\label{mf2}
\ee
which can be re-written as
\be \phi_\eta=\frac{1}{h^{-2}(\lambda x-b)^2 + 1-b^2}
=\frac{h^2}{b^2(\lambda x+ R(b))^2}\ ,
\qquad b^a=i \eta L^a\ ,\ee
where $R$ is the reflection map.
For $\beta>0$ and $k\neq 0$, we have
\be 
|k|=1\ , \quad \beta>0\ :\quad \phi^{(1)}=\sum_{i=\pm}\nu_i \phi_{\eta_i}\ ,
\ee
where $\eta_\pm=-\gamma\pm\sqrt{\e+\gamma^2}$, with 
$\gamma=\frac{i\alpha}{\lambda\beta}$ and $\e=L^a L_a$,
such that that $R^a(b_\pm)=-b^a_\mp$, and 
$\nu_+:=\mu$ and $\nu_-=\bar\mu$ for $\e k=-1$.

In the limit $\beta\rightarrow 0$, one has $\eta_-\rightarrow \infty$ and $\eta_+\rightarrow 0$, 
and hence
\be \beta=0\ :\quad \phi^{(1)}=\nu_+ h^2\ .\ee

The case of $k=0$ arises in the limit \eq{L1} and \eq{L2}. More directly, from 
\be
\phi^{(1)} = \oint_{C} \frac{d\eta}{2\pi i} \widetilde{\Phi}'(\eta) \phi_\eta\ ,
\ee
using \eq{mf2} and \eq{c2} readily gives
\be
k=0 :\qquad \phi^{(1)} = \frac{1-\lambda^2 x^2}{1+2 i \lambda \sqrt{-\epsilon} L^a x_a +\lambda^2 x^2}
\left[ \nu +\frac{2\widetilde\nu (i \lambda \sqrt{-\epsilon} 
L^a x_a+\lambda^2 x^2)}{1+2 i \lambda \sqrt{-\epsilon} L^a x_a +\lambda^2 x^2}\right]\ ,
\ee
where $\lambda=i\ell^{-1}$ and $\epsilon=-1$ for $\mathfrak{iso}(3)$, and $\lambda=\ell^{-1}$ and $\epsilon=1$ for $\mathfrak{iso}(1,2)$. In the planar coordinate system defined in \eq{plan1} and \eq{plan2}, this solutions takes the simple form given in \eq{scalarSolution} which will be discussed in more detail in 
Section \ref{SecCosm}.

The iso-scalar surfaces $S_i(c_i)$ ($i=\pm$) are defined by
\be \phi_{\eta_i}|_{S_i(c_i)}=c_i\ ,\qquad c_i\in\Comp\ ,\ee
which are $\mathfrak{g}_6$ invariant, and complexified for $\e k=-1$;
in particular, $S_i(0)$ is the boundary.
Away from the boundary, we have
\be \left.(1+b_i^2-2\lambda b^a_i x_a-(c_i^{-1}+b_i^2)h^2)\right|_{S_i(c_i)}=0\ ,\qquad c_i\,,\ h^2\neq 0.\ee
It follows that 
\be S_+(c_+)=S_-(c_-)\quad \Leftrightarrow \quad \frac{\eta_-}{c_+}+\eta_+=\frac{\eta_+}{c_-}+\eta_-\ ,\label{relatec}\ee
as can be seen first by eliminating $h^2$, which yields 
\be (c_+^{-1}+b_+^2)(1+b_+^2-2\lambda b^a_+ x_a)=
(c_-^{-1}+b_-^2)(1+b_-^2-2\lambda b^a_- x_a)\ ,\ee
that must hold identically for all $x^a$ modulo $\eta_- b_+^a=\eta_+ b_-^a$, 
$\eta_-(1+b_+^2)=\eta_+(1+b_-^2)$, and $\eta_+\eta_-=-\e$,
provided that $c_\pm$ obey the relation in \eqref{relatec}.
Taking the limit $|c_\pm|\rightarrow \infty$, it follows 
from \eqref{relatec} that if $k\neq 0$, so that $\eta_+\neq \eta_-$,
then $S_+(\infty)\cap S_-(\infty)=\emptyset$ away from the boundary; 
conversely, requiring $S_+(\infty)\cap S_-(\infty)\neq \emptyset$ 
implies that
\be h^2=0\ ,\qquad L^a x_a=\frac{i\e}2 (\eta_++\eta_-)\ ,\ee
\emph{i.e.} the two singular surfaces coincide on a two-dimensional
subspace of the boundary.
If $k= 0$, then $\eta_+= \eta_-$, and
it follows from \eqref{relatec} that $S_+(c)=S_-(c)$
for all $c$.
Moreover, if $\e k=-1$, then $S_\pm(\infty)\cap M_4^{(0)}=\emptyset$,
while if $\e k=0,1$, then $S_\pm(\infty)\cap M_4^{(0)}$ is the cone
\be 
\widetilde{x}^2_\pm \ = \ 0 \ , \qquad \widetilde{x}^a_\pm := \l x^a +R^a(b_\pm)\ .
\label{cones}
\ee 
%

\subsection{(Ir)regularity of Weyl zero-form}

As the description of the solutions in terms 
of Fronsdal fields is reliable
only at weak coupling, we resort to the full master 
fields close to the surfaces $S_\pm(\infty)$.
The Weyl zero-form (see \eq{phiL}, \eq{PsiL} and \eq{psiL})
is given by
\be \Phi^{(L)} \ = \ 2\pi{\cal O}\,\d^2(Ay+B\bar{y})\star \k_y \ ,\label{uniWeyl}\ee
which is regular on $M_4^{(0)}$ for $\e k=-1$, and degenerates on
the cones in Eq. \eq{cones} for $\e k=0,1$.
From \eq{symbols} and \eq{ABMstereo}, at the apexes and for $\e k=1$, we have  
\be\label{AB}
\begin{split}
A_\a{}^\b|_{\widetilde x^a_\pm=0,\ \eta=\eta_\pm}&= h_\pm^{-1}\left(\d_\a{}^\b + (\slashed{b}_\pm \bar{\slashed{R}}(b_\pm))_\a{}^\b \right) \ = \ h_\pm^{-1}\left( 1-\frac{b^2_\pm}{b^2_\pm} \right)\d_\a{}^\b \ = \   0\ ,
\\
\qquad B_\a{}^{\dot{\beta}}|_{\widetilde x^a_\pm=0,\ \eta=\eta_\pm} &= h_\pm^{-1}(-\slashed{R}(b_\pm)-\slashed{b}_\pm)_\a{}^{\dot{\beta}} \ = \  h_\pm^{-1}\left(\frac{1}{b^2_\pm}-1\right) (\slashed{b}_\pm)_\a{}^{\dot{\beta}}\ = \ -\sqrt{1-\frac{1}{b^2_\pm}} \,(\slashed{b}_\pm)_\a{}^{\dot{\beta}} \ ,
\end{split}
\ee
from which it follows that
\be 
\e k=+1\ :\quad \Phi^{(L)}_\pm|_{\widetilde x^a_\pm=0}=\frac{\nu_\pm}{1+\epsilon \eta^2_\pm}\,\k_y\bar{\k}_{\bar{y}}\ ,
\label{apexlimit}
\ee
which are regular $\mathfrak{so}(1,3)$-invariant elements in ${\cal A}({\cal Y}_4)$.
For $k=0$, on the other hand, the Weyl zero-form diverges
at the apex, which now resides at the boundary.

At the light cones $\tilde x^2_\pm=0$, away from the apexes, 
we can treat the case $\epsilon k=+1$ by writing
\be 
\lambda \slashed{\tilde x}_{\pm \a\dot\a} = \tau  u^{(+)}_\a \bar u^{(+)}_{\dot\a}\ ,\qquad 
\slashed{b}_{\a\dot\a}=i\eta (u^{(+)}_\a \bar u^{(+)}_{\dot\a}-\e  u^{(-)}_\a \bar u^{(-)}_{\dot\a})\ ,
\ee
where $\tau=\e \tau^\dagger$ is a linear coordinate along the 
lightcone and $u^{(\pm)}_\a$ is a normalized $x$-dependent spin frame, \emph{viz.}
\be 
u^{(+)\alpha } u^{(-)}_\alpha=1\ ,\qquad \delta_\a^\b=u^{(-)}_\a u^{(+)\b} 
-u^{(+)}_\a u^{(-)\b} \ ,\qquad \delta^2(v_\a)=\delta (u^{(+)\a} v_\a)\delta (u^{(-)\a}v_\a)\ ,
\ee
and $\bar{u}^{(\pm)}_{\dot\a}=(u^{(\pm)}_\a)^\dagger$. 
It follows from \eq{cones} that 
\be 
\lambda x_{\a\dot\a}|_{\tilde x^2_\pm=0}= \left(\tau-\frac{i\e}{\eta_\pm}\right) u^{(+)}_\a \bar u^{(+)}_{\dot\a}
+\frac{i}{\eta_\pm}u^{(-)}_\a \bar u^{(-)}_{\dot\a}\ ,\qquad
h^2|_{\tilde x^2_\pm=0}\equiv h^2_\pm=\frac{1+\epsilon \eta^2_\pm+i\epsilon \eta_\pm\tau}{\epsilon \eta^2_\pm}\ ,
\ee
and hence
\bea
A_\a{}^\b|_{\tilde x^2_\pm=0\,,\,\eta=\eta_\pm}&=&-i  \e \eta_\pm \tau 
h^{-1}_\pm  u^{(-)}_\a u^{(+)\b}\ ,
\nn\\
B_\a{}^{\dot\b}|_{\tilde x^2_\pm= 0\,,\,\eta=\eta_\pm}&=& h^{-1}_\pm \left(\tau  u^{(+)}_\a \bar u^{(+)\dot\b}
-i \left(\eta_\pm+\frac{1}{\e\eta_\pm}\right)
\left(u^{(+)}_\a \bar u^{(+)}\dot\b-\e  u^{(-)}_\a 
\bar u^{(-)\dot\b}\right)\right)\ .
\eea
Using
\be 
\int_{\Real^2} ds dt \exp(ias t)=\frac{2\pi}{a}\ ,
\ee
which provides a normalization of the real analytic delta function 
in two variables, corresponding to the delta sequence
\be \lim_{a\to 0} a\exp(iast)= 2\pi \delta(s)\delta(t)\ ,\ee
and staying away from the apex, \emph{i.e.} taking $\tau\neq 0$,
we find that
\bea 
\Phi_\pm^{(L)}|_{\tilde x_\pm^2=0\,,\, \tau \neq 0}
&=&\nu_\pm h^2_\pm \int d\xi^{(+)} d\xi^{(-)} e^{i(y^{(-)}\xi^{(+)}-y^{(+)}\xi^{(-)})}
\delta\left(-i\epsilon \left(\eta_\pm \tau \xi^{(+)} - (\eta_\pm+\frac{1}{\epsilon\eta_\pm})\bar y^{(-)}\right)\right)
\nn\\
&& 
\qquad\quad\qquad   \times \delta\left(\tau \bar y^{(+)}-i(\eta_\pm+\frac{1}{\e\eta_\pm})\bar y^{(+)}\right)
\ ,
\nn\ww2
&=&  \frac{ 2\pi\nu_\pm }{\epsilon \eta_\pm^2\tau}\, 
\delta(y^{(+)})\delta(\bar y^{(+)})\, \exp\left[i\tau^{-1}\left(1+\frac{1}{\e\eta_\pm^2}\right)
y^{(-)}\bar y^{(-)}\right]\ ,
\label{k1d}
\eea
whose apex limit is indeed in agreement with
\eq{apexlimit}, \emph{viz.}
\be 
\epsilon k=+1\,:\quad \lim_{\tau\rightarrow 0} \Phi_\pm^{(L)}|_{\tilde x_\pm^2=0}=
\frac{\nu_\pm}{1+\epsilon \eta_\pm^2}\kappa_y\bar\kappa_{\bar y}\ .
\ee
In general, one needs to distinguish between singularities that 
are gauge artifacts and genuine singularities showing up in
higher spin invariants.
In particular, the zero-form charges of \cite{Sezgin:2005pv,Sezgin:2011hq} 
are higher spin invariants built directly from the Weyl zero-form
and the twistor space connection;
in the present case, the simplest such charges takes the form
\be
{\cal I}_n = \int_{{\cal Z}_4\times {\cal Y}_4} [\Phi^{(L)} \star 
\pi (\Phi^{(L)})]^{\star n}\star J\star \overline J\star 
d^4Y=[\Phi^{(L)} \star 
\pi (\Phi^{(L)})]^{\star n}|_{Y^{\ua}=0}\ ,\ee
which are formally de Rham closed on ${\cal X}_4$ on shell.
We have checked that indeed
\be \epsilon k=+1\,:\quad
\Phi_\pm^{(L)}|_{{\tilde x}_\pm^2=0} \star \pi 
(\Phi_\pm^{(L)}|_{{\tilde x}_\pm^2=0})=(\Phi_\pm^{(L)} 
\star \pi (\Phi_\pm^{(L)}))|_{{\tilde x}_\pm^2=0}=
\frac{(\nu_\pm)^2}{1+\e (\eta_\pm)^2}\ ,\ee
while it remains to compute $\Phi_\pm^{(L)}|_{{\tilde x}_\pm^2=0} \star \pi 
(\Phi_\mp^{(L)}|_{{\tilde x}_\pm^2=0})$; if the 
latter quantity vanishes, which is our 
expectation, then the Weyl zero-form 
$\Phi$ is regular on $M_4^{(0)}$ in the sense that 
${\cal I}_n$ if well-defined on $M_4^{(0)}$.

As for $k=0$, on the other hand, we can treat the case
$\tilde \nu=0$ by taking a limit, leading to
\be k=0\,:\quad 
\Phi_\pm^{(L)}|_{\tilde x_\pm^2=0\,,\, \tau \neq 0\,, k=0\,, \tilde \nu=0}=
-\frac{2\pi\nu}{\tau}\, 
\delta(y^{(+)})\delta(\bar y^{(+)})\ ,
\label{k=0,cone}
\ee
that indeed diverges as $\tau\to 0$, in agreement with 
the separate analysis at the apex given above, and 
yields divergent values for ${\cal I}_n$.
Thus, as ${\cal I}_n$ vanishes away from the cone 
$\tilde x_\pm^2=0$, it follows that if $k=0$, then
the zero-form charges are not smooth on ${\cal M}_4^{(0)}$.

For $k=0$, one may instead seek to remove the
singularities by quotienting $M_4^{(0)}$ by discrete
symmetries, as to restrict the exact solution 
to a subregion $M_4\subset M_4^{(0)}$.
In particular, if $\Lambda>0$, this can be done 
by restriction to the causal patch
\be M_4=(dS_4\setminus S(\infty))/\mathbb{Z}_2\ ,\ee
where $\mathbb{Z}_2=\{e,\gamma\}$ is defined by
\be (X^0,X^i,X^5)(\gamma(p))=(-X^0,X^i,-X^5)(p)\ ,\qquad p\in dS_4\ ,\ee
using embedding space coordinates, and $S(\infty)$
is the surface where the Weyl zero-form blows up.
As $S(\infty)$ coincides with the set of fixed points 
of $\gamma$, it follows that $M_4$ is a smooth
manifold, on which thus the Weyl zero-form is 
well-defined.
Whether there exists a similar construction for
the $k=0$ domain wall when $\Lambda<0$, remains 
to be analyzed.

\section{The \texorpdfstring{$\mathfrak{iso}(3)$}{iso(3)} invariant solution and cosmology}
\label{SecCosm}

\subsection{The solution at linear order in deformation parameter}

In this section we take a closer look at the $\mathfrak{iso}(3)$ invariant solution and compare it with those which arise in standard inflationary cosmologies. Cosmological aspects will be discussed further in the conclusions. The linearized solution for the scalar field $\phi (x) = \Phi(x,y,\bar y)\vert_{y=\bar y=0}$,  can be obtained from \eq{Gsc} by setting $y=\bar y=0$, and using \eq{Phicontour}, \eq{c2} and \eq{ABMstereo}. One thus finds
\be
\phi (x) = \frac{1+x^2}{1+2x^0 -x^2}
\left[ \nu - \frac{2\widetilde\nu (-x^0+x^2)}{1+2x^0 -x^2}\right]\ ,
\label{sol2}
\ee
where $x^2=x^\mu x^\nu \eta_{\mu\nu}$, and we have chosen $L^a= (1,0,0,0)$, $\lambda=i$. The rotational symmetries  generated by the Killing vectors $K^\mu_{rs}$ are manifest, but not the translational symmetries generated by $K^\mu_r$, which are worked out in Appendix \ref{SecCoord}; see \eq{trt}. In planar coordinates defined in \eq{pc} (with $\varsigma=\sigma=$``+'' and $(t,y^i)=(t_+,y^i_+)$ ), the metric becomes 
\be
ds^2 = -dt^2 + e^{2t} \sum_i  (dy^i)^2\ ,
\label{pcm}
\ee
and the scalar field takes the form
\be
\phi(x) = \left( \nu +\widetilde\nu \right) e^{-t}  
-   \widetilde\nu e^{-2t}\ .
\label{scalarSolution1}
\ee
In the conformal coordinate system, obtained by the coordinate transformation $\tau= -e^{-t}$, the
solution reads
\be
\phi(x) = -\left( \nu +\widetilde\nu \right) \tau  - \widetilde\nu \tau^2\ .
\label{scalarSolution}
\ee
As usual in cosmology, we will call $t$ the cosmic time and $\tau$ the conformal time. 
While the solution for the scalar field given above arises as the solution of the scalar field equation arising in higher spin theory at linear order in deformation parameters, we observe that it is also the solution of Klein-Gordon field equation with conformal mass, i.e. $(\Box^{(0)} +2 \lambda^2 )\phi = 0$, 
subject to the condition that $\phi$ depends only on $\tau$, by $\mathfrak{iso}(3)$ symmetry. Note that this differs from the vacuum solution of the same equation if we require $\mathfrak{so}(1,4)$ invariance by which the scalar field vanishes. Since the metric in this case is de Sitter, we expect that the solution for the metric will be a deformation from de Sitter metric starting at second order in the deformation parameter. 
As for the fields with spins $s>2$, they vanish in the background solution at lowest order in the deformation parameter. Whether they arise in higher orders remains to be determined.

\subsection{Comparison with standard cosmological backgrounds}

In order to compare with standard inflationary models, let us quickly summarize the behavior of the inflaton. In standard slow roll inflation, one studies a solution for which the metric is close to de Sitter, with deviations parametrised by the slow-roll parameters\footnote{These are the slow-roll parameters defined in terms of the Hubble parameter and the derivatives of the field. The slow-roll parameters defined in terms of the potential $\epsilon_v := \frac{M_{pl}}{2}\left(\frac{V'(\phi)}{V(\phi)}\right)^2$, $\eta_v:= M_{pl}^2 \frac{V''(\phi)}{V(\phi)}$ are related to these by $\epsilon \approx \epsilon_v$ and $\eta \approx \eta_v - \epsilon_v$ to leading order in slow-roll.}
\begin{equation}
\epsilon := -\frac{\dot{H}}{H^2}\,,\quad \eta := -\frac{\ddot{\phi}}{H\dot{\phi}}\ ,
\end{equation}
where $H := \dot{a}/a$ is the Hubble parameter, and a dot denotes a derivative with respect to cosmic time $t$. The metric is de Sitter when $H$ is constant, and by definition inflation happens when $\ddot{a} >0$ or $\epsilon < 1$. The solutions for the scalar field and the metric to next-to-leading order in slow-roll, and valid for $t - t_\ast \lesssim \phi/\dot\phi$, are given in terms of conformal time $\tau$ by
\be
a \approx a_\ast \left(\frac{\tau_\ast}{\tau}\right)^{1/(1+\epsilon)}\,,\quad \phi \approx \phi_\ast + 2\sqrt{\epsilon}\ln\frac{\tau}{\tau_\ast}\ ,
\label{slowRollSolution}
\ee
where at some fixed time $\tau_\ast$ we have imposed that the scale factor and the scalar field take on some given values $a_\ast$ and $\phi_\ast$. For the simplest potential $V(\phi) = m^2 \phi^2/2$, the slow-roll parameters are $\epsilon_v = 2\phi^{-2}$ and $\eta_v = 2\phi^{-2}$, with $\phi$ being the background solution. The slow-roll approximation then requires $\phi \gg 1$ in Planck units. It follows that $\phi_\ast = 6 H/m$, which means that we must have $m \ll H$. If one perturbs around this background, writing $\phi(x) = \bar\phi(\tau) + \varphi(x)$, one obtains in Fourier space after choosing the Bunch-Davies vacuum
\be
\varphi(k, \tau) = e^{i(2\kappa + 1)\pi/4}\left(-\frac{\pi H^4 \tau^3}{8 \epsilon}\right)^{1/2}H^{(1)}_\kappa(-k\tau)\,, \quad \kappa = \frac{3}{2} + 2\epsilon - \eta\,,
\ee
where we see that the slow-roll parameters appear in the index of the Hankel functions. These solutions are long-lived in the sense that $\varphi \sim \tau^{3/2 - \kappa}$ at late times $\tau \rightarrow 0$. For a recent review, see \cite{Baumann:2009ds}.

In the approaches to cosmology with higher spin fields in \cite{Arkani-Hamed:2015bza,Lee:2016vti,Kehagias:2017cym}, the scalar field is taken to be an inflaton, whose background behaves as described above, and the background metric field is taken to be de Sitter spacetime; all higher spin fields are taken to vanish in the background. However, these approaches are not derived from a theory with higher spin symmetry.

In comparing the background solutions summarised above with ours, we note that the linearised Vasiliev scalar field satisfies the equation $(\Box^{(0)} +2 \lambda^2 )\phi = 0$, subject to $\mathfrak{iso}(3)$ invariance as discussed earlier. As such it has a potential with $m \sim H$ in the notation employed above, which is incompatible with the slow-roll approximation. Indeed, it does not have the shape of the solution for standard slow-roll inflation \eqref{slowRollSolution}, and even goes to zero at late times. 

Even if the scalar behaves differently from the standard slow-roll inflaton at the level we are working, our solution is still inflationary in the sense that the metric is close to de Sitter. The time-dependent deformation away from de Sitter may in principle lead to an end of the accelerated phase. This would require $\ddot{a}<0$ which is far from de Sitter, and may in principle be achieved by summing all orders in the deformation parameter.

Though the calculation of the fluctuations in higher spin gravity is beyond the scope of this paper, let us discuss their expected behavior by considering fluctuations of a conformally coupled scalar field around de Sitter. CMB observations have fixed primordial fluctuations sourced by scalar fluctuations, to be nearly scale-invariant (in this context this is defined as their 2-point function in Fourier space behaving as $1/k^3$). They are also observed to have a larger amplitude than fluctuations sourced by the graviton. This is different from what would be generated by a conformally coupled scalar field: The behaviour of its 2-point function in Fourier space in the limit $\tau \rightarrow 0$ goes like $~1/k$. Furthermore, the amplitude of this 2-point function is suppressed by positive powers of $\tau$, so one can say that they are ``short lived'' and suppressed with respect to the fluctuations of a massless graviton (which go as $\tau^0$ in that limit, and are thus ``long lived''). We expect that the corrections to this behaviour of the scalar field during inflation will be suppressed by the deformation parameter. 

We can envisage two mechanisms by which the behavior of the scalar in higher spin theory may be ``long lived''. One possibility is that an exact FRW-like solution, i.e. to all orders in the deformation parameter, may lead to a behavior of the metric for which it takes an infinite proper time to reach a critical value of the conformal time. Whether this leads to long-lived scalar fluctuations remains to be seen. Another possible mechanism is to consider a coupling with a massive higher spin multiplet that contains a massive scalar with conformal dimension zero. Indeed, this arises in $6$-fold product of the fundamental representation of de Sitter group. A long-lived scalar field would arise in this scenario even though the coupling of massive higher spin multiplets with Vasiliev higher spin gravity is a formidable task which has hardly been studied so far. We should also require the scalar two-point function to be approximately, but not exactly, scale invariant. Since our solution is close to de Sitter, but not exactly, such behaviour can emerge. 

Assuming that one resolves the problem described above, the amplitude of fluctuations produced should agree with observations, in particular the CMB data. Clearly, since observations are made at very late times, when the characteristic energy scales are small, higher spin symmetry should be broken. In a conservative scenario, one may assume that at such low energies physics is well described by the Standard Model coupled to gravity in a gravitational background inherited from inflation. However, it remains to be seen whether the details of the higher spin symmetry breaking gives rise to novel interactions in the effective action. For inflation to be described by the unbroken phase of higher spin gravity, the symmetry breaking should happen at energy scales smaller than $\sim 10^{15}$ GeV according to the upper bounds on graviton (tensor mode) fluctuations\footnote{In particular, the amplitude of the two-point function of massless graviton (tensor mode) fluctuations should be smaller than $\sim 10^{-11}$. If the two-point function is similar to that of standard inflation, which goes as $H^2/M_{pl}^2$, this means that $H \lesssim 10^{-4} M_{pl}$.} \cite{Ade:2015lrj}.
If the dependence of the graviton two-point function deviates from the $H^2/M_{pl}^2$ behavior significantly, this scale will change accordingly.

\subsection{Towards perturbation theory around exact solutions}
\label{SecPertExact}

Given the $\mathfrak{g}_6$-invariant solutions constructed above, 
it is natural to study fluctuations around them.
This can be facilitated using the factorization method, with zero-form initial data
\be 
\Psi=\Psi_{\rm bg} + \Psi_{\rm fl}\ ,
\ee
and treating $\Psi_{\rm bg}$ exactly while keeping only the first order 
in $\Psi_{\rm fl}$. 
In what follows, we shall make the stronger assumption that 
$\Psi_{\rm bg}, \Psi_{\rm fl}\in{\cal A}({\cal Y}_4)$,
\emph{i.e.} we assume that both background and fluctuations 
belong to the same algebra, such that $\Psi^{\star n}\in{\cal A}({\cal Y}_4)$,
which can then be expanded separately in background as well as 
fluctuation parameters.

Thus, in order to construct a concrete model, we need to 
choose ${\cal A}({\cal Y}_4)$ in accordance with the 
dual boundary conditions in twistor space and spacetime.
As a concrete example, let us take $\Lambda<0$, and 
consider fluctuations around 
\be 
\Psi_{\rm bg}=\Psi_{\rm FRW^{(AdS)}_-}\ ,
\ee
\emph{i.e.} the (unique) FRW-like solution in the case of 
negative cosmological constant.
On physical grounds, we take ${\cal A}({\cal Y}_4)$ to
consists of deformations of the cosmological background, 
which correspond to spacetime mode functions that cannot 
be localized\footnote{As for holographic interpretations,
while the particle and black hole-like states can be mapped to 
operators of dual conformal theories, it is natural
to associate the non-localizable modes to operators in a phase 
of the boundary field theory in which conformal
invariance is spontaneously broken; for the case 
of the flat domainwall in anti-de Sitter spacetime, 
see \cite{Bardeen:2014paa}.}, and particle and black hole-like states,
corresponding to localizable spacetime mode functions.
Thus,
\be 
{\cal A}({\cal Y}_4)={\cal A}_{\rm nl}\oplus {\cal A}_{\rm pt}({\cal Y}_4)
\oplus {\cal A}_{\rm bh}({\cal Y}_4)\ ,
\label{fr}
\ee
given, respectively, by the orbits of the higher spin algebra $\mathfrak{hs}_1(4)$ (obtained by repeated action 
with constant $\mathfrak{hs}_1(4)$ parameters) of $\Psi_{\rm FRW^{(AdS)}_-}$, denoted by ${\cal A}_{\rm nl}$, 
and the identity operator; the massless scalar particle ground state 
(with anti-de Sitter energies $\pm 1$ and vanishing spin); and 
the black hole-like solution with vanishing anti-de Sitter energy and spin
\cite{Didenko:2009td,Iazeolla:2011cb}. The higher spin algebra $\mathfrak{hs}_1(4)$ is simply the algebra of even order polynomials in $Y^{\ua}$ with respect to the commutation rule \eq{yza}.
Using the regular presentation 
(see Eq. \eq{3.17} and Appendix \ref{App:Lemmas}), the star products of these ground states are well-defined, 
leading to the following fusion rules:
\bea 
&& {\cal A}_{\rm nl} \star {\cal A}_{\rm nl}\subseteq {\cal A}_{\rm nl}\ ,\qquad\ \  
{\cal A}_{\rm nl} \star {\cal A}_{\rm pt}\subseteq {\cal A}_{\rm pt}\ ,\qquad\ \ 
{\cal A}_{\rm nl} \star {\cal A}_{\rm bh}\subseteq {\cal A}_{\rm bh}\ ,
\ww2
&& {\cal A}_{\rm pt} \star {\cal A}_{\rm pt}\subseteq {\cal A}_{\rm bh}\ ,\qquad\quad
{\cal A}_{\rm pt} \star {\cal A}_{\rm bh}\subseteq {\cal A}_{\rm pt}\ ,\qquad
{\cal A}_{\rm bh} \star {\cal A}_{\rm bh}\subseteq {\cal A}_{\rm bh}\ ,
\eea
where we note the interesting facts that the non-localizable modes and the
localizable modes form two self-interacting subsystems, and that
non-localizable modes undergo stimulated decay to localizable modes.
The system of self-interacting system of particles and black holes
has been studied in \cite{Iazeolla:2017vng}, where a fully non-linear 
solution space was obtained by superposing rotationally invariant
scalar particle and black hole-like states, which form a
subalgebra of ${\cal A}_{\rm pt}\oplus {\cal A}_{\rm bh}$ spanned
by projectors and twisted projectors. 
Put into equations, letting $E$ denote the energy operator, one has
\bea 
\Psi_{\rm pt}  &:=& \sum_{\mu=\pm 1, \pm 2, ...} \m_n P_n\star \kappa_y  \ ,\qquad \mu_{-n}=\mu_n^\ast \ ,\\
\Psi_{\rm bh}  &:=& \sum_{\mu=\pm 1, \pm 2, ...} \nu_n P_n  \ ,\qquad\qquad \nu_{n}=\nu_n^\ast \ ,
\eea
where
\bea
P_n(E) &=& 2(-1)^{n-1}\varepsilon \oint_{C(\varepsilon)}\frac{d\eta}{2\pi i}\left(\frac{\eta+1}{\eta-1}\right)^n e^{-4\eta E} \ , \qquad \varepsilon := \frac{n}{|n|} \ .
\eea
and 
\be 
\Psi_{FRW_-} =\oint_{C(i)} \frac{d\eta}{2\pi i} \frac{\nu_+}{\eta-i}e^{-4\eta E}\star\kappa_y+\oint_{C(-i)} \frac{d\eta}{2\pi i} \frac{\nu_-}{\eta+i}e^{-4\eta E}\star\kappa_y\ , 
\ee
with $C(\varepsilon)$ and $C(\pm i)$ being small contours encircling 
$\varepsilon$ and $\pm i$, respectively.
Using the star product lemmas in Appendix \ref{App:Lemmas} and contour
integration techniques, it follows that  
\be 
P_n\star P_m=\delta_{m,n}P_n\ ,\qquad (P_n)^\dagger=P_n\ ,\qquad
P_n\star \kappa_y\star\bar{\kappa}_{\bar y}=(-1)^n P_n\ ,
\ee
and indeed
\be 
\Psi_{\rm pt}\star \Psi_{\rm pt}\in{\cal A}_{\rm bh}\ ,\quad
\Psi_{FRW_-}\star \Psi_{\rm bh}\in{\cal A}_{\rm bh}\ ,\quad
\Psi_{FRW_-}\star \Psi_{\rm pt}\in{\cal A}_{\rm pt}\ ,
\ee
in accordance with the fusion rules given above.

The following remarks are in order: 

\noindent 1. The particle states
form Hilbert spaces with $\mathfrak{hs}_1(4)$-invariant sesqui-linear forms
that are isomorphic to direct product of two singletons, the 
black hole-like states belong to a real vector space 
with $\mathfrak{hs}_1(4)$-invariant Euclidean bilinear forms
that are isomorphic to the direct product of a singleton and 
an anti-singleton \cite{Iazeolla:2017vng,Iazeolla:2008ix}; it remains to be seen whether
the non-localizable modes admit any such first-quantized
interpretation.

\noindent 2. The above considerations apply to other 
$\mathfrak{g}_6$ invariant solutions with $\Lambda<0$ as well, 
while for $\Lambda>0$ the star product realization of 
particle and black hole-like states need further study.

\noindent 3. In the case of $\Lambda>0$, we let 
${\cal A}_{\rm nl}$ 
stand for the $\mathfrak{hs}_1(4)$ orbit generated from the 
$\mathfrak{so}(4)$-invariant and the identity operator, and
${\cal A}_{\rm pt}$ for the orbit of 
the $\mathfrak{iso}(3)$ invariant solutions.
Using Eq. \eq{3.17} and the regular presentation,
it follows that 
\be {\cal A}_{\rm nl}\star {\cal A}_{\rm nl}\subseteq {\cal A}_{\rm nl}
\ ,\quad
{\cal A}_{\rm nl}\star {\cal A}_{\rm pt}\subseteq 
{\cal A}_{\rm pt}\ ,\quad {\cal A}_{\rm pt}
\star {\cal A}_{\rm pt}=0\ .\ee 
Thus, if it is possible to equip  
${\cal A}_{\rm pt}$ with a basis plane waves 
normalized on Dirac delta functions that permits
an interpretation in terms of localizable particle
states such that the two first fusion rules
remain intact, while possibly 
${\cal A}_{\rm pt}\star {\cal A}_{\rm pt}$ 
may become nontrivial, then we would have
a mechanism in the case of $\Lambda>0$
analogous to that given above in the case of
$\Lambda<0$.

\section{Conclusions}
\label{SecConclusions}

We have constructed classes of exact solutions of 
Vasiliev's bosonic higher spin gravities
with Killing symmetries given the enveloping of
six-dimensional subalgebras of the (anti-)de Sitter
symmetry algebras.
In order to construct the solutions we have used
the fact that Vasiliev's equations form a integrable
system on an enlargement of spacetime by an internal
non-commutative twistor space.
As the integrability is of Cartan type, we can solve 
the integrable system transforming a particular holomorphic 
solution $(\Phi',A'_{\ua})$ in twistor space into a physical 
solution $(\Phi^{(G)},A^{(G)}_{\ua},A^{(G)}_\mu):=G^{-1}\star
(\Phi',A'_{\ua}+\partial^{(Z)}_{\ua},\partial_\mu)\star G$ 
using a gauge transformation generated by a 
gauge function $G=L\star H$ that is large in the
sense that it alters the asymptotic behaviour of the 
master fields in both spacetime and 
twistor space.
The resulting chain of maps take the following form:
\be (\Phi',A'_{\ua})\stackrel{L}{\mapsto}(\Phi^{(L)},A^{(L)}_{\ua},A^{(L)}_\mu)\stackrel{H}{\mapsto}
(\Phi^{(G)},A^{(G)}_{\ua},A^{(G)}_\mu)\ ,\ee
where $L$ is a vacuum gauge function and $H$ is a field dependent
gauge transformation.
The role of $L$ is to switch on the dependence of the fields on the spacetime coordinates and to create a finite region of spacetime in which 
\be
(\Phi^{(L)},A^{(L)}_{\ua},A^{(L)}_\mu):=L^{-1}\star
\left(\Phi',A'_{\ua}+\partial^{(Z)}_{\ua},\partial_\mu \right)\star L
\ee
are  real analytic in the twistor $Z$ space.
The latter property permits the perturbative construction of $H$,
whose role is to create asymptotic Fronsdal fields.
The symmetries of the solution are encoded into the particular
solution, which is chosen to be invariant under parameters in the enveloping algebra generated from the six-dimensional symmetry Lie algebra $\mathfrak g_6$, \emph{viz.}
\be D'\epsilon'=0\ ,\qquad [\epsilon',\Phi']_\pi=0\ ,\ee
where $\epsilon'$ are constants built from star products of the generators of $\mathfrak g_6$.
As a result, the solutions in $L$-gauge and the physical gauge are invariant under gauge transformations generated by the rigid gauge parameters $\epsilon^{(L)}=L^{-1}\star\epsilon'\star L$ and
$\epsilon^{(G)}=H^{-1}\star\epsilon^{(L)}\star H$, respectively.
In the holomorphic and $L$-gauges, we have given the master fields to all orders, involving an expression for the twistor space connection given by two parametric integrals.
In the physical gauge, we have given the solution to first order, and proposed a perturbative scheme for continuing to higher orders based on dual boundary conditions in spacetime and twistor space.
It remains to push the gauge function method to higher orders of perturbation theory in the physical gauge, which we hope to report on in a future work. 
We expect this to generate physically interesting domain wall solutions and FRW--like solutions.

A strong motivation for this work has been the prospects for a higher spin cosmology by a direct approach 
based on finding its accelerating solutions and studying the cosmological perturbations around them. As a first step in this direction, we have constructed the FRW-like solutions and described a framework for studying the fluctuations around them, with the unusual feature that they involve black hole like states as well. Our solutions are exact in holomorphic and L-gauges. While the higher spin transformations that put these solutions in Vasiliev gauge are implemented at the leading order here, nontrivial consequences can still be extracted by studying the cosmological perturbations  around solution, just as the study of such perturbations in standard cosmology where slow roll approximation is made for the background. 

In a realistic higher spin cosmology, matter couplings and internal symmetry will need to be introduced. The requirement of higher spin symmetry puts severe constraints in doing so. The Vasiliev higher spin theory we have considered here is a universal sector of any higher spin theory, just as the graviton, dilaton and Kalb-Ramond two-form potential form a universal sector of any string theory. Assuming that the universal higher spin gravity sector dominates the physics of the inflation, it has the advantage of being unique, thereby avoiding the excessive freedom in choosing field content, interactions and parameters. For example, in the favored approach to standard inflationary scenario, Einstein gravity is coupled to a real scalar with a potential that is picked by hand to satisfy suitable `slow-roll' conditions. Moreover, the origin of the scalar field in a fundamental theory is not known. In the higher spin theory based inflation scenario envisaged here, however, the scalar field is necessarily part of the spectrum for the consistency of the higher spin theory, and the inflation is not driven solely by the energy stored in a slowly varying scalar field. Indeed, there is a frame in which the only contact term for the scalar field is a mass term \cite{Sezgin:2003pt}. Note, however, that the theory comes with infinite derivative couplings even at a given order in weak fields. Given that there is no mass scale at our disposal to argue that those couplings will be suppressed, they are all equally important. Thus, the inflationary solution to the higher spin theory will be driven by the higher spin invariant, higher derivative couplings.

While the problem of matter couplings and breaking of higher spin symmetry will need to be ultimately attended to, at present the more pressing problems to tackle seem to be the determination of the higher order terms in the FRW background in Vasiliev gauge, carrying out the cosmological perturbation theory along the line described in Section \ref{SecPertExact} and seeking possible holographic interpretation of the results.

\subsection*{Acknowledgements}

We would like to thank Gary Gibbons, William Linch, Yi Pang, Mitya Ponomarev, Chris Pope, Andy Royston, Sav Sethi and Andy Strominger for helpful discussions. 
ES would like to thank Universidad San Sebasti\'an and Pontifica Universidad Cat\'olica de Valparaiso for hospitality. 
The work of ES is supported in part by NSF grant PHY-1214344 and in part by Conicyt MEC grant PAI 80160107. 
CI would like to thank Universidad Andres Bello for kind hospitality during the final stages of this work.
The work of CI was supported in part by the Russian Science Foundation grant 
14-42-00047 in association with the Lebedev Physical Institute 
in Moscow. This work was also partially funded by grants FONDECYT 1151107, 1140296, and 1171466. PS and RA would like acknowledge DPI20140115 for some financial support.
The work of YY is supported by the FONDECYT Postdoc Project 3150692.

\newpage

\begin{appendix}

\section{Conventions and definitions}
\label{SecConvention}

Using conventions in which $(z_\alpha)^\dagger=- \bar z_{\dot\alpha}$, the star product 
can be realized using a normal ordering scheme as follows:
\bea
 &&f(y,\bar y,z,\bar z)~\star~ g(y,\bar
y,z,\bar z)
\label{star}\\[5pt]
&=&\ \int \frac{d^2\xi d^2\eta d^2\bar\xi
d^2\bar\eta}{(2\pi)^4}~ e^{i\eta^\a\xi_\a+
i\bar\eta^{\dot\a}\bar\xi_{\dot\a}}  f(y+\xi,\bar y+\bar
\xi,z+\xi,\bar z+ \bar \xi) g(y+\eta,\bar y+\bar
\eta,z-\eta,\bar z- \bar \eta)\ .
\nn
\eea
Defining 
\be
\partial_\alpha^{s_1s_2} := s_1 \frac{\partial}{\partial z^\a} 
+ s_2 \frac{\partial}{\partial y^\a}\ ,\qquad
\partial_{\dot\alpha}^{s_1s_2} :=  s_1 \frac{\partial}{\partial \bar{z}^{\dot\a}} 
+ s_2 \frac{\partial}{\partial \bar{y}^{\dot\a}}\ ,
\ee
where $s_1$ and $s_2$ are $+1$ or $-1$, and given a function $f(y,\bar y, z,\bar z)$, one finds
\begin{align}
y_\a \star f & = y_\a f +i\partial_\a^{-+} f\ , &   f \star y_\a &= y_\a f +i\partial_\a^{--} f\ ,
\nn\\
z_\a \star f &= z_\a f +i\partial_\a^{-+} f\ , &  f \star z_\a &= z_\a f +i\partial_\a^{++} f\ ,
\nn\\
\bar y_{\dot\a} \star f & = \bar y_{\dot\a} f +i\partial_{\dot\a}^{-+} f\ , &   
f \star \bar y_{\dot \a} &= \bar y_{\dot\a} f + i\partial_{\dot\a}^{--} f\ ,
\nn\\
\bar z_{\dot\a} \star f &= \bar z_{\dot\a} f +i\partial_{\dot\a}^{-+} f\ , &  
f \star \bar z_{\dot\a} &= \bar z_{\dot\a} f +i\partial_{\dot\a}^{++} f\ ,
\end{align}
Frequently used quantities in the body of the paper are defined as follows:
\begin{align}
z^{\pm} &= u^{\pm\a} z_\a\ , &  w &= i z^+ z^-\ ,    
&  \xi &= (1-\tau)/(1+\tau)\ ,
\nn\\
u^{\a\b} &= 2u^{+(\a} u^{-\b)}\ , &  b^a &= i\eta L^a\ , & \slashed{b} &= b^a \left( \sigma _{a}\right) _{\a\dot\a}\ . 
\nn\\
h &= \sqrt{1-\lambda^2 x^2}\ ,&  x^{2} &=x^{a}x_{a}\ ,& \slashed{x}_{\a \dot\a} &= x^{a}\left( \sigma _{a}\right) _{\a\dot a}\ ,
\label{symbols}
\end{align}
\bea
\left[\begin{array}{c} y^L_\alpha\\\bar y^L_{\dot\alpha}\end{array}\right] &=& 
L^{-1}\star \left[\begin{array}{c} y_\alpha\\\bar y_{\dot\alpha}\end{array}\right]\star L
= \left[\begin{array}{cc} L_\a{}^\beta& K_\a{}^{\dot\beta}\\ 
\overline K_{\dot\a}{}^\beta & \overline L_{\dot\alpha}{}^{\dot\beta}\end{array}\right] 
\left[\begin{array}{c} y_\b\\\bar y_{\dot\beta}\end{array}\right]\ ,\qquad  
\widetilde y_\a = y_\a + M_a{}^{\dot\b} \bar y_{\dot\b}\ ,
\nn\ww2
A_\a{}^\b & =& L_\a{}^\b-b_a(\sigma^a \overline K)_\a{}^\b\ , \qquad
B_\a{}^{\dot\b} :=K_\a{}^{\dot\b}-b_a(\sigma^a \overline L)_\a{}^{\dot\b}\ , \qquad
M  = A^{-1} B\ .
\nn
\eea
The matrices $A,B,M$ and $\det A$ are given in sterographic coordonates in \eq{ABMstereo}, and in planar coordinates in \eq{ABMplanar}.

We use the convention in which $(\sigma^a)_{\alpha\dot\alpha}= (1,\vec \sigma)$ where $\vec\sigma$ are the Pauli matrices, and $(\bar\sigma)_{\dot\alpha\alpha}$ is complex conjugate of $(\sigma^a)_{\alpha\dot\alpha}$.
%
%
Furthermore we define  $\left( \sigma _{ab}\right) _{\alpha \beta }=\left( \sigma _{\lbrack
a}\right) _{\alpha }^{\ \ \dot{\gamma}}\left( \bar{\sigma}_{b]}\right) _{ 
\dot{\gamma}\beta }$ and $\left( \bar{\sigma}_{ab}\right) _{\dot{\alpha}\dot{\beta}}=\left( \bar{
\sigma }_{[a}\right) _{\dot{\alpha}}^{\ \ \gamma }\left( \sigma _{b]}\right)
_{\gamma \dot{\beta}}$. The spinor-indices are raised or lowered by 
$ \epsilon_{\alpha\beta}=\epsilon^{\alpha\beta}=\epsilon_{\dot\alpha\dot\beta}=
\epsilon^{\dot\alpha\dot\beta}=i\sigma^2$, using the NW-SE convention.

\section{Coordinate systems and Killing vectors}
\label{SecCoord}

\subsection{Embedding space} 

The metric 
space $(M_4^{(0)},(ds_4^2)^{(0)})\equiv (A)dS_4$ 
with inverse radius $|\lambda^{-1}|$ can be defined
as the surface in the five-dimensional plane
$(\Real^5,ds_5^2)$ with global Cartesian coordinates $X^M$
and constant metric $ds_5^2=dX^M dX^N\eta_{MN}$,
where $\eta_{MN}=(\eta_{ab},-{\rm sign}(\lambda^2))$,
described by the constraint
\be E:= X^M X^N\eta_{MN}+\lambda^{-2}\approx 0\ .\ee
Formally, this amounts to a smooth map 
$f:M_4^{(0)}\hookrightarrow \Real^5$ that
obeys $E\circ f\equiv 0$ and that 
is invertible on $f(M_4^{(0)})$, that is,
there exists an inverse $f^{-1}:f(M_4^{(0)})\rightarrow
M_4^{(0)}$ that is a diffeomorphism.
The intrinsic metric on $M_4^{(0)}$ is defined
by
\be (ds_4^2)^{(0)}):=f^\ast ds_5^2\equiv (dX^M dX^N\eta_{MN})|_{E\approx 0}\ ,\ee
where $f^\ast$ denotes the pull-back operation,
that is, in terms of coordinates $x^\mu$ on 
$M_4^{(0)}$, one has
\be g^{(0)}_{\mu\nu} = 
\partial_\mu X^M \partial_\nu X^N \eta_{MN}\ .
\ee
%

\subsection{Killing vector fields}
%
The symmetry algebra $\mathfrak{g}_6$ of the solutions under
consideration is embedded via \eq{embed} into the isometry 
algebra $\mathfrak{g}_{10}$ of $(M_4^{(0)},(ds_4^2)^{(0)})$.
The latter is inherited from the isometry algebra of
$(\Real^5,ds_5^2)$, namely as its subalgebra 
\be \mathfrak{l}_{10}:=
\left\{ \vec L:\ {\cal L}_{\vec L}
ds_5^2=0\ ,\quad \vec L E\approx 0\ \right\}\cong \mathfrak{g}_{10}\ ,\ee
with Lie bracket induced from the Schouten bracket,
and a well-defined action on the ring of 
equivalence classes $[\Phi]=\left\{\Phi'\in C^\infty(\Real^5): (\Phi'-\Phi)\circ f\equiv0\right\}$ with product 
$[\Phi_1][\Phi_2]:=[\Phi_1\Phi_2]$, given by 
$\vec L [\Phi]:=[\vec L\Phi]$. 
This ring is isomorphic to $C^\infty(M_4^{(0)})$ via 
$\phi_{[\Phi]}:=\Phi\circ f$.
Thus, each $\vec L\in \mathfrak{l}_{10}$ induces an intrinsic
Killing vector field $\vec K_{\vec L}$ on $M_4^{(0)}$ defined by
$\vec K_{\vec L}\phi_{[\Phi]}:=\phi_{\vec L[\Phi]}\equiv (\vec L\Phi)\circ f$.
Hence, letting $p\in M_4^{(0)}$ and assuming that
$\Phi$ is smooth close to $f(M_4^{(0)})$), we have
the Killing vector relation
\be f_\ast (\vec K_{\vec L}|_p) \Phi\equiv \vec K_{\vec L}|_p(\Phi\circ f)
=\vec K_{\vec L}|_p \phi_{[\Phi]}=[\vec L \Phi]|_{f(p)}=
(\vec L \Phi)|_{f(p)}\ ,\ee
where $f_\ast$ denotes the push-forward operation, that is 
\be f_\ast (\vec K_{\vec L}|_p)=\vec L|_{f(p)}\ ,\ee
or $K^\mu_{\vec L}|_p f_\ast(\vec\partial_\mu|_p) = 
L^M \vec\partial_M|_{f(p)}$, in terms of an intrinsic 
coordinate $x^\mu$ at $p$. 
In the global coordinate basis, 
\be 
\vec L_{MN} = X_M \vec\partial_N - X_N\vec\partial_M\ ,
\ee
inducing the intrinsic Killing vectors fields
\be \vec K_{MN}=K^\mu_{MN} \vec\partial_\mu\ ,\qquad
K^\mu_{MN} \partial_\mu X^P  = 2 X_{[M}\delta_{N]}^P\ ,\ee
with components 
\be
K^\mu_{MN} = 2g^{\mu\nu} X_{[M} \partial_\nu X_{N]}\ .
\ee
It follows that the intrinsic Killing vector fields 
associated with the $\mathfrak{g}_6$ generators $M_{rs}$ 
and $T_r$ defined in \eq{embed} are given by
\bea
K^\mu_{rs} &=& 2g^{\mu\nu} L_{[r}{}^a L_{s]}{}^b X_a\partial_\nu X_b\ ,
\ww2
K^\mu_r &=&  g^{\mu\nu} L_r{}^a \left( \alpha L^b X_a\partial_\nu X_b
- \alpha L^b X_b\partial_\nu X_a
-\beta \ell^{-1} X_a\partial_\nu X_5 + \beta \ell^{-1} X_5\partial_\nu X_a\right)\ .
\label{kv1}
\eea
Thus, under the $\mathfrak{g}_6$ transformations defined in \eq{embed}, we have
\be
\delta x^\mu = \xi^{rs} K^\mu_{rs} + \xi^r K^\mu_r\ ,
\ee
with constant parameters $(\xi^{rs}, \xi^r)$.

\subsection{Global \texorpdfstring{$dS_4/S^3$}{dS4/S3} and \texorpdfstring{$AdS_4/AdS_3$}{AdS4/AdS3} foliations 
(iso-scalar leafs for \texorpdfstring{$\e k=-1$}{epsilon k = -1}).}

Coordinates adapted to the solutions with $\epsilon k=-1$, 
that is, the FRW$^{({\rm dS})}_+$ and DW$^{({\rm AdS)}}_-$ solutions, 
can be obtained by foliating $dS_4$ and $AdS_4$ with $S^3$ and $AdS_3$ 
iso-scalar leafs, respectively,\emph{viz.}
\bea dS_4&:&\eta_{AB}=(-,\delta_{IJ})\ ,\quad X^0\approx\ell \tau\ ,\quad 
X^{I}\approx\ell\sqrt{1+\tau^2} n^{I}\ ,\quad n^{I}n^{J}\delta_{IJ}=1\ ,\\
AdS_4&:&\eta_{AB}=(\eta_{RS},+)\ ,\quad 
X^3\approx\ell \s\ ,\quad X^{R}\approx\ell
\sqrt{1+\s^2} n^{R}\ ,\quad n^{R}n^{S}\eta_{RS}=-1\ ,\qquad\eea
where $\delta_{IJ}={\rm diag}(+,+,+,+)$ and $\eta_{RS}={\rm diag}(-,-,+,+)$.
Here we have used the labeling $I=1,2,3,5$ and $R=0,5,1,2$.
The resulting global parametrizations of the induced metrics are given by
\bea
dS_4&:&(ds_4^2)^{(0)}
=\ell^2\left(-\frac{d\tau^2}{1+\tau^2}+(1+\tau^2)dn^2|_{n^I n^J\delta_{IJ}=1}\right)\ ,
\qquad \tau\in \Real\ ,
\label{dsm2}\\
AdS_4&:&(ds_4^2)^{(0)}=
\ell^2\left(\frac{d\s^2}{1+\s^2}+(1+\s^2)dn^2|_{n^R n^S\eta_{RS}=-1}\right)\ ,
\qquad  \s\in \Real \ .
\label{adsm2}
\eea

\subsection{Bifurcating \texorpdfstring{$(A)dS_4/\{dS_3,H_3\}$}{(A)dS4/dS3,H3} foliations (iso-scalar leafs for
instantons)}

Coordinates adapted to the instanton solutions can be obtained
by decomposing $dS_4$ and $AdS_4$ into subregions foliated by 
$dS_3$ and $H_3$ leafs as follows:
\bea 
dS_4&:& X^a \approx \ell \xi_\pm n^a\ ,\quad n^a n^b \eta_{ab}=\pm 1\ ,
\quad |X^{5}|\approx \ell \sqrt{1\mp \xi_\pm^2}\ ,
\ww2
AdS_4&:& X^a\approx\ell\xi_\pm n^a\ ,\quad n^a n^b \eta_{ab}=\pm 1\ ,\quad
|X^5|\approx\ell\sqrt{1\pm \xi_\pm^2}\ ,
\eea
where $a=0,1,2,3$ and  $\eta_{ab}={\rm diag} (-+++)$. The resulting induced metrics are
\bea
dS_4 &:& (ds_4^2)^{(0)}=\ell^2\left(\pm \frac{d\xi^2_\pm}{1\mp\xi_\pm^2}+
\xi_\pm^2dn^2|_{n^a n^b \eta_{ab}=\pm 1}\right)\ ,\qquad \quad 0\leqslant\xi_+<1<\xi_-\ ,
\ww2
AdS_4 &:& (ds_4^2)^{(0)}=\ell^2\left(\pm \frac{d\xi^2_\pm}{1\pm \xi_\pm^2}+
\xi^2_\pm dn^2|_{n^a n^b \eta_{ab}=\pm 1}\right)\ ,
\qquad \quad 0\leqslant \xi_-<1<\xi_+\ ,
\eea
where the upper (lower) sign corresponds to $dS_3 (H_3)$ foliations.

\subsection{Planar coordinates for \texorpdfstring{$(A)dS_4$}{(A)dS4} (iso-scalar leafs for \texorpdfstring{$\e k=0$}{epsilon k = 0}).}
%
Coordinates adapted to the solutions with $\e k=0$ can be obtained 
by foliating $dS_4$ ($\e=-1$) and $AdS_4$ ($\e=+1$) using Euclidean and Lorentzian three-planes, respectively, as follows:
\bea
dS_4&:& X^0 \approx \sigma\ell (\sinh t +\tfrac12 r^2 e^{t})\ ,\qquad
X^5 \approx \sigma'\ell(\cosh t -\tfrac12 r^2 e^{ t})\ ,
\label{plan1}\\
&&  X^i \approx \ell e^{ t} y^i\ ,\qquad r^2:=y^iy^j\delta_{ij}\ ,\qquad i,j=1,2,3
\nn\nonumber\\
AdS_4&:& X^5 \approx \sigma\ell (\cosh t +\tfrac12 r^2 e^{t})\ ,\qquad
X^3 \approx \sigma'\ell(\sinh t -\tfrac12 r^2  e^{ t})\ , 
\label{plan2}\\
&& X^i \approx \ell e^{ t} y^i\ ,\qquad r^2:=y^iy^j\eta_{ij}\ , \qquad i,j=0,1,2\ ,
\nonumber
\eea
where $\eta_{ij} ={\rm diag}(-++)$, $(\sigma,\sigma')=(\pm1,\pm1)$ and $(y^i,t)\in \Real^4$,
which provides four charts $U_{\sigma,\sigma'}\cong \Real^4$,
referred to as Poincar\'e patches, on which the intrinsic metrics 
\be
(ds^2_4)^{(0)}|_{U_{\sigma,\sigma'}} =\ell^2 \left(\e dt^2 + e^{2t} dy^2\right)\ ,
\ee
and an embedding space light-cone coordinate has a definite sign, \emph{viz.}
\bea 
dS_4&:& (\sigma X^0+\sigma' X^5)|_{U_{\sigma,\sigma'}}=\ell e^t>0\ ,
\\
AdS_4&:& (\sigma X^3+\sigma' X^5)|_{U_{\sigma,\sigma'}}=\ell e^t>0\ ,\quad
\eea
such that 
\be dS_4=U_{+,+}\cup U_{-,-}\cup (\Real \times S^2)\ ,\qquad AdS_4=
U_{+,+}\cup U_{-,-}\cup (\Real \times AdS^2)\ ,\ee
where $U_{+,+}\cap U_{-,-}=\emptyset$, 
with equivalent expressions using $U_{\pm,\mp}$.
Adding $U_{\pm,\mp}$ one obtains an atlas with nontrivial
transition functions given by
\bea 
\quad U_{\sigma,\pm}\cap U_{\sigma,\mp}&:& 
e^{\tilde t}=(r^2 e^{2 t}+\e) e^{- t}\ ,\qquad
\tilde y^i=\frac{e^{2t}}{r^2 e^{2t}+\e}y^i\ ,\qquad r^2>-\e \,e^{-2t}
\ ,\\
U_{\pm,\sigma}\cap U_{\mp,\sigma}&:& 
e^{\tilde t}=-(r^2 e^{2 t}+\e) e^{- t}\ ,\qquad
\tilde y^i=-\frac{e^{2t}}{r^2 e^{2t}+\e}y^i\ ,\qquad r^2<-\e \,e^{-2t}\ .
\eea

In the case of $dS_4$, each Poincar\'e patch provide a geodesically
complete spacetime, with time flowing in the direction of $X^0$ 
and $-X^0$ on $U_{+,\pm}$ and $U_{-,\pm}$, respectively.
If one introduces conformal time
\be \tau=-e^{-t}\in \Real_-\ ,\ee
then $\tau\rightarrow 0^-$ at the future (or past) boundary, 
and the metric takes the form
\be
dS_4\ :\quad (ds^2_4)^{(0)}|_{U_{\sigma,\sigma'}} =\ell^2 \frac{-d\tau^2 + dy^2}{\tau^2}\ .
\ee
In terms the inverse conformal time, which can be extended from $\Real_-$ to
$\Real$, the transition function between $U_{+,+}$ and 
$U_{-,-}$ is given by 
\be (\tau^{-1}+ \widetilde \tau^{-1})|_{U_{+,+}\cap U_{-,-}}=0\ .\ee

In the case of $AdS_4$, the conformal radius 
\be z=e^{-t}\in \Real_+\ ,\ee
obeys $z\rightarrow 0^+$ at the boundary, and the metric takes the form
\be
AdS_4\ :\quad (ds^2_4)^{(0)}|_{U_{\sigma,\sigma'}} =\ell^2 \frac{dz^2 + dy^2}{z^2}\ ,
\ee
in each Poincar\'e patch; the two Poincar\'e patches can be glued
together using 
\be (z^{-1}+ \widetilde z^{-1})|_{U_{+,+}\cap U_{-,-}}=0\ .\ee
%

\subsection{Stereographic coordinates.}
%
A convenient set of coordinates, that facilitate a unified
description of all solutions, are the stereographic coordinates 
$\{x^a_\pm\}_{a=0,1,2,3}$ that arise via the parameterization
\be
X^M|_{U_\pm}\approx \left(\frac{2x^a_\pm}{1-\lambda^2 x^2_\pm},\pm \ell\frac{1+\lambda^2 x^2_\pm}{1-\lambda^2 x^2_\pm}\right)\ ,\qquad -1<\lambda^2 x^2_\pm <1\ ,\qquad x^2=x^a x^b\eta_{ab}\ ,
\label{sc}
\ee
with inverse
\be x^a_\pm =\left.\frac{X^a}{1+\sqrt{1+\lambda^2 X^b X_b}}\right|_{U_\pm}\ ,\ee
where $U_\pm$ denotes the two stereographic coordinates charts.
Each chart covers one half of $(A)dS_4$, and can be 
continued smoothly into $\lambda^2x_\pm^2<-1$;
the resulting transition function is given by
\be x^a_\pm = \lambda^{-2}R^a(x_\mp)\ ,\qquad \lambda^2x_\pm^2<0\ ,\ee 
where the reflection map
\be R^a(v):=-\frac{v^a}{v^2}\ .\ee
The charts can also be extended (non-smoothly) into 
$\lambda^2x_\pm^2>1$ as follows: as a point $p\in(A)dS_4$ 
approaches a point $p_0$ the subspace $\lambda^2x_\pm^2(p_0)=1$ 
from the inside, \emph{i.e.} $\lambda^2x_\pm^2(p)<1$, the 
reflected image $R(p)$ approaches the point $R(p_0)$
with $x^\mu_\pm(R(p_0))=-x^\mu_\pm(p_0)$ from the outside,
\emph{i.e.} $\lambda^2x_\pm^2(R(p))>1$.
Thus, one may cover all of $(A)dS_4$ using a single 
stereographic coordinate, that we shall take to be 
$x^a\equiv x^a_+$, defined on four-dimensional 
Minkowski space minus the subspace $\lambda^2x^2=1$.
The boundary is given two copies of the surface 
$\lambda^2x^2=1$; an outer sheet with normal
pointing inwards and an inner sheet with normal
pointing outwards.

In the $AdS_4$ case, the surface $\lambda^2x_\pm^2=1$ 
has the topology of $dS_3\cong \Real\times S^2$, while
its two-sheeted counterpart can be glued together using the
reflection map into a single surface with $S^1\times S^2$ 
topology, \emph{i.e.}
\be \partial (AdS_4)\cong S^1\times S^2\ ,\ee
as can be seen by taking a tour around the boundary using
reflection maps as follows: Start at a point $p_1$ 
on the outer sheet at large negative (stereographic) time; 
move up to a point $p_2$ on the same sheet at large positive 
times; cross over to $R(p_2)$, which is a point at the 
inner sheet at large negative times; move up to a point $p_3$
on the same sheet at large positive times; finally, cross 
back to $R(p_3)=p_1$, thereby closing a time-like curve.

In the $dS_4$ case, the boundary consists of two two-sheeted 
surfaces; one with $x^0>0$ and another one with $x^0<0$.
Using the reflection map, these four sheets, each of which 
thus has the topology of an hyperbolic three-plane, form
two pairs, each of which can be glued together into a
three-sphere, \emph{i.e.}
\be \partial (dS_4)\cong S^3_-\cup  S^3_+\ ,\ee
where $x^0<0$ on $S^3_-$ and $x^0>0$ on $S^3_+$.

In stereographic coordinates, the metric takes the form 
\be
(ds^2_4)^{(0)}=\frac{4 dx^2}{(1-\lambda^2 x^2)^2}\ .
\ee
Using the plus-branch, where $X^5= \ell (1+\lambda^2 x^2)/(1-\lambda^2 x^2)$, the $\mathfrak g_6$ Killing vector fields read
\bea
K^\mu_{rs} &=& 2 L_{[r}{}^a L_{s]}{}^b ( x_a\, \delta_b{}^\mu )\ ,
\nonumber\\
K^\mu_r &=&  L_r{}^a \left\{ \alpha L^b ( x_a \delta_b^\mu - x_b \delta_a^\mu)
-\beta \left[ -\tfrac12 (1+\lambda^2 x^2) \delta_a^\mu 
+\lambda^2 x_a x^\mu\right] \right\}\ .
\label{kv2}
\eea
Comparing \eq{plan1} with \eq{sc}, one finds the following transition
functions on $U_\varsigma\cap U_{\sigma,\sigma}$ ($\varsigma,\sigma=\pm$):
\be
\begin{split}
dS_4 &: x_{\varsigma }^{0} = 
\sigma\ell \frac{\sinh t+\frac{1}{2}r^{2}e^{t}}{1+\varsigma \sigma(\cosh t- 
\frac{1}{2}r^{2}e^{t})}\ ,
\qquad 
 x_{\varsigma }^{i} = \ell\frac{e^{t}\
y^{i}}{1+\varsigma\sigma( \cosh t- \frac{1}{2}r^{2}e^{t})}\ , \qquad
i=1,2,3\ ,
\\
AdS_4 &: \ x_{\varsigma }^{3} = 
\sigma \ell\frac{\sinh t-\frac{1}{2}r^{2}e^{t}}{1+\varsigma \sigma(\cosh t+ 
\frac{1}{2}r^{2}e^{t})}\ ,
\qquad 
x_{\varsigma }^{i} = \ell\frac{e^{t}\
y^{i}}{1+\varsigma\sigma( \cosh t+ \frac{1}{2}r^{2}e^{t})}\ , \qquad
i=0,1,2\ ,
\label{pc}
\end{split}
\ee
and
\be 
\lambda^2 x_{\varsigma}^{2} = 1-\frac{2}{
1+\varsigma\sigma \left[ \cosh t - \frac{1}{2}{\rm sign}(\lambda^2)\, r^{2}e^{t}\right]}\ .
\label{coordtransf}
\ee

In particular, in checking that the solution \eq{sol2} is invariant under the translations with parameter $\xi^r$, it is useful to note that (taking $\alpha=\beta, (L^a, L_r{}^a) = (\delta_0^a, \delta_r^a)$ and setting $\lambda=i$)
\be
\delta x^0= \alpha \xi^r x_r (1+x^0)\ ,\qquad 
\delta x^2= \alpha \xi^r x_r (1+x^2)\ .
\label{trt}
\ee
%


\section{Gauge functions}
\label{SecGaugeFunc}


\subsection{Stereographic coordinates} 

Given the Lie algebra $\mathfrak{so}(p,q)$ 
in the basis over $\Real$ spanned by $P_I$, $I=1,\dots,p+q-1$, 
and $M_{IJ}=-M_{JI}$ obeying 
\be [M_{IJ},M_{KL}]_\star =4i\eta_{[K|[J} M_{I]|L]}\ ,\quad
[M_{IJ},P_K]_\star=2\eta_{K[J} P_{I]}\ ,\quad 
[P_I,P_J]=i\lambda^2 M_{IJ}\ ,\ee
where $\eta_{IJ}$ has signature $(p',q')$ and 
$\lambda^2\in\Real\setminus\{0\}$, the gauge function
\be L:=\exp_\star (i\xi^I P_I)\ ,\qquad \xi^I\in\Real\ ,\ee
yields a Maurer--Cartan form
\be L^{-1}\star dL=i e^I P_I+ \frac{1}{2i} \omega^{IJ}M_{IJ}\ ,\ee
with components
\be e^I=2 h^{-2}dx^I\ ,\qquad \omega^{IJ} = 4 h^{-2} x^{[I} dx^{J]}\ ,
\ee
where 
\be \xi^I=4 \Upsilon x^I\ ,\qquad \Upsilon=\frac{1}{\sqrt{1-h^2}}\tanh^{-1}
\sqrt{\frac{1-h}{1+h}}\ ,\ee
\be h=\sqrt{1-\lambda^2 x^2}\ ,\qquad x^2=x^I x^J\eta_{IJ}\ ,\ee
which is defined for $\lambda^2 x^2<1$.

\subsection{Stereographic coordinates for \texorpdfstring{$(A)dS_4$}{(A)dS4}} 
\label{SecGaugeFuncStereo}
%
The gauge function 
\be
L =\exp_\star (4i\Upsilon  x^{a}P_{a})
=\frac{2h}{1+h} \exp \left(\frac{i\lambda \slashed{x}_{\alpha \dot{\alpha}}y^{\alpha}
\bar{y}^{\dot{\alpha}}}{1+h}\right)\ ,\qquad
\slashed{x}_{\alpha\dot{\alpha}}=x^{a}\left(\sigma _{a}\right)_{\alpha \dot{\alpha}}\ ,
\ee
which is defined for $\lambda^2 x^2<1$, 
yields the Maurer--Cartan form for $(A)dS_4$ in the stereographic 
coordinates, with components
\be e^a=2h^{-1} dx^a\ ,\qquad \omega^{ab}=4h^{-1} x^{[a} dx^{b]}\ ;\ee
which can be extended into $\lambda^2 x^2>1$.
Thus, whereas the Maurer--Cartan form can be described
globally on $(A)dS_4$ using a single stereographic
coordinate, the usage of vacuum gauge functions requires 
patching.
Using the two charts $U_\pm$ defined above, with 
$1>\lambda^2x_\pm^2|_{U_\pm}\geqslant -1$, and 
letting $L_\pm$ denote the corresponding locally 
defined gauge functions, it follows from  
\be L_\pm|_{\lambda^2x_\pm^2=-1}=L_\mp|_{\lambda^2x_\mp^2=-1}\ ,\ee
that the transition function 
\be T_\pm^\mp := (L_\pm\star (L_\mp)^{-1})|_{\lambda^2x_\pm^2=-1}=1' ,\ee
\emph{i.e.} the procedure of gluing together $U_+$ and $U_-$ into 
$(A)dS_4$ does not refer to any choice of structure group.

From \eq{KL} one finds
\be  
\left[\begin{array}{cc} L_\a{}^\beta& K_\a{}^{\dot\beta}\\ 
\overline K_{\dot\a}{}^\beta & \overline L_{\dot\alpha}{}^{\dot\beta}\end{array}\right] =
h^{-1}\left[\begin{array}{cc}\delta_\a^\b&\lambda \slashed{x}_\a{}^{\dot\b}\\ \lambda\slashed{\bar x}_{\dot\a}{}^{\beta}&\delta_{\dot\a}^{\dot\b}\end{array}\right]\ ,
\ee
From \eq{AB1} it then follows that 
\bse
\label{ABMstereo}
\bea
A &=& h^{-1}(1- \lambda \slashed{b}\slashed{\bar x})\ ,
\qquad B = h^{-1}(\lambda \slashed{x}-\slashed{b} )\ ,
\ww2
M &=& A^{-1}B =  \frac{1}{\det A} 
\left( \frac{1-2\lambda b^a x_a +b^2}{h^2}\lambda \slashed{x}- \slashed{b} \right)=\lambda\slashed{x}+\frac{\lambda b^2\slashed{x}-\slashed{b}}{\det A}\ ,
\ww3
{\det A} &=& h^{-2}(1-2\lambda b_a x^a +b^2\lambda^2 x^2) \ ,
\eea
\ese
where $b^a=i\eta L^a$; on the poles of ${\cal O}$ it is imaginary
if $\epsilon k=-1$ and real if $\epsilon k=0$ or $1$.

\subsection{Global foliation coordinates for 
\texorpdfstring{$\epsilon k=-1$}{epsilon k = -1}} 
%
The gauge function
\be L=\exp_\star (i\xi^r T_r)\star \exp_\star (i\ell  \rho P)\ ,\qquad
(\xi^r,\rho)\in\Real^4\ ,\ee
yield the Maurer--Cartan form
\be L^{-1}\star dL=i\ell d\rho P+ i \check e^r\left(\cosh(\sqrt{\e} \lambda\ell \rho)  T_r +
\frac{\lambda}{\epsilon} \sinh(\sqrt{\e} \lambda\ell \rho)B_r\right)
+\frac{1}{2i} \check\omega^{rs} M_{rs}\ ,\ee
where $B_r=L^a L^b_r M_{ab}$ and
\be \check e^r= 2\check h^{-2}dx^r\ ,\qquad \check\omega^{rs}=4\check h^{-2}x^{[r}dx^{s]}\ ,\qquad
\check h=\sqrt{1-\lambda^2 x^2}\ ,\qquad x^2=x^r x^s\eta_{rs}\ ,\ee
with $\lambda^2 x^2<1$.
The global foliation coordinates in $dS_4$ and $AdS_4$ are obtained by taking
\bea dS_4&:&  \e=-1\ ,\qquad \tau=\sinh^{-1}\rho\ ,\qquad n^I=\left(\frac{1+\lambda^2 x^2}{1-\lambda^2 x^2},\frac{2\ell^{-1} x^r}{1-\lambda^2 x^2}\right)\ ,\\
AdS_4&:& \e=+1\ ,\qquad \sigma=\sinh^{-1}\rho\ ,\qquad
n^R=\left(\frac{1+\lambda^2 x^2}{1-\lambda^2 x^2},\frac{2\ell^{-1} x^r}{1-\lambda^2 x^2}\right)\ .\eea

\subsection{Planar coordinates for \texorpdfstring{$(\e,k)=(\pm1,0)$}{(epsilon, k) = (+- 1, 0)}} 
\label{SecGaugeFuncPlanar}
%
If $k=0$, then 
\be \e={\rm sign}(\lambda^2)\ ,\qquad \beta=\ell\alpha\ ,\qquad 
\gamma\equiv \frac{i\alpha}{\lambda\beta}= i\e\ell \lambda=\left\{\begin{array}{ll} 1& \mbox{if $\e=-1$}\ ,\\
i& \mbox{if $\e=+1$}\ ,\end{array}\right.\ee
and the gauge function
\be
L=\exp_\star\left(i \a^{-1} y^r T_r\right)\star \exp_\star(-i\e \ell t P)\ ,\qquad
[P,T_r]_\star=-i \e \frac{1}{\ell} T_r\ ,\ee
\label{C8}
gives rise to the Maurer--Cartan form
\be L^{-1}\star dL=i\ell( -\e dt P +  \beta^{-1} e^{t}dy^r T_r)\ ,
\ee
with components
\be e^a=\ell(-\e L^a dt+e^t L^a_r dy^r)\ ,\qquad
\omega^{ab} =2 \ell e^t L^{[a} L^{b]}_r dy^r\ .\ee
As for the corresponding adjoint action, we have
\be
L^{-1}\star Y_{\ua}\star L=\underline{L}_{\ua}{}^{\ub}Y_{\ub}\ ,  \qquad
\underline{L} =\underline{L}_{T} \underline{L}_{P}\ ,
\label{LYL}
\ee
where 
\be \left(\underline{L}_{T}\right)_{\ua}{}^{\ub} Y_{\ub}:=e^{-i\a^{-1}y^{r} \mathrm{Ad}_{\star }\,T_{r} }
Y_{\ua}\ ,\qquad \left(\underline{L}_{P}\right)_{\ua}{}^{\ub} Y_{\ub} := e^{i\e \ell t\mathrm{Ad}_{\star }\, P } Y_{\ua}\ .\ee
From\footnote{It also follows that $T_{r_1}\star\cdots \star T_{r_n}=T_{r_1}\cdots T_{r_n}$.}
\bea T_{r}&=&\frac{1}{8}\left(\underline{T}_{r}\right)_{\underline{\a\b}}Y^{\ua}Y^{\ub}\ ,
\qquad \underline{T}_{r}=L^a_r(-\alpha \Gamma_{ab} L^b +\beta\lambda \Gamma_{a})\ ,\qquad
\underline{T}_{r}\underline{T}_{s}=0\ ,\\
P&=&\frac{1}{8}\left(\underline{P}\right)_{\underline{\a\b}}Y^{\ua}Y^{\ub}\ ,\qquad
\underline P= \lambda L^a \Gamma_a\ ,\qquad \underline{P}^2=\e \lambda^2\underline{\mathbf 1}\ ,\eea
it follows that 
\bea
\underline{L}_{T}&=&
1+\frac{1}{2} \alpha^{-1}
y^{r}\underline{T}_{r} =1+\frac{1}{2} y^{a} ( -\Gamma_{ab}L^b
+i \gamma^{-1} \Gamma_a)\ ,\qquad y^a:= y^r L^a_r\ ,\\
\underline{L}_{P}&=&\cosh\frac{t}2 -\e\ell \underline{P}\sinh \frac{t}2=
\cosh\frac{t}2 +i\gamma L^a \Gamma_a \sinh \frac{t}2\ ,\eea
and hence
\be 
\underline L= \cosh\frac{t}2 +i\gamma L^a \Gamma_a \sinh \frac{t}2+
\frac12 e^{t/2}y^a ( -\Gamma_{ab}L^b +i\gamma^{-1}  \Gamma_a)\ .
\ee
From the definitions made in \eq{AB} and \eq{M}, it follows that 
\bse
\label{ABMplanar}
\bea 
A&=&\cosh\frac{t}2+\epsilon\gamma\eta \sinh \frac{t}2-\frac12 (1+\frac{\eta}{\gamma}) 
e^{t/2}y^a L^b \sigma_{ab}\ ,\\
B&=&i(\gamma \sinh \frac{t}2-\eta\cosh \frac{t}2) L^a \sigma_a + \frac{i}2 e^{t/2}(\gamma^{-1}-\e\eta) y^a \sigma_a\ ,
\\
M &=& A^{-1}B=
\frac{i\gamma}{\det A}\left\{ 
\left[
 \det A+
\frac{\e}2 \left( ( \eta + \gamma )^{2}e^{-t}-\eta ^{2}-\epsilon
\right)
\right] 
L^a\sigma_a 
+\frac{1}2 \left(\eta +\gamma\right) ^{2}y^{a} \sigma _{a} 
\right\}\ ,\quad
\\
\det A &=&
\left(\cosh \frac{t}{2} + \e \gamma\eta \sinh \frac{t}{2}\right)^2 
-\frac{1}{4} \left(\eta+\gamma \right)^2 y^a y_a e^t \ .
\eea
\ese
The resulting form of the scalar field $\phi $ is given by
\be
\phi =  \left. \left( \nu +\gamma\widetilde{\nu}\frac{\partial }{\partial \eta }
\right) \frac{1}{\mathrm{det}A}\right\vert _{\eta =-\gamma}
= \left( \nu +\widetilde{\nu}\right) e^{-t}-\widetilde{\nu}e^{-2t}\ , \label{phiadapt}
\ee
which in the $\mathfrak{iso}(3)$ case agrees with \eqref{scalarSolution1}.

\section{Analysis of integrability condition on \texorpdfstring{$H^{(1)}|_{Z=0}$}{H(1) |Z=0}}
\label{App:H1}

From $D^{(0)} V^{(0)}_{\alpha}(\eta)=0$ it follows that
\be \left.\left( D^{(0)} \frac1{{\cal L}_{\vec Z}} 
z^\a V^{(0)}_{\alpha}(\eta)\right)\right|_{Z=0}= \left.\left(\left[D^{(0)}, \frac1{{\cal L}_{\vec Z}} 
z^\a\right] V^{(0)}_{\alpha}(\eta)\right)\right|_{Z=0}=\frac{1}{4i} \Omega^{\underline{\a\b}}
\left.\left(\left[{\rm ad}^\star_{Y_{\ua}Y_{\ub}}, \frac1{{\cal L}_{\vec Z}} 
z^\a\right] V^{(0)}_{\alpha}(\eta)\right)\right|_{Z=0}\ ,\ee
where 
\begin{align}
\left.\left(\left[{\rm ad}^\star_{y_{\a}y_{\b}}, \frac1{{\cal L}_{\vec Z}} 
z^\gamma\right] V^{(0)}_{\gamma}(\eta)\right)\right|_{Z=0}
&=4\partial^{(y)}_{(\a}V^{(0)}_{\beta)}(\eta)|_{z=0}\ , 
\\[5pt]
\left.\left(\left[{\rm ad}^\star_{y_{\a}\bar y_{\dot\a}}, \frac1{{\cal L}_{\vec Z}} 
z^\gamma\right] V^{(0)}_{\gamma}(\eta)\right)\right|_{Z=0}&=2\partial^{(\bar y)}_{(\dot\a}V^{(0)}_{\a)}(\eta)|_{z=0}\ ,
\\[5pt]
\left.\left(\left[{\rm ad}^\star_{\bar y_{\dot\a}\bar y_{\dot\b}}, \frac1{{\cal L}_{\vec Z}} 
z^\gamma\right] V^{(0)}_{\gamma}(\eta)\right)\right|_{Z=0}&=0\ ,
\cr
\end{align}
using the holomorphicity properties of $V^{(0)}_{\alpha}(\eta)$.
Thus, adding also the contributions from the anti-holomorphic connection, we find
\be 
U^{(G,1)}|_{Z=0} = D^{(0)} \left(H^{(1)}|_{Z=0}\right)+i{\cal O}
\left.\Omega^{\underline{\a\b}} \partial_{\ua}^{(Y)} 
V^{(0)}_{\ub}(\eta)\right|_{Z=0}\ .
\ee
From $(D^{(0)})^2=0$ it follows that the singularities of the second 
term at $Y=0$ can be cancelled by the first term only if
\be D^{(0)} (U^{(G,1)}|_{Z=0})= i{\cal O}
D^{(0)} \left(\left.\Omega^{\underline{\a\b}} \partial_{\ua}^{(Y)} 
V^{(0)}_{\ub}(\eta)\right|_{Z=0}\right)\ee
is real analytic at $Y=0$, which is thus a necessary condition
for the existence of $H^{(1)}|_{Z=0}$.
To demonstrate this, we use once more $(D^{(0)})^2=0$, on the form $R^{\underline{\a\b}}:=d\Omega^{\underline{\a\b}}-\Omega^{\underline{\a\gamma}}\wedge\Omega_{\uc}{}^{\ub}=0$,
and also $D^{(0)}V^{(0)}_{\underline\b}(\eta)=0$, to compute
\begin{align} & D^{(0)}\left(\left.\Omega^{\underline{\a\b}} \partial_{\ua}^{(Y)} 
V^{(0)}_{\ub}(\eta)\right|_{Z=0}\right)\cr
&=\left(d+\Omega^{\underline{\gamma\d}} Y_{\uc}\partial^{(Y)}_{\ud}\right)
\left(\left.\Omega^{\underline{\a\b}} \partial_{\ua}^{(Y)} 
V^{(0)}_{\ub}(\eta)\right|_{Z=0}\right)\cr
& =\left.d\Omega^{\underline{\a\b}} \partial_{\ua}^{(Y)} V^{(0)}_{\ub}(\eta)\right|_{Z=0}
+\Omega^{\underline{\gamma\d}} \wedge\Omega^{\underline{\a\b}} \left[Y_{\uc}\partial_{\ud}^{(Y)},\partial_{\ua}^{(Y)} \right]
\left.V^{(0)}_{\ub}(\eta)\right|_{Z=0}-\Omega^{\underline{\a\b}}\wedge \partial_{\ua}^{(Y)} 
D^{(0)}\left(\left.V^{(0)}_{\ub}(\eta)\right|_{Z=0}\right)\cr
& =\left.R^{\underline{\a\b}} \partial_{\ua}^{(Y)} V^{(0)}_{\ub}(\eta)\right|_{Z=0}
-\Omega^{\underline{\a\b}} \partial_{\ua}^{(Y)}\left. \left(
D^{(0)}V^{(0)}_{\ub}(\eta)-i\Omega^{\underline{\gamma\d}}\wedge \partial_{\uc}^{(Y)} \partial_{\ud}^{(Z)}
V^{(0)}_{\ub} \right)\right|_{Z=0}\cr
&=i\Omega^{\underline{\a\b}}\wedge \Omega^{\underline{\gamma\d}} \partial_{\ua}^{(Y)} 
\left.\partial_{\uc}^{(Y)} \left(\partial_{[\ud}^{(Z)}
V^{(0)}_{\ub]}\right)\right|_{Z=0}\ .
\end{align}
Using also ${\cal O}V_{\ua}^{(0)}=A_{\ua}^{(L,1)}$ (see Eq. \eq{scodd}), we arrive at
\be D^{(0)} (U^{(G,1)}|_{Z=0})=
-\Omega^{\underline{\a\b}}\wedge\Omega^{\underline{\gamma\d}} \partial_{\ua}^{(Y)} 
\left.\partial_{\uc}^{(Y)} \left(\partial_{[\ud}^{(Z)}
A^{(L,1)}_{\ub]}\right)\right|_{Z=0}\ ,\ee
where $\partial_{[\ud}^{(Z)}
A^{(L,1)}_{\ub]}$ is a linear combination of  
$\epsilon_{\a\b}\Phi^{(L,1)}\star\kappa$ and its Hermitian
conjugate, which are real analytic in $Y$ space.
In the space of forms $f(x,dx,Y)$ in $X$-space 
that are functions of $Y$, the background
exterior derivative $D^{(0)}$ commutes to
the Euler derivative
\be \vec E_Y := Y^{\ua} \vec \partial_{\ua}^{(Y)}\ ,\ee
\emph{viz.} $(\vec E_Y D^{(0)}-D^{(0)} \vec E_Y) f(x,dx,Y)=0$.
Moreover, the spectrum of $\vec E_Y$ in this space 
is given by $\{0,1,2,\dots\}$.
Thus, letting
\be P^{(1)}:=  i {\cal O} \Omega^{\underline{\a\b}} \partial_{\ua}^{(Y)} V^{(0)}_{\ub}|_{Z=0}\ ,\ee
we have
\be (\vec E_Y + 2 ) P^{(1)} = 0\ ,\qquad  
(\vec E_Y + 2 ) D^{(0)} P^{(1)} = 0\ ,\ee
which together with the already established real analyticity of 
$D^{(0)} P^{(1)}$ in $Y$ space implies that
\be D^{(0)} P^{(1)} = 0\ ,\ee
as can also be seen directly using the fact that 
$\partial_{[\ud}^{(Z)} A^{(L,1)}_{\ub]}|_{Z=0}$ is 
a linear function of $Y$, hence annihilated
by $\Omega^{\underline{\a\b}}\Omega^{\underline{\gamma\d}} \partial_{\ua}^{(Y)} 
\partial_{\uc}^{(Y)}$.
Finally, conjugation by $L$ yields 
\be 
d(L\star P^{(1)}\star L^{-1})=0\quad \Rightarrow\quad
L\star P^{(1)}\star L^{-1} = dQ^{(1)}\ ,
\ee
that is
\be P^{(1)} = L^{-1} \star dQ^{(1)}\star L = D^{(0)} (L^{-1}\star Q^{(1)}  \star L)\ .\ee
Hence, choosing
\be H^{(1)} =  -  L^{-1}\star Q^{(1)}  \star L\ ,\ee
we arrive at $U^{(G,1)}|_{Z=0}= 0$, which is the desired result
in view of the fact that the Weyl zero-form
consists of scalar modes.
As for the explicit form of $H^{(1)}$ we refer to
a more general analysis including solutions with 
four and two Killing symmetries to appear elsewhere.

\section{Lemmas}
\label{App:Lemmas}

\subsection{A Twistor space distribution}

At the first order in the $\nu$-expansion of $a_\a$ given above,  one encounters the integral (see \eq{aa2})
\be
I^\pm(z) \ := \ 2z^\pm\int_{-1}^1 \frac{d\tau}{(\tau +1)^2}\, 
\exp\left(i\frac{\tau-1}{\tau +1}z^+z^-\right)\ .
\label{I}
\ee
Using the delta sequence 
\be \lim_{\e\rightarrow 0^+} e^{-\frac{i}{\e}z^+z^-}=0\ ,\ee
one finds
\be 
I^\pm(z)\ = \ \frac{1}{iz^\mp}\ .\label{adist}\ee
The linearized equations of motion require
\be
\partial_\pm I^\pm=\kappa_z\ .
\label{propI}
\ee
In order to differentiate $I^\pm(z)$, we must first rewrite it as a distribution that is differentiable at $z^\mp=0$, for which we use 
\be
\partial_\pm I^\pm=
\partial_\pm \left(\int_0^{z^\pm} dz^{\prime \pm}\lim_{\e\to 0^+}\frac{1}{\e}e^{-\frac{i}{\e}z^{\prime \pm}z^\mp}\right) =
2\pi\partial_\pm\left(\int_0^{z^\pm} dz^{\prime \pm}\d(z^{\prime \pm})\d(z^\mp)\right) \ = \ 2\pi\delta(z^\pm)\d(z^\mp)\ .
\ee

\subsection{Fusion rules}

Denoting the generators of the complexified Weyl algebra ${\cal W}$ by $(I,u,v)$, where $I$ is central and $[u,v]_\star=I$, we factor out the ideal generated by $I-\hbar {\rm Id}_{\cal W}$, and set $\hbar=1$, leading to a graded associative algebra degree map given by the monomial degree and $\mathfrak{osp}(1|2)$ subalgebra $(u,v;u\star u,\frac12\{u,v\}_\star,v\star v)$.
Letting 
\be w:=\frac12\{u,v\}_\star\ ,\qquad  g_\xi:=\exp_\star (\xi w)\ ,\qquad
\xi\in \Comp\ ,\ee
one has $g_{\xi}\star g_{\xi'}=g_{\xi+\xi'}$.
Going to Weyl order, one finds the symbols
\be
g_\xi=\frac1{\cosh \xi}\exp[\tanh (\xi w)]\ , 
\ee
which are real analytic except for $\xi \in (\mathbb Z +\frac12)\pi i$
in which case they are phase space delta functions defined using delta 
sequences.
It follows that 
\be 
E_\eta\star E_{\eta'}=\frac1{1+\eta\eta'}E_{\frac{\eta+\eta'}{1+\eta\eta'}}\ ,
\qquad E_\eta:=\exp(-2\eta w)\ ,
\ee
whose star product we extend to all values of $\eta$ 
using the closed contour regularization scheme defined 
in Section \ref{regs}.
In particular, for $\eta=\pm 1$ we recover the Fock space and anti-Fock space ground state projectors $P_\s:=2E_{\s}$, $\s=\pm 1$, thus 
obeying $u\star P_+=v\star P_-=0$ and 
\be (P_\sigma\star P_{\sigma'})|_{\rm reg}=\delta_{\sigma,\sigma'}P_\sigma\ .\ee
The $\mathfrak{g}_6$-invariant solutions make use of the elements
$E_{\pm i}$, which thus obey
\be (E_{\sigma i}\star E_{\sigma' i})|_{\rm reg}=
\frac12 \delta_{\sigma,-\sigma'}\ ,\ee
that is, they close on the identity.
We note that $E_{\pm i}=\frac12 g_{\pm \pi/4}$, that is, 
the regularization amounts to discarding the 
non-real analytic group elements $g_{\pm \pi/2}$.
We also remark that $E_{\eta}$ gives rise to an 
${\rm Env}(\mathfrak{osp}(1|2))$ orbit obtained
by left- and right-action by polynomial elements
in $\mathfrak{osp}(1|2)$, that is, by $u$ and $v$;
for $\eta=\pm1$ these are simply the algebras of 
endomorphisms of the Fock and anti-Fock spaces.
Taking instead $\eta=\pm i$ and restricting to 
even elements, the resulting ${\rm Env}(\mathfrak{sp}(1|2))$ 
orbits of $E_{\pm i}$ are of use in considering 
fluctuations around the $\mathfrak{g}_6$-invariant solutions;
see Section \ref{SecPertExact} for an outline.


\end{appendix}

\newpage

\bibliography{biblio}

\end{document}